\documentclass[prd,twocolumn,tightenlines,showpacs,nofootinbib,superscriptaddress]{revtex4-1}%draft
\usepackage{bm,dcolumn,amsmath,graphicx,amsfonts,amssymb}

\usepackage{hyperref} 
\usepackage{xcolor}
\definecolor{cite}{rgb}{0.,0.,0.5}   % more subtle than the default blue
\hypersetup{ 
	colorlinks,
	linkcolor={cite},
	citecolor={cite},
	urlcolor={cite}
}

\newcommand{\abs}[1]{\ensuremath{\left |#1\right |}}
\newcommand{\bra}[1]{\ensuremath{\langle #1|}}	%Dirac Bras
\newcommand{\ket}[1]{\ensuremath{|#1\rangle}}	%Dirac Kets
\newcommand{\braket}[1]{\ensuremath{\langle #1\rangle}}	%Dirac BraKets
\newcommand{\Exp}[1]{\ensuremath{e^{#1}}}
\newcommand{\threej}[6]{\ensuremath{\begin{pmatrix}#1&#2&#3\\#4&#5&#6\end{pmatrix}}}	% 3-j symbol
\renewcommand{\v}[1]{\ensuremath{\boldsymbol{#1}}}		%bold-math for vectors
\newcommand{\E}[1]{\ensuremath{\times10^{#1}}}	%exponent:    A x 10^y
\newcommand{\twocomp}[2]{\ensuremath{\begin{pmatrix}#1\\#2\end{pmatrix}}}	% 2-component wavefunction

\renewcommand{\d}{\ensuremath{{\,\rm d}}}	

\newcommand{\en}{\ensuremath{\mathcal{E}}}  %single-particle energies!
\newcommand{\g}{\ensuremath{\gamma}} %for dirac/pauli matrices. 
\renewcommand{\a}{\ensuremath{\alpha}} 
\newcommand{\s}{\ensuremath{\sigma}} 
\renewcommand{\k}{\ensuremath{\kappa}} 
 
\newcommand{\e}{\ensuremath{\varepsilon}} 

\newcommand{\un}[1]{\ensuremath{\,{\rm{#1}}}}

\definecolor{orange}{rgb}{1,0.2,0}   % 
\begin{document}

\title{%
Dark matter scattering on electrons:\\
Accurate calculations of atomic excitations and implications for the DAMA signal}

\author{B. M. Roberts} \email[]{benjaminroberts@unr.edu}
%	\affiliation{School of Physics, University of New South Wales, Sydney, New South Wales 2052, Australia} % Old address
	\affiliation{Department of Physics, University of Nevada, Reno, 89557, USA} %Current address
\author{V. A. Dzuba}   %\email[]{dzuba@phys.unsw.edu.au}
	\affiliation{School of Physics, University of New South Wales, Sydney, New South Wales 2052, Australia}
\author{V. V. Flambaum} %\email[]{v.flambaum@unsw.edu.au}
	\affiliation{School of Physics, University of New South Wales, Sydney, New South Wales 2052, Australia}
	\affiliation{Mainz Institute for Theoretical Physics, Johannes Gutenberg University Mainz, D 55122 Mainz, Germany}
\author{M. Pospelov}
	\affiliation{Department of Physics and Astronomy, University of Victoria, Victoria, British Columbia V8P 5C2, Canada}
	\affiliation{Perimeter Institute for Theoretical Physics, Waterloo, Ontario N2J 2W9, Canada}
\author{Y. V. Stadnik}
	\affiliation{School of Physics, University of New South Wales, Sydney, New South Wales 2052, Australia}

\date{ \today }

%-------------------------------------------------------------------------
\begin{abstract}
We revisit the WIMP-type dark matter scattering on electrons that results in 
atomic ionization, and can manifest itself in a variety of 
existing direct-detection experiments. 
Unlike the WIMP-nucleon scattering, where current 
experiments probe typical interaction strengths much smaller than the Fermi constant, the scattering on electrons 
requires a much stronger interaction to be detectable, which in turn requires new light force carriers. We account for such new forces 
explicitly, by introducing a mediator particle with scalar or vector couplings to dark matter and to electrons. 
We then perform state of the art numerical calculations of atomic ionization relevant to the existing experiments. 
Our goals are to consistently take into account the atomic physics aspect of the problem (e.g., 
the relativistic effects, which can be quite significant), and to 
scan the parameter space: the dark matter mass, the mediator mass, and the effective coupling strength, to see 
if there is any part of the parameter space that could potentially explain the DAMA  modulation signal. 
While we find that the modulation fraction of all events with energy deposition above $2$~keV in NaI can be quite significant,
reaching $\sim 50$\%, the relevant parts of the parameter space are excluded by the XENON10 and XENON100 experiments.

\end{abstract}


\maketitle

%31.15.A-	Ab initio calculations
%31.15.am	Relativistic configuration interaction (CI) and many-body perturbation calculations
%95.35.+d	Dark matter (stellar, interstellar, galactic, and cosmological)
%95.30.Cq	Elementary particle processes
%95.36.+x    Dark energy
%34.80.Dp	Atomic excitation and ionization

%34.10.+x	General theories and models of atomic and molecular collisions and interactions (including statistical theories, transition state, stochastic and trajectory models, etc.)
%34.80.Dp	Atomic excitation and ionization
%34.80.Gs	Molecular excitation and ionization
%31.15.A-	Ab initio calculations
%31.15.am	Relativistic configuration interaction (CI) and many-body perturbation calculations
%95.35.+d	Dark matter (stellar, interstellar, galactic, and cosmological)
%95.30.Cq	Elementary particle processes
%95.36.+x    Dark energy
%34.80.Dp	Atomic excitation and ionization

%~\\[0.15cm]{\tiny *Table of contents just for editing purposes}\\[-1.75cm]
%{\scriptsize
%\tableofcontents
%~\\
%}

%---------------------------------------------------------
\section{Introduction}

The evidence for the existence of dark matter (DM) is overwhelming, extending over many orders of distance scales, from the Universe's horizon down to the scale of dwarf galaxies.
This realization drives a comprehensive scientific effort to uncover the nature of DM and its connection to a relatively well-understood 
world of subatomic particles. 
The most sustained effort to date is the search for so-called weakly interacting massive particles (WIMPs)
through their possible non-gravitational interactions with matter fields of the Standard Model (SM)~\cite{Cushman2013}. 
Despite this, no conclusive terrestrial observation of DM has yet been reported, and its 
identity remains one of the most important outstanding problems facing physics today.

One intriguing claim of 
potential detection was made by the DAMA Collaboration, which uses a NaI-based scintillation detector to search for possible DM interactions within the crystal in the underground laboratory at the Gran Sasso National Laboratory, INFN, Italy \cite{Bernabei2013} (see also \cite{Bernabei2008b,Bernabei2008,*Bernabei2010,Bernabei2013a,Bernabei2014,Bernabei2015},
and references therein).
The data from the combined DAMA/LIBRA and DAMA/NaI experiments (to which we will collectively refer as DAMA for concision) indicates an annual modulation in the event rate at around 3 keV electron-equivalent energy deposition with a 9.3$\sigma$ significance  (the low-energy threshold for DAMA is $\sim2$ keV) \cite{Bernabei2008,*Bernabei2010,Bernabei2013}.
The phase of this modulation agrees very well with the assumption that the signal is due to the scattering of DM particles (e.g., WIMPs) present in the DM galactic halo.
The annual modulation is one of the key expected observables for WIMP dark matter detection and is expected due to the motion of the earth around the sun which results in an annual variation of the DM flux (and incident energy) through a detector; see, e.g., Refs.~\cite{Freese2013,Lee2015}.
The DAMA result stands as the only enduring DM direct-detection claim to date.

There are, however, several reasons to doubt that the signal observed by the DAMA Collaboration is due to WIMPs.  Null 
results from several other sources such as, for example, the 
XENON100 \cite{Aprile2012},
LUX \cite{Akerib2014,*Akerib2015}, 
and SuperCDMS \cite{AgneseSCDMS2014} experiments, all but rule out 
the possibility that the DAMA signal is due to an elastic WIMP--nucleus interaction (see also Refs.~\cite{Freese2013,Farina2011,Savage2009}). 
For example, for the 10 GeV WIMP spin-independent scattering on nucleons, the LUX exclusion limits extend four orders of magnitude below the parameter
space favoured by the DAMA signal. 
A very limited possibility may remain for the inelastic DM scattering (scattering with an excitation of a close in mass excited WIMP state)
due to a magnetic moment operator~\cite{Chang2010,Barello2014}. The upcoming new experiments may close this remaining loophole soon. 

An alternative route, DM interaction with electrons, has been long entertained as a possible cause of the DAMA modulation signal. 
While the DAMA experiment is sensitive to scattering of DM particles off both electrons and nuclei, most other DM detection experiments reject pure electron events in order to search for nuclear recoils with as little background as possible.
The suggestions that an absorption of a few keV mass axion-like particles \cite{BernabeiSPS2006} as the origin of the DAMA signal had been promoted by the collaboration itself \cite{Bernabei2008}. 
Unfortunately, this idea was based on erroneous calculations, and subsequent work \cite{Pospelov2008} has shown that the absorption signal quite generally leads to $\sigma_{\rm abs} v_{\rm DM} \propto {\rm const}$, and therefore is not modulated (see also Ref.~\cite{An2014}). 

The remaining option, WIMP-electron scattering, 
 could potentially explain the DAMA modulation without being ruled out 
by the other null results (see, e.g., Refs.~\cite{Savage2009,Gondolo2005}).
This possibility has been investigated previously in the literature---see, e.g., Refs.~\cite{Bernabei2008a,Kopp2009,Feldstein2010,Dedes2010,Foot2014,Lee2015}.
It is also possible (see, e.g.,~Refs.~\cite{Fox2009,Cao2009,Ibarra2009,Bell2014}) that leptonically interacting 
WIMP models may be behind other anomalies in the indirect detection, such as select results from 
AMS~\cite{Aguilar2013},
ATIC~\cite{Chang2008},
Fermi~\cite{Ackermann2012}, 
and 
PAMELA~\cite{Adriani2009,*Adriani2013} 
experiments. Gradually, the scattering of WIMPs on electrons became the topic of increased 
interest, and it is viewed as a new opportunity for extending the existing and future direct detection experiments 
and technologies to DM masses well below the GeV scale 
\cite{Essig2012,EssigPRL2012,Graham2012,Essig2015,Hochberg2016,*Hochberg2015,Lee2015}.

The most common signature of elastic scattering of WIMPs on electrons is atomic excitation or ionization. 
The latter is basis for the DM detection in many existing experiments. One thing to keep in mind is that while massive DM 
($m_{\rm DM} \gg m_e$) has enough energy to ionize the atom, its velocity is quite low, leading to a 
rather insignificant momentum exchange between 
DM and electrons. The emergent ionization electron moves with velocities $v_e \sim (2 \Delta E / m_e)^{1/2}$ 
that can be quite high for energy deposition $\Delta E$ of few~keV.  
Consequently, the scattering probes deep inside the bound state wave function, with main 
distances at play being much smaller than the characteristic Bohr radius of an atom. 
In such a situation the neglect of relativistic effects can lead to large errors in the predicted ionization cross sections~\cite{RobertsAdiabatic2016}.

Another important aspect of WIMP-electron scattering is the need for new forces that mediate such an interaction. 
Indeed, if the interaction between some DM state $\chi$ and light SM fermion fields $\psi$ (e.g., electron or quark) 
is parametrized by a contact operator, e.g., $G_{\chi\psi}(\bar\chi \gamma_\mu \chi)(\bar\psi \gamma_\mu \psi)$,
one can immediately discover that scattering of WIMPs on quarks probes interaction strengths $G_{\chi q} \ll G_F$, while 
scattering on electrons is less sensitive, $G_{\chi e} \gg G_F$. Consequently, the scattering on electrons require some additional 
new physics {\em below} the weak scale. At these scales many additional particle physics probes of such mediators exist \cite{EssigCSS2013}, 
and it is highly desirable to put the mediator force into the model explicitly
(as it is done in many recent works \cite{Essig2012,Hochberg2016,*Hochberg2015,Lee2015}), rather than staying at the level of effective operators. 

In this paper, we pursue two main goals. The first goal is to consider DM-electron scattering with $O(10~{\rm eV} - 10~{\rm keV})$ 
energy deposition and include relevant atomic physics effects for calculating the ionization rate. The second goal is to investigate in some detail the 
DAMA modulation signal and compare it with constraints on electron recoil imposed by other experiments.
Going away from a simple parametrization by a constant cross section $\sigma_{\chi e}$ or 
by a contact operator, we shall scan the parameter space of the DM mass, mediator mass, and the coupling constant to determine ``the region of interest'' (ROI) that could be consistent with the DAMA signal. We then calculate ionization signal predicted for the  the same ROI for  other DM experiments that currently see null results. 

Reviewing the literature on WIMP-electron scattering, we notice that one of the original papers, Ref.~\cite{Kopp2009}, employs a 
contact operator approach to write down effective interactions between DM particles and electrons, 
and deals with simplified atomic physics. This work also makes a connection between tree-level 
$\chi-e$ interactions and loop induced $\chi-N$ interactions, which helps to set new constraints on
some forms of interactions using nuclear recoil type measurements \cite{Aprile2012,Akerib2014,AgneseSCDMS2014}.
Another notable series of works have used the ``ionization only'' signal of XENON10 to set constraints on 
DM scattering with relatively low energy deposition $\Delta E$, and as a consequence record low 
DM masses \cite{Essig2012,EssigPRL2012}. These papers also use non-relativistic treatment of electrons.  
Finally, recent analyses of data from the XENON100 experiment have also investigated WIMP-induced electron-recoil events~\cite{AprileAxion2014,XENONcollab2015,XENONcollab2015a}.
This experiment has also observed modest evidence for an annual modulation (at the $2.8\sigma$ level).
However, based on their analysis of the average unmodulated event-rate, the contact WIMP-electron interaction 
via an axial vector coupling was excluded as an explanation for the DAMA result at the $4.4\s$ level \cite{XENONcollab2015}.
By assuming the DAMA result was due to an pseudovector coupling, and using the theoretical analysis from Ref.~\cite{Kopp2009}, the corresponding modulation amplitude that would be expected in the XENON100 experiment was calculated in Ref.~\cite{XENONcollab2015a}. 
The observed amplitude was smaller than this by a factor of a few, and it was concluded that the XENON100 results were inconsistent with the DAMA results at the $4.8\s$ level \cite{XENONcollab2015a}.
We note, however, that this analysis needs to be repeated for other types of interactions, and a 
rigorous relativistic analysis of atomic structure effects is desirable.
We also note that there is no {\em a priori} reason to believe that the fraction of the modulated signal should be small or proportional to the fractional annual change in the DM velocity distribution.
In fact, the scattering amplitude is very highly dependent on the values of momentum transfer involved, which depend on the velocity of the DM particles.
As we shall show, electron relativistic effects must be taken into account properly to recover the correct momentum-transfer dependence of the cross section, which is a significant point.

\begin{figure*}
	\begin{center}
		\includegraphics[width=0.49\textwidth]{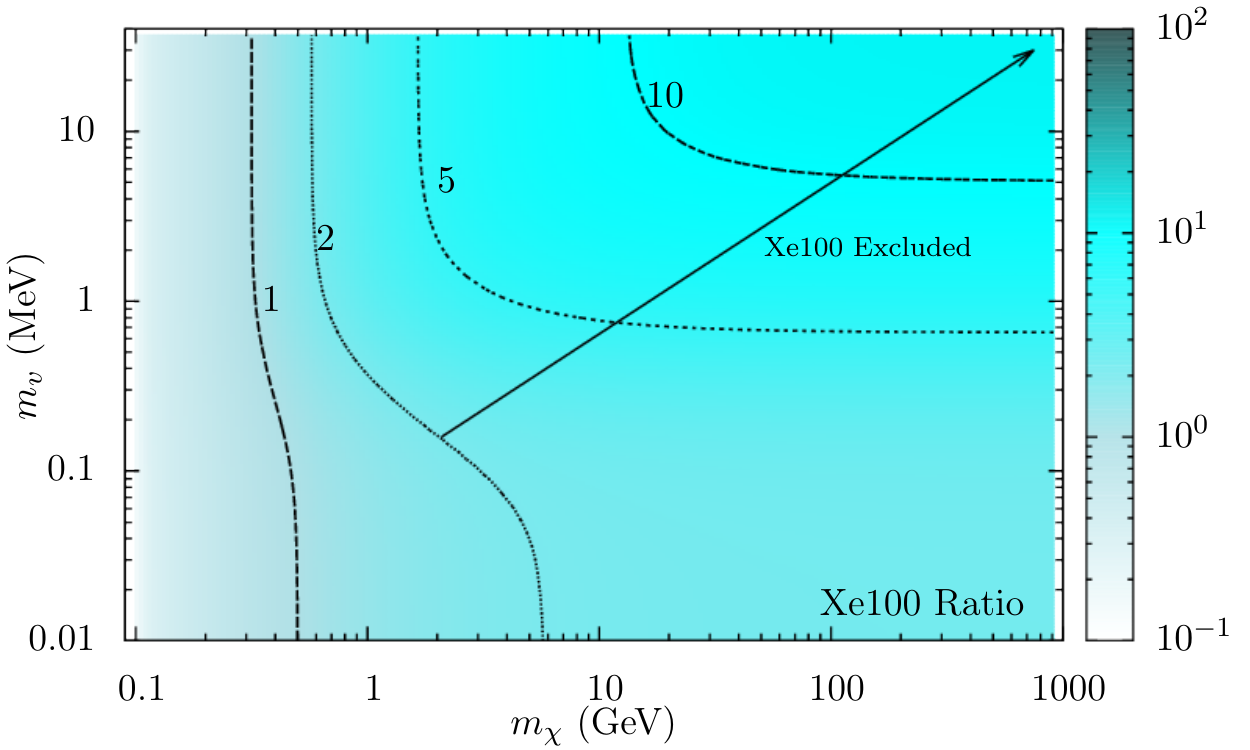}
		\includegraphics[width=0.49\textwidth]{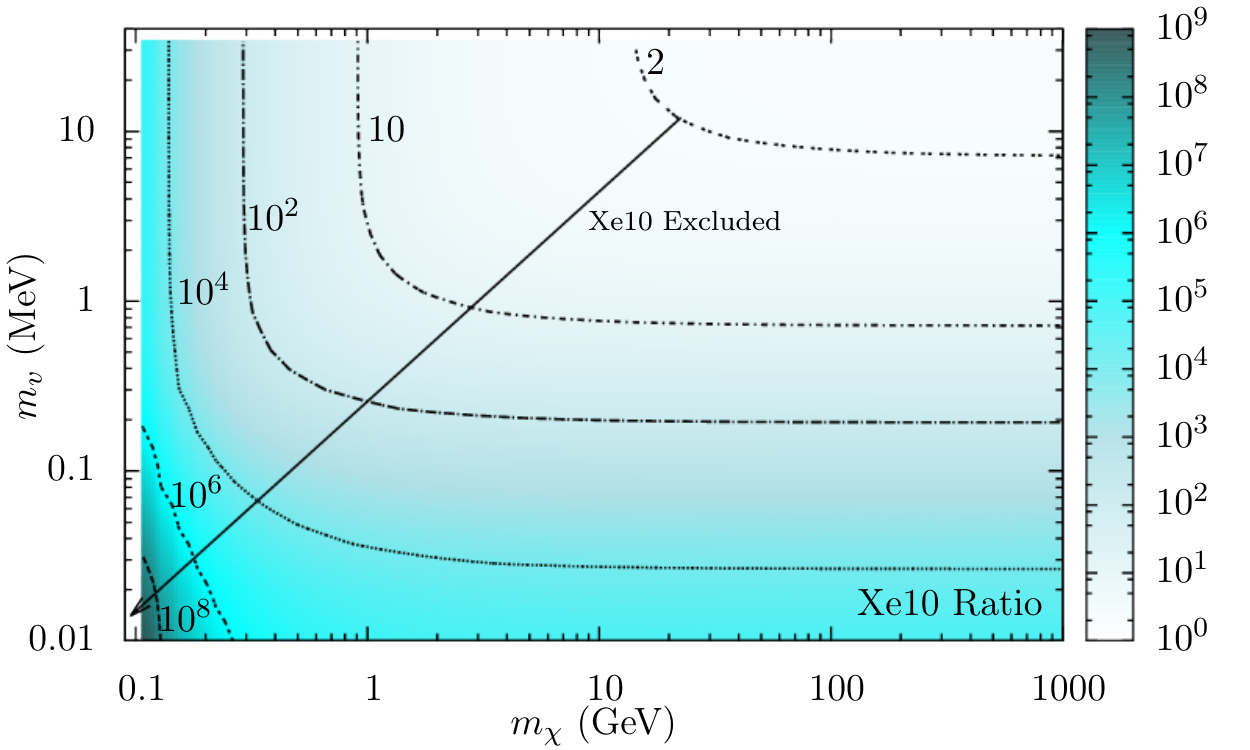}
		\caption{
Summary of results: Ratio of the expected signal for XENON100 (search of electron recoil with $\Delta E > 1$ keV following  Ref.~\cite{XENONcollab2015,XENONcollab2015a}, {\em left}) and XENON10 (limit on ionization with arbitrary energy deposit \cite{Angle2011}, {\em right})---assuming DAMA to be a positive WIMP detection---to the 90\% confidence-level limits from those respective experiments, as a function of the DM mass, $m_\chi$, and the mediator mass, $m_v$. The contours denote the level of exclusion; for example, the line marked `2' contours the part of the parameter space for which the expected signal in XENON100/10 is two times larger than the signal that has been ruled out (90\% c.l.). By combining both plots, it is seen that all regions of the parameter space are excluded.
}
		\label{fig:summary}
	\end{center}
\end{figure*}

A rigorous {\em ab initio} relativistic treatment of the atomic structure has not yet been implemented in the existing literature, and---as was demonstrated in Ref.~\cite{RobertsAdiabatic2016} (and confirmed in this work)---is crucial.
 Ref.~\cite{RobertsAdiabatic2016} demonstrated that the ionization cross section due to a WIMP--electron interaction when (several) keV of energy is exchanged is actually dominated by relativistic effects.
This is due to the non-analytic cusp-like behavior of the Coulomb-like wave functions at very small radial distances, and the significant difference in the small-distance radial dependence of the Dirac wave functions compared to the Schr\"odinger wave functions.
The implication is that the suppression from the electron matrix elements may not be as strong as previously assumed, meaning the nonrelativistic calculations may significantly underestimate the cross section. 
Furthermore, as several new experiments designed to test the DAMA results are currently under way \cite{Bernabei2015,Amare2015b,*Amare2015a,*Amare2015,Xu2015,*Froborg2016,XMASS2015,Angloher2016,BarbosadeSouza2016}, it is crucial that the relevant theory required for their interpretation is correct, and adjustable to the choice of mediator and interactions.

In this paper, we employ the relativistic Hartree-Fock method to calculate model-independent cross sections and event rates for the atomic and molecular ionization induced by the interaction of atomic electrons with DM for several systems of experimental interest.
Atomic ionization has been considered previously for the case of absorption of light particles such as massive axions 
\cite{Derevianko2010,*DzubaPRD2010}.
By performing the atomic structure calculations in an {\em ab initio} manner including all relativistic effects we are able to 
minimize errors from the atomic structure to the point of their irrelevance -- as more important sources of errors are now associated with other factors, such as the DM velocity distribution.
Our current analysis is performed for the scalar and vector interactions, but can be easily generalized to other types of interactions. 
It is our hope that as well as providing a direct analysis, our calculations may be useful to others who can insert our calculations of the electron structure part of the cross section (the ``atomic kernel'') into the cross section for various DM models including other galactic density and velocity distributions. 
It is in this sense that our calculations are model independent.

By assuming the DAMA modulation is due to electron-interacting WIMPs, we calculate the event rates that would be expected in the XENON100 \cite{XENONcollab2015,XENONcollab2015a} and XENON10 \cite{Angle2011} experiments, and compare the results to the limits set by those experiments.
(The XENON10 experiment places a limit on ionization with an arbitrary energy 
deposition; XENON100 searches for electron recoil with $\Delta E \gtrsim 1$ keV.)
The details of the calculations are presented in the coming sections, but for convenience, our results are summarized in Fig.~\ref{fig:summary}. We conclude that there is no region of the parameter space for which the electron-interacting WIMP hypothesis remains a viable explanation for the DAMA modulation.

In Sec.~\ref{sec:theory}, we derive the scattering cross sections and other relevant quantities, and discuss how the approach taken in this paper differs from the previous investigations into this matter.
In Sec.~\ref{sec:calculations}, we outline the techniques we utilize for the accurate relativistic atomic calculations, and we go on in Sec.~\ref{sec:DAMA} to present our results and to discuss the implications of these for the interpretation of the DAMA  annual modulation in terms of DM interactions with atomic electrons.
Finally, in Sec.~\ref{sec:concl} we present our summary and conclusions.

%\newpage
%===================================
\section{Theory}
\label{sec:theory}

%----------------------------------------------------------------------------------------------
\subsection{Scattering cross-section and event rate}

\begin{figure}%[h]
	\begin{center}
		\includegraphics[width=0.4\textwidth]{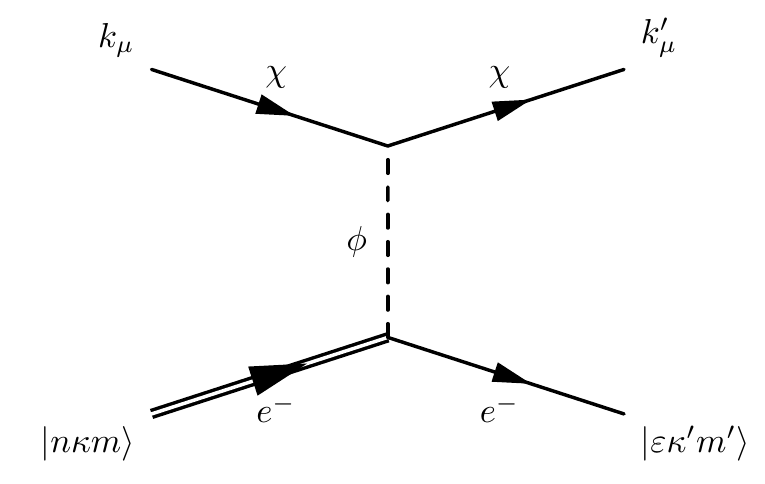}
		\caption{Example diagram for interaction of a dark matter particle ($\chi$), with an electron via the exchange of mediator $\phi$. Double line denotes a bound atomic electron ($E_{n\k}<0$), and the single electron line denotes a continuum state electron with energy $\varepsilon>0$.}
		\label{fig:DM-e-diagram}
	\end{center}
\end{figure}

We consider the case in which the scattering of an incident DM particle from the galactic halo off the atomic electrons leads to the ionization of the atom or molecule. 
This situation is shown diagrammatically in Fig.~\ref{fig:DM-e-diagram}. Therefore, for most cases, such process would correspond to the energy deposition of $O(1-10)$ eV or higher. Many of the existing detectors have a substantially higher thresholds, with typical energy depositions of interest being at the level of $O({\rm keV})$.

First, we assume the well-motivated case that the DM particle interacts with electrons via the exchange of a vector-boson mediator with 
 a mass $m_v$.
For a Dirac-type fermionic DM, the amplitude for such process can be parametrized as
\begin{equation}
-\bar \chi \gamma_{\mu} \chi 
	\frac{\sqrt{4 \pi \alpha_\chi}}{q_\rho q^\rho-m_v^2}
\bar e \gamma^\mu e 
\approx
\bar \chi \gamma_{\mu} \chi 
	\frac{\sqrt{4 \pi \alpha_\chi}}{\v{q}^2+m_v^2}
\bar e \gamma^\mu e ,
\end{equation}
where $q_\mu$ is the momentum transfer, which for the most part satisfies the $|q^0| \ll |\v{q}|$ condition. The coupling constant 
of electrons and DM fermions $\chi$ is denoted by $\alpha_\chi$. 
In non-relativistic approximation, this four-fermion amplitude corresponds to an effective Yukawa-type potential,
\begin{equation}
\hat h_{\rm int}=\alpha_\chi\frac{\Exp{-m_v r}}{r},
\label{eq:yukawa}
\end{equation}
where $r$ is the distance between the electron and DM particle. The same form of the non-relativistic 
Hamiltonian will result from the scalar exchange interaction. 

We then apply the Born approximation to write down the cross section on the whole atom \cite{RobertsAdiabatic2016}.
In the limit that $m_v\to 0$, the interaction (\ref{eq:yukawa}) corresponds to a Coulomb-like interaction, and in the opposite limit ($m_v\to \infty$) it corresponds to a purely contact (delta-function--type) interaction.
Treating the DM nonrelativistically, the partial differential cross section corresponding to the ejection of a bound electron initially in the state $a$ to a state in the continuum is given by
\begin{equation}
\d \s_a = 
		\frac{8\pi \alpha_\chi^2}{v_{\chi}^2}	
		\int_{q_-}^{q_+} 
		\frac{q\d q }{(q^2+m_v^2c^2)^2} 
		\abs{\bra{\e}e^{i \v{q}\cdot\v{r}}\ket{a}}^2 \frac{p^2\d p\d \Omega_p}{(2\pi)^3},
\end{equation}
where $q = |{\bf q}|$, 
$q_\pm = k\pm\sqrt{k^2-2m_\chi\Delta E}$, \ket{\e} is an atomic state in the continuum with energy $\e \simeq p^2/2m_e$, the state \ket{a} is a bound atomic state, and $\Omega_p$ denotes the momentum-space angular variables for the outgoing electron.
The total energy deposition, $\Delta E \equiv \e-E_a$, is related to the change in energy of the DM particle and to the energy of the ejected electron:
\begin{equation}
\Delta E = \frac{k^2-k'^2}{2m_\chi} = I_a+\e,
\label{eq:dE}
\end{equation}
where $I_a$ is the ionization potential for the state \ket{a}. One point to note is that even though we can drop the relativistic 
corrections on top of the interaction (\ref{eq:yukawa}), it is important to keep relatvistic form for the initial and final wave functions.

\begin{figure}%[h!]
	\begin{center}
		\includegraphics[width=0.49\textwidth]{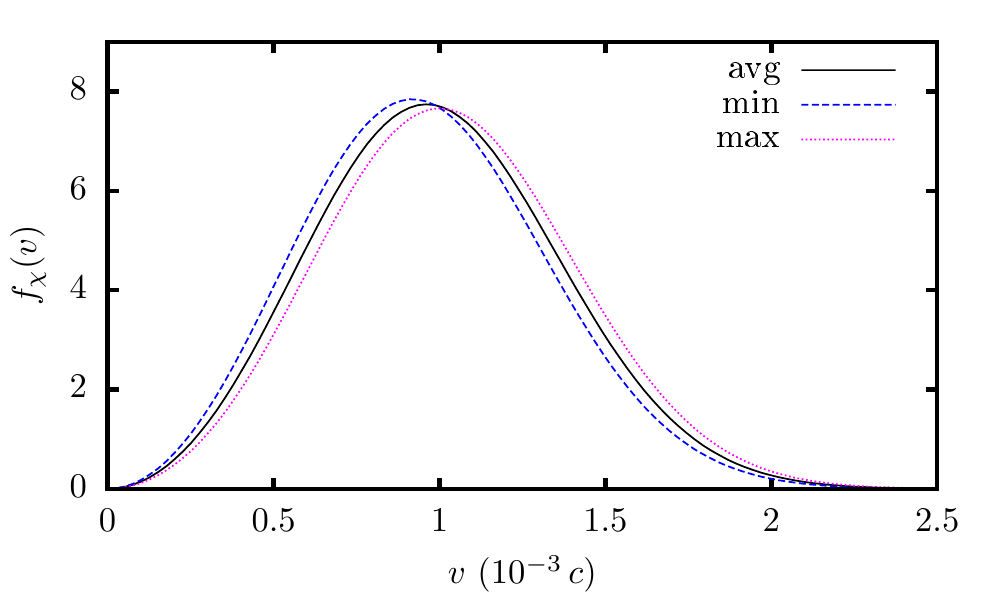}
		\caption{Normalized distributions for the DM velocity in the earth frame [see Eq.~(\ref{eq:veldistro})].
The solid black line (avg) corresponds to the DM velocity distribution in the solar frame, and the dotted blue (min) and dashed magenta (max) lines refer to the distributions in the earth frame around December 2 and June 2, respectively.}
		\label{fig:velocity}
	\end{center}
\end{figure}

The event rate is proportional to the function ${\sigma v_\chi}$, which must be averaged over the distribution for the DM particle velocity $v_\chi$:
\begin{equation}
\braket{\sigma v_\chi} = \int_0^\infty f_{\chi}(v) \,\sigma\, v \d v
\end{equation}
(note that the cross section $\sigma$ itself depends strongly on $v_\chi$, since this sets the incident energy of the DM particles).
We take the velocity distribution to be pseudo-Maxwellian (see, e.g., Ref.~\cite{Freese2013}):
\begin{equation}
f_{\chi}(v) \propto v^2 \int_{-1}^1 \exp\left[{-\frac{3(\v{v}+\v{v}_e)^2}{2v_{\rm rms}^2}}\right]\d(\cos\g)\, \Theta,
\label{eq:veldistro}
\end{equation}
where,  
$v_{\rm rms}$ is the root-mean-square (rms) velocity of the DM particles in the galactic frame, 
$\g$ is the angle between $\v{v}$ and $\v{v}_e$,
and
$\v{v}_e$ is the velocity of the earth in the galactic frame:
\begin{equation}
v_e^2\simeq v_\odot^2+v_\oplus^2+2v_\odot v_\oplus\cos \beta \cos(\omega t),
\label{eq:vearth}
\end{equation}
where
$v_\odot$ is the speed of the sun in the galactic frame,
$v_\oplus$ is the orbital speed of the earth in the solar frame,
and $\beta\approx60^\circ$ is the inclination of the earth's orbit relative to the galactic plane.
A more precise modelling of the  earth's motion through the galactic halo can be used if needed.  
Time $t=0$ is when the velocities of the earth and sun add maximally in the galactic frame (corresponding to around June 2), and $\omega=\frac{2\pi}{T}$ with $T\sim1\,\rm{yr}$.
The Heaviside-theta function $\Theta$ could be a rather blunt approximation, but it enforces the appropriate escape velocity ($v_{\rm esc}$) cut-off (the maximum allowed velocity of the DM particles in the galactic frame). 
The proportionality constant is determined from the normalization condition: $\int_0^\infty f_{\chi}(v)\d v=1$.
We take 
$v_{\rm rms}=0.73\E{-3}\,c$, 
$v_\odot=0.77\E{-3}\,c$,
$v_{\rm esc}=2.2\E{-3}\,c$, 
and  
$v_\oplus=0.10\E{-3}\,c$ \cite{Freese2013}.
The particular distributions of interest are shown in Fig.~\ref{fig:velocity}.
We note, however, that the above Maxwell distribution (\ref{eq:veldistro}) is not the only candidate; in fact non-Maxwellian distributions are well-motivated, and, in certain circumstances, may have a significant impact on the modulation rate \cite{Green2001} (see also Ref.~\cite{Green2012} and references therein).
Partly due to this reason, in Appendix~\ref{sec:scaling} we present detailed plots of the atomic structure calculations showing the energy-deposition and momentum-transfer dependence for several systems of experimental interest.
These calculations can then be used to form a simple parametric model for the atomic structure factor that can be inserted into a general formula for the ionization cross section.

In a typical  DM detection experiment,  $\e$ and $I_a$ are 
difficult to measure individually; instead it is the combination $\Delta E$ (\ref{eq:dE}) that is important.
The number of ``single-hit'' events in certain energy intervals are recorded; only the single-hit rate is recorded, since the likelihood that a double-hit event would be caused by a DM interaction is vanishingly small.
Therefore, the quantity of interest is
\begin{align}
\braket{{\rm d}{\s} \,v_\chi } =& 
\frac{4 \alpha_\chi^2}{\pi}
\int_0^\infty{\rm d}v  \frac{f_\chi(v) }{v}
			\int_{q_-}^{q_+}{\rm d} q \frac{q }{(q^2+m_v^2c^2)^2} 
			\times\notag\\&
			\sum_{n,\k}m_e\sqrt{2m_e(\Delta E-I_{n\k})}K_{n\k}
			\, \d (\Delta E)
			,
\label{eq:dsv-de}
\end{align}
where $\k=(l-j)(2j+1)$ is the Dirac quantum number\footnote{$\k=-1$ for $s_{1/2}$, $\k=1$ for $p_{1/2}$, $\k=-2$ for $p_{3/2}$, etc.} 
with $l$ and $j$ the orbital and total (single-electron) angular momentum quantum numbers, respectively, and the ``atomic kernel'' is defined
\begin{equation}
K_{n\k}(\Delta E,q)
=\sum_{\k'}\sum_{m,m'}
\abs{\bra{\e\k'm'}\Exp{i\v{q}\cdot\v{r}}\ket{n\k m}}^2.
\label{eq:AK}
\end{equation}
Here, $m$ is the projection of $\v{j}$ onto the axis of quantization.
Full formulas for the evaluation of the atomic kernel, including for the other Lorentz structures, are given in Appendix~\ref{sec:appendix}.

Then, the differential event rate per unit energy per kilogram, is given by
\begin{equation}
\mathcal{R}_{m_\chi,m_v,\alpha_\chi}(\Delta E)=\frac{n_{\rm A} \rho_{\chi}}{m_\chi}\frac{\braket{{\rm d}{\s} \,v_\chi }}{\d (\Delta E)},
\label{eq:diffEventRate}
\end{equation}
where $\rho\approx0.4\,$GeV\,cm$^{-3}$ is the assumed local DM energy density, and $n_{\rm A}$ is the number of target atoms per kilogram.
The total average event rate per kilogram in the energy interval $\Delta E \in [a, b]$ is given by
\begin{equation}
R_{a\to b}=\frac{1}{\Delta E_b-\Delta E_a}\frac{n_{\rm A}  \rho_{\chi}}{m_\chi}\int_{a}^{b}\braket{{\rm d}{\s} \,v_\chi },
\end{equation}
which is expressed in units of counts per day (cpd) per kg/keV.
Of course, the event rate that is actually observed in the experiment depends on a number of other factors, including the detector efficiencies and energy resolution. These factors depend on the design of the apparatus, so we save our discussion of these effects until we consider specific experiments.

%----------------------------------------------------------------------------------------------
\subsection{Comments on relativistic structure of the cross-section}\label{sec:relcrossec}

In general, in the presence of multiple mediators and/or broken descrete symmetries, 
the amplitude for the DM-electron scattering can be expressed as a linear combination of terms of the form
\begin{equation}
(\bar \chi \Gamma_{\mu\chi} \chi )\times 
	(\bar e \Gamma^\mu_e e ),
\end{equation}
where 
\begin{equation}
\Gamma^{(\mu)} =  g_S ; ~ i g_{PS}\g_5 ; ~ g_{V}\g^\mu ; ~ g_{PV}\g^\mu g_5 ; ~ \ldots
\end{equation}

A detailed study of the relevant Lorentz structure combinations can be found in Ref.~\cite{Kopp2009}.
The number of possible structures would shrink if $\chi$ is a scalar or a Majorana fermion. 
The wave functions for the incident and outgoing DM particles are taken as Born plane waves, and the initial and final electron wave functions are the bound and continuum atomic wave functions, respectively.
The (spin-independent) structure of the electron matrix elements for the different Lorentz structures are given in Appendix~\ref{sec:appendix}.

For the highest velocity the DM particles can reasonably be expected to have, $v_\chi/c\sim10^{-3}$, which is small. It can therefore reasonably be expected that relativistic expansion in $v/c$ for  DM particles works very well, and taking into account the leading terms 
usually suffices to get a reliable answer.
For the deep inner-shell atomic electrons, however, we have $v_e/c\sim Z\alpha$, which is not so small. 
For iodine ($Z=53$) and xenon ($Z=54$) $Z\a\approx0.4$, so electron relativistic effects may be important.
In fact, as we shall demonstrate in the next section, electron relativistic effects are crucial, actually giving the the dominant contribution to the amplitude.

While the general analysis is perhaps also of interest, we will concentrate on two cases, $(\bar \chi  \chi )
	(\bar e  e )$ and $(\bar \chi  \gamma_\mu \chi )
	(\bar e \gamma_\mu e )$ proportional amplitudes, which both lead to the Yukawa potential (\ref{eq:yukawa}).

%==========================================
\section{Calculations}
\label{sec:calculations}

%----------------------------------------------------------------------------------------------
\subsection{Importance of electron relativistic effects}

In Ref.~\cite{RobertsAdiabatic2016} it was demonstrated that relativistic effects give the dominant contribution to the cross section for atomic and molecular ionization by scattering of slow, heavy particles (such as WIMPs) off the atomic electrons when the momentum transfer to the electron is large in atomic units.
This means that nonrelativistic calculations may greatly underestimate the amplitude.

Here, we remind briefly the reason for the relativistic enhancement.
Consider the ejection of an electron from an atomic orbital $nl$ to a final state (in the continuum) with energy $\e$. 
Assuming the wave functions can be well-described by nonrelativistic Schr\"odinger functions, the contribution to the cross section coming from the final-electron partial wave $l'$ is proportional to the square of the radial integral
\[
\int _0^\infty R_{\e l'} (r) R_{nl}(r)j_L(qr)r^2\d r,
\]
where $R_{nl}(r)$ and $R_{\e l'}(r)$ are the radial wave functions of the initial and final states, $j_L(x)$ is the spherical Bessel function, the values of $l$, $l'$ and $L$ must satisfy the triangle inequality, and $l+l'+L$ must be even due to parity selection (see Appendix~\ref{sec:appendix}). 
The leading contribution to this integral at large $q$ comes from small $r\sim 1/q$, where the radial functions behave as $R(r) \sim r^{l}$.
It therefore appears that the leading contribution to the amplitude for large $q$ is proportional to
\[
\int _0^\infty r^{l+l'+2}j_L(qr)\d r,
\]
however, this integral is identically zero (see Ref.~\cite{RobertsAdiabatic2016}).
The next lowest-order in $r$ correction for either $R_{nl}(r)$ or $R_{\e l'}(r)$, is proportional to $Z$, and leads to the integral $\int _0^\infty r^{l+l'+3}j_L(qr)\d r$, which is nonzero. 
Therefore, at large $q$, the amplitude is dominated by the following term:
\begin{equation}
\label{eq:rad2}
\int_0^\infty R_{\e l'} (r) R_{nl}(r)j_L(qr)r^2\d r\propto \frac{Z}{q^{l+l'+4}}.
\end{equation}
Conventional wisdom would suggest that the ionization probability for such a process should be exponentially small (see, {\em e.g.} the corresponding discussion in Ref.~\cite{LLVol3}).
The power, instead of the exponent, emerges due to the Coulomb singularity of the electron wave function at the nucleus.
Singularity for the $s$-wave electrons is stronger than in higher partial waves, which translates to the least amount of suppression
for the $s$-states.

The situation becomes different if instead we consider the relativistic Dirac wave functions.
At small distances, the radial function of the large Dirac component behave as 
$
{f(r)}/{r} \sim r^{\g-1},
$
where $\g=\sqrt{\k^2-(Z\a)^2}$ (see Ref.~\cite{LLvol4} and Eq.~(\ref{eq:psi}) in the Appendix).
In the nonrelativistic limit, this of course reduces to exactly the same situation as above.
However, the corrections in $\g=|\k |-(Z\a)^2/2|\k|+\dots $ actually change the power of $r$ that appears in the low-$r$ expansion. 
As a result, the contribution to the scattering amplitude from the lowest-order in $r$ term, which vanished in the nonrelativistic case, now becomes
\begin{equation}
\int _0^\infty r^{\g+\g'}j_L(qr)\d r
=\frac{2^{2\g-1}}{q^{2\g+1}}\sqrt{\pi }
\frac{\Gamma\left[\frac{1}{2}(L+\g+\g'+1)\right]}{\Gamma\left[ \frac{1}{2}(L-\g-\g'+2)\right]},
\label{eq:rint-rel1}
\end{equation}
which is different from zero.
For example, taking initial and final states as $s$-waves [$\k=-1$, $\g=\g '\simeq 1-(Z\a)^2/2$], we have
\begin{equation}
\label{eq:rint-rel2}
\int _0^\infty r^{2\g }j_0(qr)\d r
\simeq \frac{\pi(Z\a)^2}{2 \, q^{3-(Z\a)^2}}. 
\end{equation}
If one considers the contribution from a $p_{1/2}$ state ($\k =1$) for either the bound or continuum electron (or both), the power of the $q$ dependence remains the same, but the coefficient is further suppressed by a power of $Z\a$ (which, for xenon and iodine, is not small).
This is true for scalar, pseudoscalar, vector, and pseudovector electron interactions (see Appendix~\ref{sec:appendix}). 
Thus we see that the electron wave function suppression is significantly weaker than that found in the nonrelativistic case.
The cross section goes as the square of the amplitude, meaning that the momentum-transfer dependence of the leading atomic structure contribution to the cross section is proportional to $q^{-6+2(Z\a)^2}$ 
(compared to $q^{-8}$ in the nonrelativistic case).

\begin{figure}
	\begin{center}
		\includegraphics[width=0.49\textwidth]{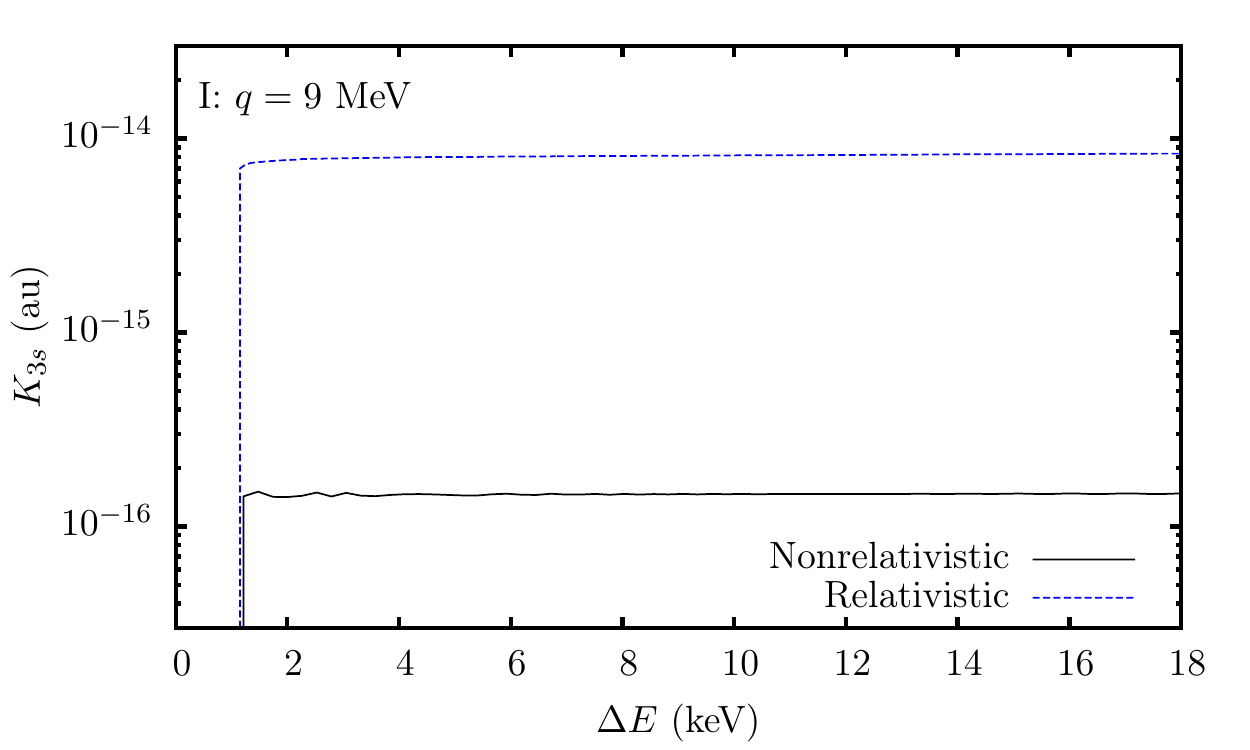}
		\includegraphics[width=0.49\textwidth]{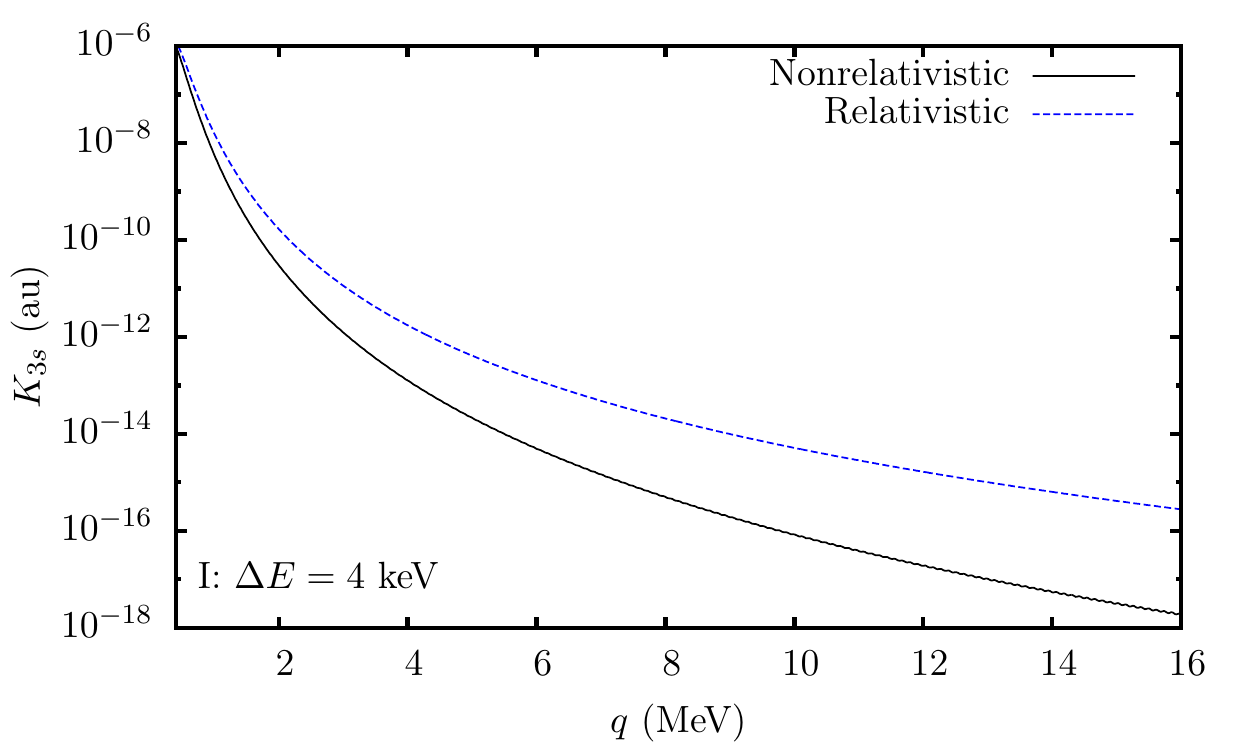}
		\caption{Comparison of the contribution of the $3s$ state to the atomic kernel of iodine in the relativistic and nonrelativistic approximations: ({\em top}) as a function of the energy deposition ($\Delta E$) for a value of the momentum transfer of $q\simeq9\,{\rm MeV}$, and ({\em bottom}) as a function of $q$ for $\Delta E\simeq 4\,{\rm keV}$.
}
		\label{fig:I-relativistic}
	\end{center}
\end{figure}

A comparison of the relativistic and nonrelativistic calculations of the atomic kernel of iodine is presented in Fig.~\ref{fig:I-relativistic} for relatively high values of the momentum transfer, $q$ (only high values of $q$ can contribute the cross section).
For consistency, the relativistic and nonrelativistic calculations are performed using the exact same methods and computer codes (the relativistic Hartree-Fock method, as described below); the nonrelativistic limit is achieved by letting the speed of light approach infinity in the code before the Dirac equation is solved.
As $q\to0$, the difference between the relativistic and nonrelativistic approaches diminishes, as expected.
It is also instructive to discuss the origin of slight numerical instabilities in the nonrelativistic calculations visible in the plots in Fig.~\ref{fig:I-relativistic} (solid black line). These instabilities are absent in the relativistic calculations.
This is because in the relativistic case the atomic kernel is dominated by a single contribution coming from very low $r$,
while the nonrelativistic case has contributions from larger $r$ which cover several oscillations of the (very rapidly oscillating) $j_L$ function. 
(Of course, the instabilities in the nonrelativistic calculations can be removed by increasing the parameters of the numerics, however, this is not necessary for the current purpose.)

%----------------------------------------------------------------------------------------------
\subsection{Calculations of the atomic kernel}
\label{sec:AK}

\begin{table}%
\centering%
\caption{Relativistic Hartree-Fock ionization energies for the core states of Na, Ge, I, Xe, and Tl in atomic units$^a$.}%
\begin{ruledtabular}%
  \begin{tabular}{l ddddd}%
%\multicolumn{1}{c}{}  \\
\multicolumn{1}{c}{Atom}	&	\multicolumn{1}{c}{Na}&	\multicolumn{1}{c}{Ge}	&	\multicolumn{1}{c}{I}	&	\multicolumn{1}{c}{Xe}&	\multicolumn{1}{c}{Tl}	\\
\multicolumn{1}{c}{$Z$}	&	\multicolumn{1}{c}{11}	&	\multicolumn{1}{c}{32}	&	\multicolumn{1}{c}{53}	&	\multicolumn{1}{c}{54}&	\multicolumn{1}{c}{81}	\\
\hline%
$1s_{1/2}$	&  40.54	&  411.1	&  1226	 	&  1277	 	&  2851 \\
$2s_{1/2}$	&  2.805	&  53.46	&  193.0	&  202.5	&  484.5 \\
$2p_{1/2}$	&  1.522	&  47.33	&  180.6	&  189.7	&  465.7 \\
$2p_{3/2}$	&  1.515	&  46.15	&  169.6	&  177.7	&  465.7 \\
$3s_{1/2}$	&  0.182	&  7.410	&  40.53	&  43.01	&  117.1 \\
$3p_{1/2}$	&  			&  5.325	&  35.34	&  37.66	&  108.2 \\
$3p_{3/2}$	&  			&  5.157	&  32.21	&  35.33	&  108.2 \\
$3d_{3/2}$	&  			&  1.616	&  24.19	&  26.02	&  91.71 \\
$3d_{5/2}$	&  			&  1.592	&  23.75	&  25.54	&  91.71 \\
$4s_{1/2}$	&  			&  0.569	&  7.759	&  8.430	&  26.88 \\
$4p_{1/2}$	&  			&  0.282	&  5.869	&  6.453	&  22.92 \\
$4p_{3/2}$	&  			&  0.273	&  5.450	&  5.983	&  22.92 \\
$4d_{3/2}$	&  			&  			&  2.342	&  2.711	&  15.65 \\
$4d_{5/2}$	&  			&  			&  2.274	&  2.634	&  15.65 \\
$5s_{1/2}$	&  			&  			&  0.876	&  1.010	&  4.617 \\
$5p_{1/2}$	&  			&  			&  0.434	&  0.493	&  3.230 \\
$5p_{3/2}$	&  			&  			&  0.390	&  0.440	&  3.230 \\
$4f_{5/2}$	&  			&  			&  			&  			&  5.784 \\
$4f_{7/2}$	&  			&  			&  			&  			&  5.784 \\	
$5d_{3/2}$	&  			&  			&  			&  			&  0.967 \\
$5d_{5/2}$	&  			&  			&  			&  			&  0.967 \\
$6s_{1/2}$	&  			&  			&  			&  			&  0.360 \\
$6p_{1/2}$	&  			&  			&  			&  			&  0.201 \\
$6p_{3/2}$	&  			&  			&  			&  			&  0.201 \\
\end{tabular}%
\tablenotetext[1]{Note: $1\,\rm{au}=27.211\,\rm{eV}$}
\end{ruledtabular}%
\label{tab:HFenergies}%
\end{table}%

To perform the atomic structure calculations we use the relativistic Hartree-Fock method, which is described briefly in Appendix~\ref{sec:methods}.
Calculations of the bound-state energies for the core orbitals of atomic Na, Ge, I, Xe, and Tl are given in Table~\ref{tab:HFenergies}.

\begin{figure}
	\begin{center}
		\includegraphics[width=0.49\textwidth]{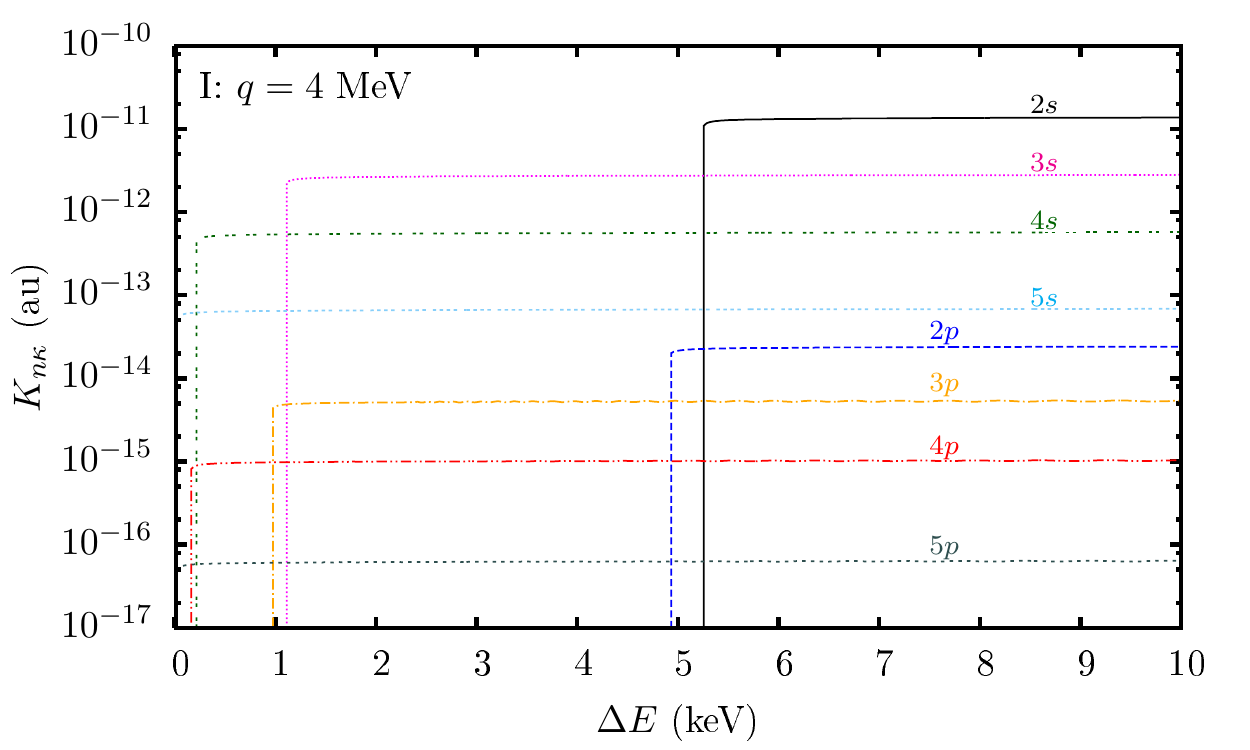}
		\caption{Core-state contributions to the atomic kernel [defined in Eq.~(\ref{eq:AK})] for I as a function of the energy deposition, $\Delta E$, at momentum transfer $q\simeq4$ MeV. 
The $s$ states dominate the amplitude; this domination only increases at larger $q$. 
The contributions from the $d$ states (not shown) are orders of magnitude smaller again. 
}
		\label{fig:I-AKdE-1000-cntm}
	\end{center}
\end{figure}

\begin{figure}
	\begin{center}
		\includegraphics[width=0.49\textwidth]{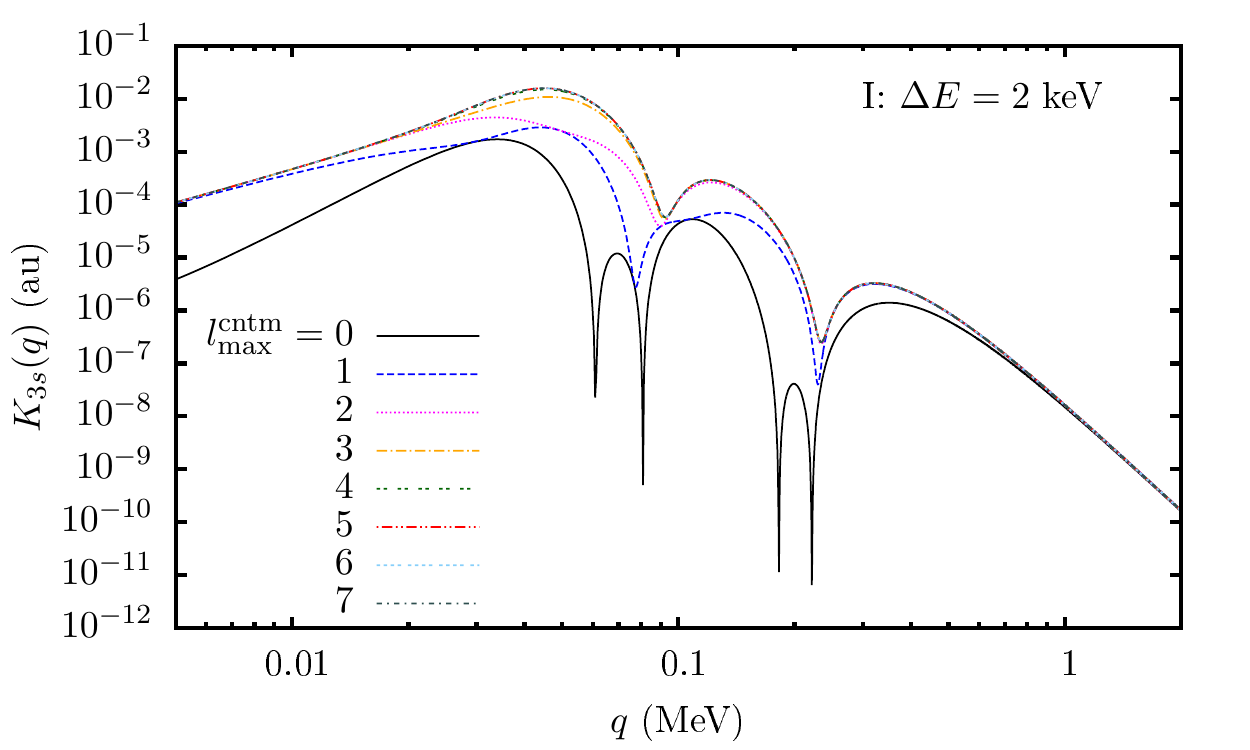}
		\caption{Contribution of the $3s$ core state to the atomic kernel for I as a function of the momentum transfer, $q$, at $\Delta E\simeq2$~keV.
Shown separately are the kernels with different values for the high-$l$ cut-off for the continuum-state electron orbital angular momentum.
The higher $l$ continuum states contribute significantly at low values of $q$, however, at the values relevant to this work ($q>\sim$MeV), they contribute negligibly.
}
		\label{fig:I-AKq-cntm}
	\end{center}
\end{figure}

In Fig.~\ref{fig:I-AKdE-1000-cntm}, we plot the contributions of the different core states to the atomic kernel (\ref{eq:AK}) for iodine as a function of the energy deposition for a fixed momentum transfer.
It is seen that the $s$-states dominate the amplitude, as expected.
In Fig.~\ref{fig:I-AKq-cntm} we plot the $3s$ core contribution to the iodine atomic kernel for different values of the maximum included continuum-state angular momentum as a function of the momentum transfer for fixed energy deposition.
For very low values of momentum transfer, only the $j=1/2$ states give significant contributions.
For intermediate values, higher angular momentum states become important.
For the high momentum transfer values, which are those relevant to the ionization problem, the higher angular momentum states contribute negligibly and only $s$-wave continuum states are important.
Note that this is a result of the relativistic effects; in the nonrelativistic limit higher angular momentum states contribute non-negligibly because the $s$-state contribution is significantly underestimated.
The general result is that in the calculations, only $s$-states need to be considered both for the bound states and for the continuum states, as suggested above; $p$-states contribute at the few-percent level. 
We have checked this in the direct calculations of the cross section as well, and it continues to hold true.
Regardless of that, in our full atomic structure calculations we keep all higher angular momentum states until the cross section converges explicitly to the $\sim0.1\%$ level.
For lower values of energy deposition $(\Delta E\lesssim1$\,keV) this condition becomes less strong. 
Though not directly relevant to the DAMA experiment, $(\Delta E\lesssim1$\,keV) range may be important for other types of electron-recoil experiments, such as the XENON10 experiment \cite{Angle2011}, and those suggested in Refs.~\cite{Essig2012,Essig2015}.

%==========================================
\section{Results}\label{sec:DAMA}

%----------------------------------------------------------
\subsection{DAMA Analysis}\label{sec:results-DAMA}

\begin{figure}
	\begin{center}
		\includegraphics[width=0.49\textwidth]{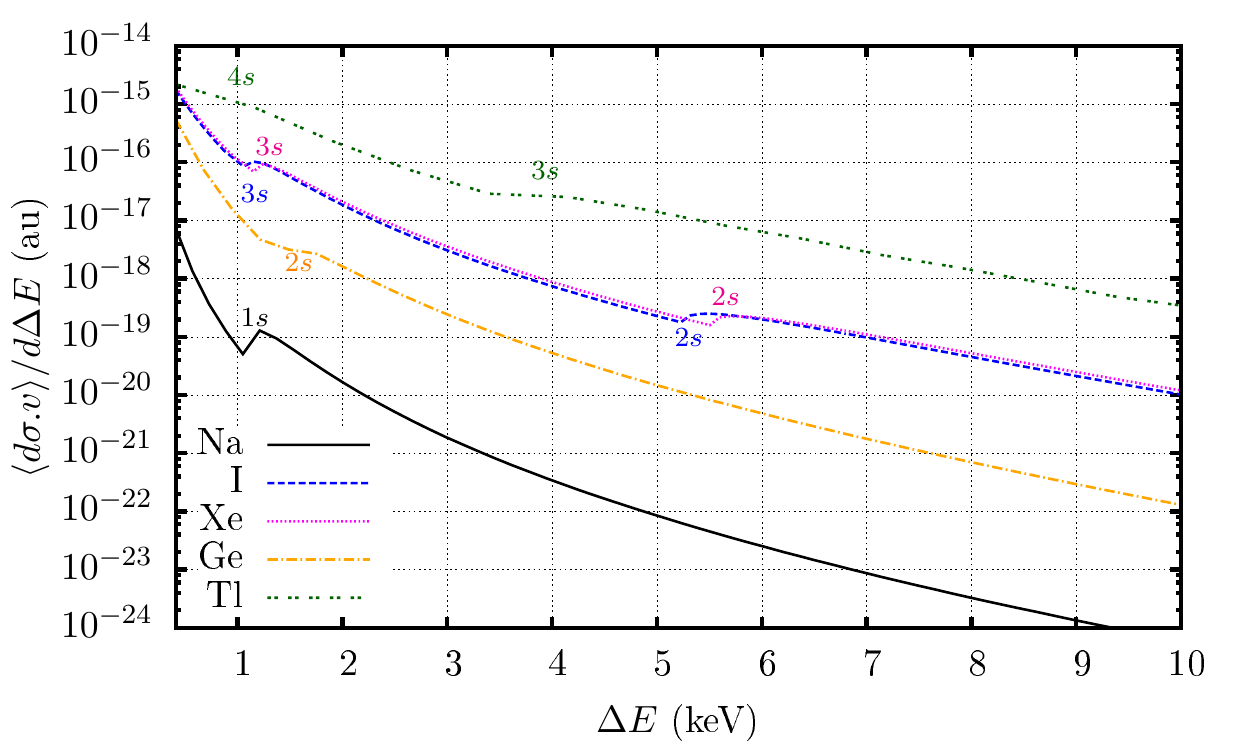}
		\caption{Plot of the differential cross section [defined in Eq.~(\ref{eq:dsv-de}), with $m_\chi=10\,$GeV, $m_v=10\,$MeV, and for simplicity $\alpha_\chi=1$] for Na, I, Xe, Ge, and Tl as a function of the total energy deposition, $\Delta E$. 
The kinks in the curves correspond to the opening of deeper atomic shells; see Table~\ref{tab:HFenergies}. 
There is a clear and significant $Z$ dependence, which is due to the low-$r$ scaling of the wave functions and the relativistic effects.
}
		\label{fig:NaXeI}
	\end{center}
\end{figure}

\begin{figure}
	\begin{center}
		\includegraphics[width=0.49\textwidth]{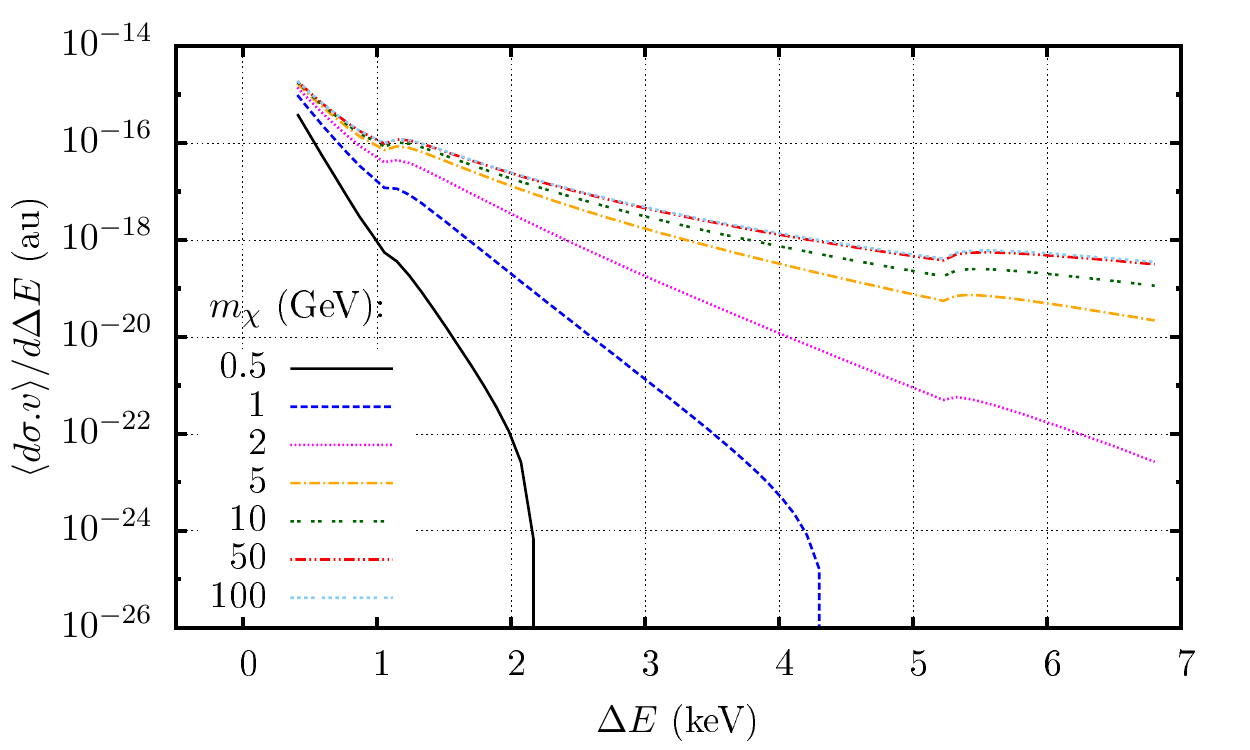}
		\includegraphics[width=0.49\textwidth]{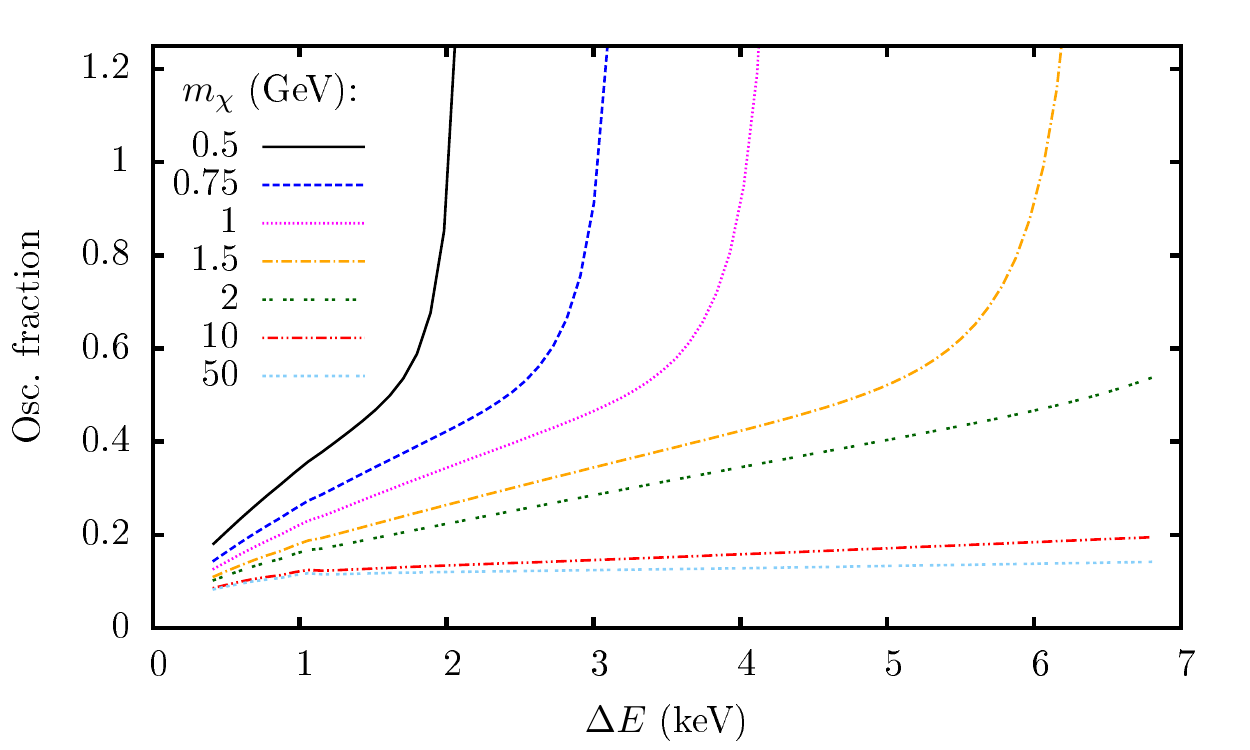}
		\caption{Plots showing the $m_\chi$ dependence of ({\sl top}) the differential cross section, and ({\sl bottom}) the oscillation fraction, for ionization of iodine as a function of the deposited energy, $\Delta E$. For the plots we have taken $m_v=10$ MeV, and $\alpha_\chi=1$.
Low values of $m_\chi$ lead to significantly lower cross-sections, however, as $m_\chi$ increases the increase in the effect wanes. The energy dependence of the oscillations increases with decreasing $m_\chi$, since in these regions only part of the DM velocity distribution can give rise to an effect.}
		\label{fig:Imx}
	\end{center}
\end{figure}

For our calculations of the atomic structure, we employ the system of atomic units $(\hbar=a_B=e=1$, $c=1/\alpha$).
The conversion factor for the total cross section from atomic units is $a_B^2\approx2.8\E{-17}\,$cm$^2$, and for the function
$\frac{\braket{{\rm d}\s\cdot v}}{{\rm d}\Delta E}$
is $a_B^2 c \a /2Ry\approx0.019\,$cm$^3$/keV/day.
We present the event rates in the standard units of counts per day (cpd) per kg/keV.

Setting $\alpha_\chi=1$ for a moment, in Fig.~\ref{fig:NaXeI}, we 
plot the differential cross section (\ref{eq:dsv-de}) for Na, I, Xe, Ge, and Tl as a function of the total energy deposition, $\Delta E$, for a specific set of DM parameters and assuming the standard halo velocity distribution (\ref{eq:veldistro}).
Note that the NaI detector in the DAMA experiment is doped with Tl. With a significantly higher atomic number, the effect arising from thallium is substantially larger than that from iodine; however, the small amount present in the detector means that the DAMA signal would still  be dominated by the iodine contribution.

To a first approximation, the expected event-rate due to scattering of WIMPs from the galactic halo can be expressed as
\begin{equation}
R(t) = R_0 + R_m\cos(\omega t),
\label{eq:cosine}
\end{equation}
where $R_0$ is the constant or average part of the event-rate, which comes from the velocity distribution of the WIMPs in the solar frame, and $R_m$ is the amplitude of the modulations in the event rate, which come from the relative motion of the earth around the sun; the factor $\omega t$ is defined in Eq.~(\ref{eq:vearth}).
We do note, however, that due to the very strong dependence of the scattering cross section on the incident energy of the DM particles (and therefore on the DM velocity), $R_m$ itself depends on the phase of the earths orbit and therefore the event rate is not purely sinusoidal.
The deviations from a sinusoidal shape, however, are modest for most of the parameter space, and do not affect the analysis substantially.

The so-called oscillation fraction, defined as
$
R_m/R_0,
$
has a strong dependence on the energy deposition, and on the mass of the DM particles.
Fig.~\ref{fig:Imx} shows the $m_\chi$ dependence of the differential cross section and the oscillation fraction for iodine as a function of the deposited energy, $\Delta E$. 
The energy dependence of the oscillations increases with decreasing $m_\chi$, since at small DM mass only the velocity 
tail of the DM velocity distribution can give rise to an effect.
%Note that in Fig.~\ref{fig:Imx} it is seen that the oscillation fraction can become larger than one. This is because of the high velocity dependence of the cross-section; for light DM ionizations cannot occur at high energy deposition when the velocity distribution does not allow an apreciable fraction of the DM particles to have enough incident energy (i.e., $R_0$ tends to zero faster than $R_m$ does). The actual event rate in this window is very small.

A possibility of such {\em very large} time-modulated fraction in dark-matter-induced atomic ionization
is interesting for the following reasons: In most models of the elastic 
DM scattering off nuclei the modulated fraction is typically much smaller, under $\sim 0.1$. In recent papers \cite{Pradler2013},
the DM-nucleus scattering of DAMA results were questioned because of a possibility of underestimated $^{40}$K 
background events around $\Delta E = 3$~keV. According to Ref.~\cite{Pradler2013}, if such background is properly subtracted, the 
remaining DAMA signal is modulated at 20\% or higher, which is incompatible with the most straightforward explanation 
based on elastic DM--nucleus scattering. Thus, the large modulated fraction of the dark-matter-induced ionization
could serve as an explanation of DAMA even with the presence of unaccounted backgrounds in the unmodulated rate.

In order to calculate the number of events detected within a particular energy range, the energy resolution of the detectors must be taken into account.
To do this, we convolute the calculated rate with a Gaussian:
\begin{equation}
\widetilde{\mathcal{R}}(\Delta E) = \int R(\e) g_{\Delta E}(\e) \d\e,
\label{eq:DAMA-Gaussian}
\end{equation}
where $g_{\Delta E}(\e)$ is a Gaussian function centred at $\Delta E$, with standard deviation 
\[\sigma=0.448\sqrt{\Delta E/{\rm keV}}+0.0091\Delta E/{\rm keV},\]
as measured by the DAMA Collaboration \cite{Bernabei2008b}.
This has the effect of ``smearing out'' the $2$ keV low threshold, allowing a small fraction of events that originate from lower energies to be accepted.
Note that since there is an almost-exponential enhancement of the event rate at lower energies (see Fig.~\ref{fig:NaXeI}) this has a significant impact on the results.
We also assume that the DAMA detectors are 100\% efficient, and importantly, that the efficiency is not a function of the energy deposition. This is the most 
conservative  assumption for the prupose of deriving limits on the DAMA signal interpretations from other experiments.

\begin{figure}
	\begin{center}
		\includegraphics[width=0.49\textwidth]{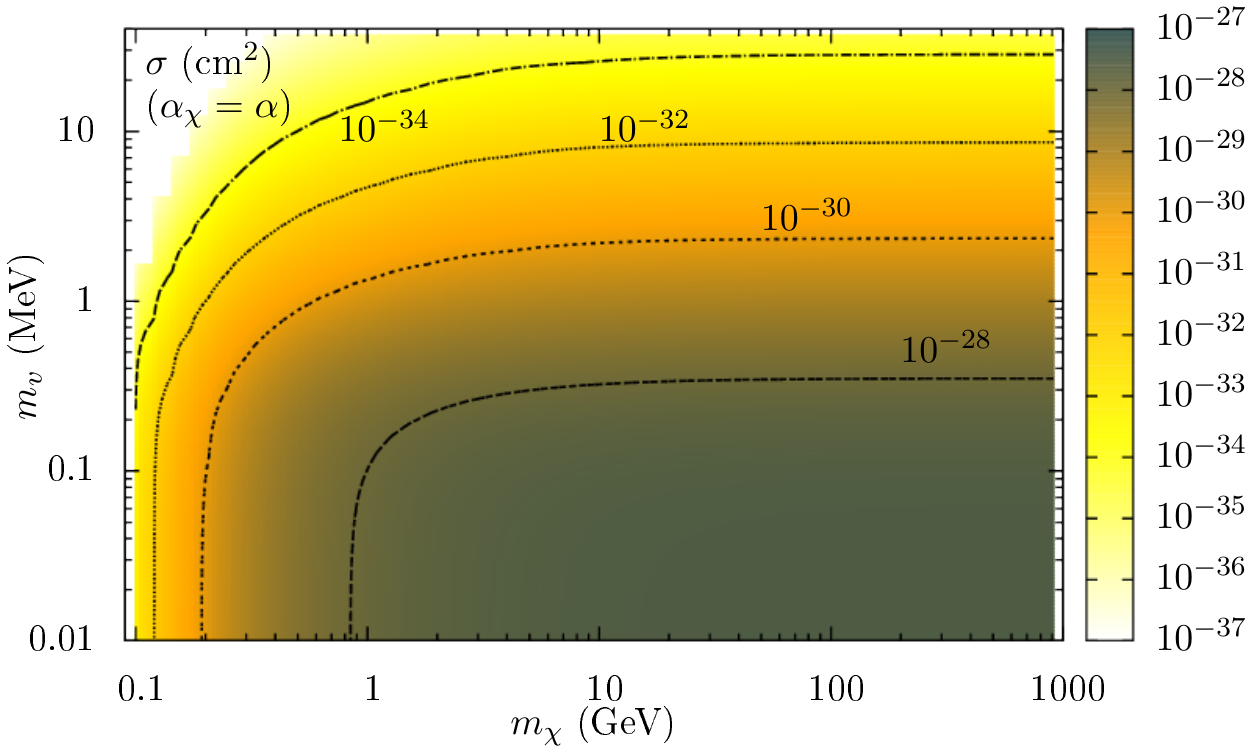}
		\caption{Total cross section (cm$^2$) for the ionization of NaI in the 2 -- 6 keV interval assuming $\alpha_\chi=\alpha$ for the average (i.e.~spring/fall) DM velocity distribution,
including the Gaussian resolution profile (\ref{eq:DAMA-Gaussian}).
}
		\label{fig:sigma-DAMA}
	\end{center}
\end{figure}

\begin{figure}
	\begin{center}
		\includegraphics[width=0.49\textwidth]{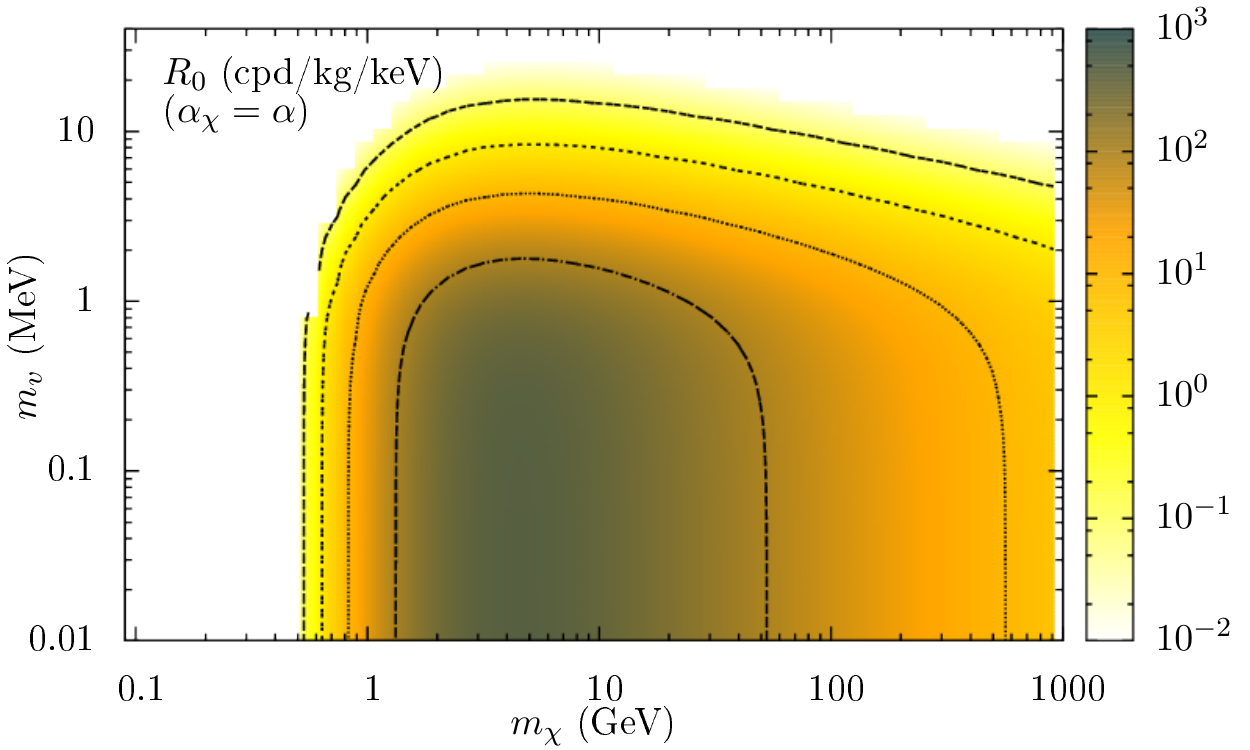}	
		\includegraphics[width=0.49\textwidth]{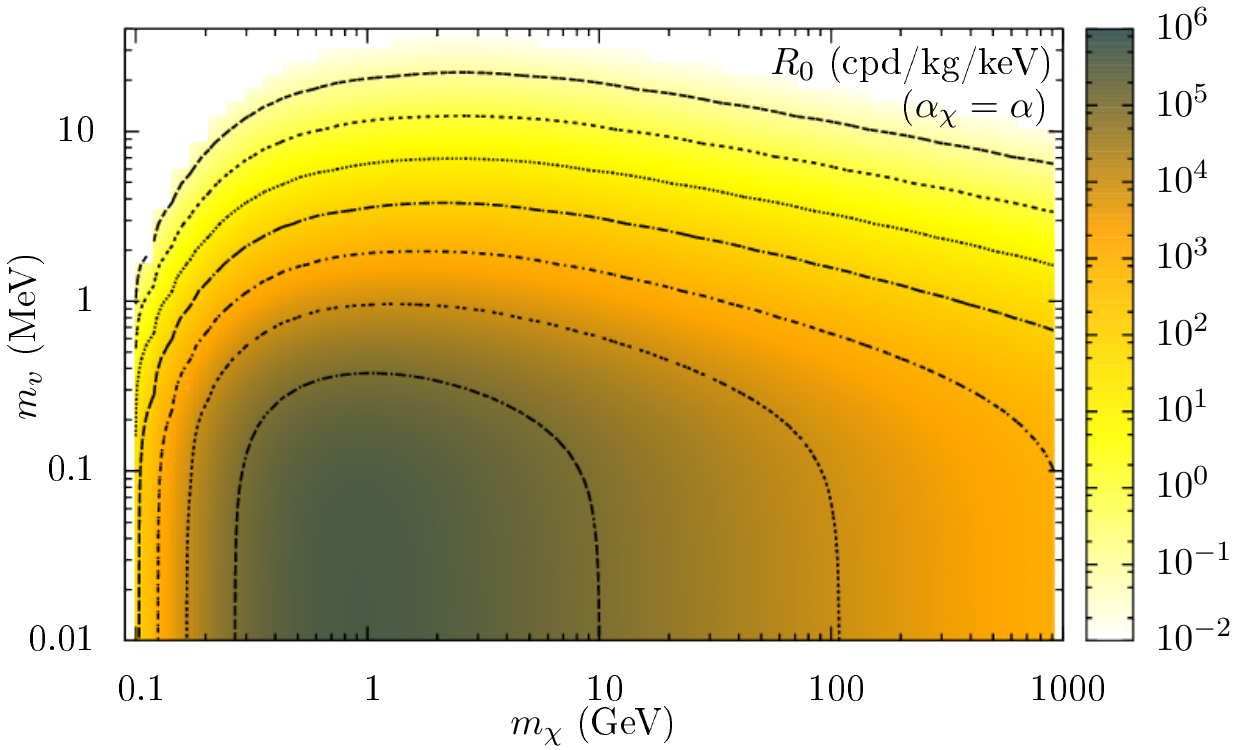}
		\caption{Unmodulated event rate $R_0$ 
for NaI in the 2 -- 6 keV interval assuming $\alpha_\chi=\alpha$ in units of cpd/kg/keV:
({\sl top}) assuming perfect detector resolution;
({\sl bottom}) including a Gaussian resolution profile (\ref{eq:DAMA-Gaussian}).
}
		\label{fig:rates-DAMAR0}
	\end{center}
\end{figure}

\begin{figure}
	\begin{center}
		\includegraphics[width=0.49\textwidth]{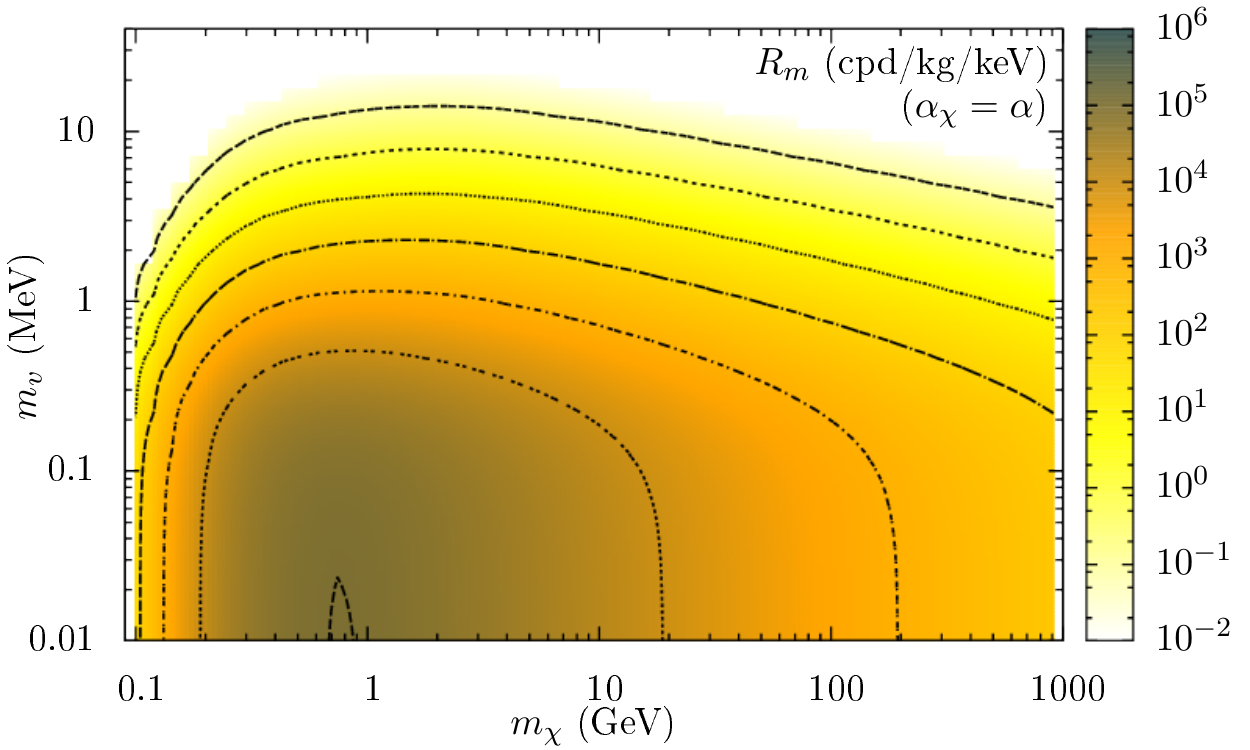}
		\caption{Modulation amplitude $R_m$ for NaI in the 2 -- 6 keV interval assuming $\alpha_\chi=\alpha$ in units of cpd/kg/keV (including the Gaussian resolution profile).
}
		\label{fig:rates-DAMARm}
	\end{center}
\end{figure}

\begin{figure}
	\begin{center}
		\includegraphics[width=0.49\textwidth]{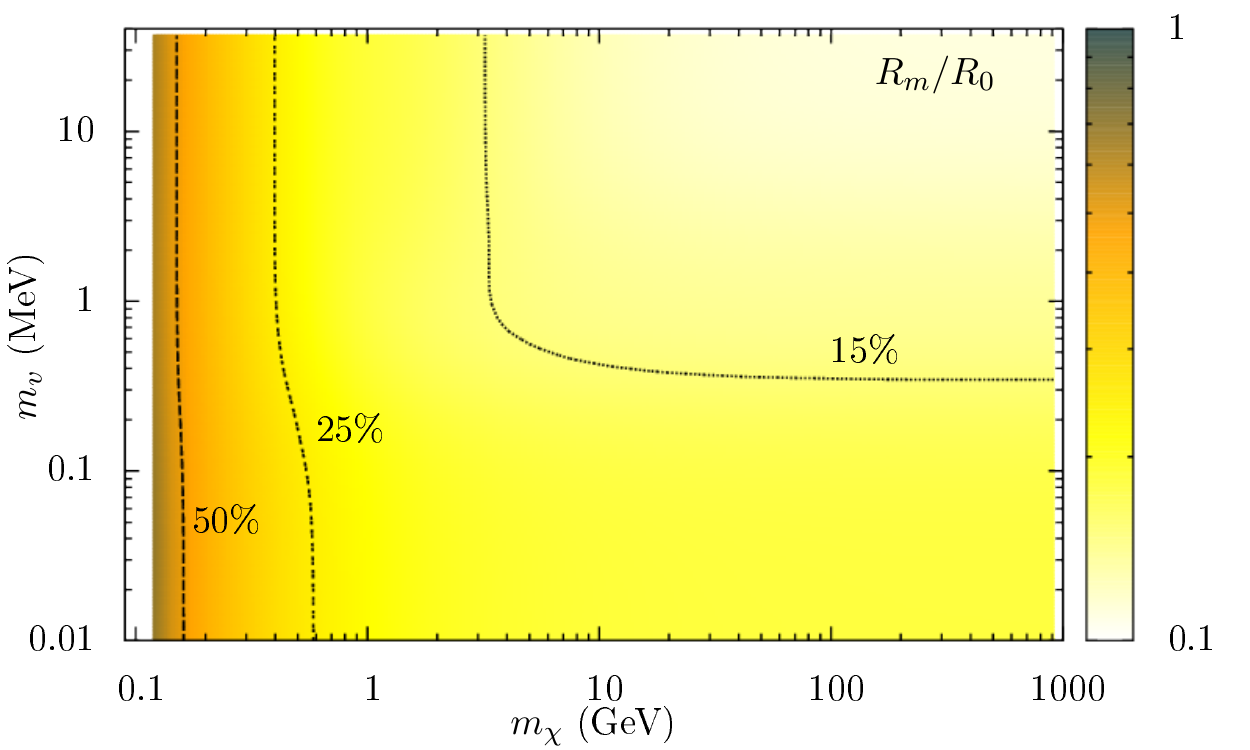}
		\caption{The calculated modulation fraction ($R_m/R_0$) expected for the scintillation signal in the 2 -- 6 keV interval for NaI (including the Gaussian resolution profile).}
		\label{fig:oscfrac-DAMA}
	\end{center}
\end{figure}

\begin{figure}
	\begin{center}
		\includegraphics[width=0.49\textwidth]{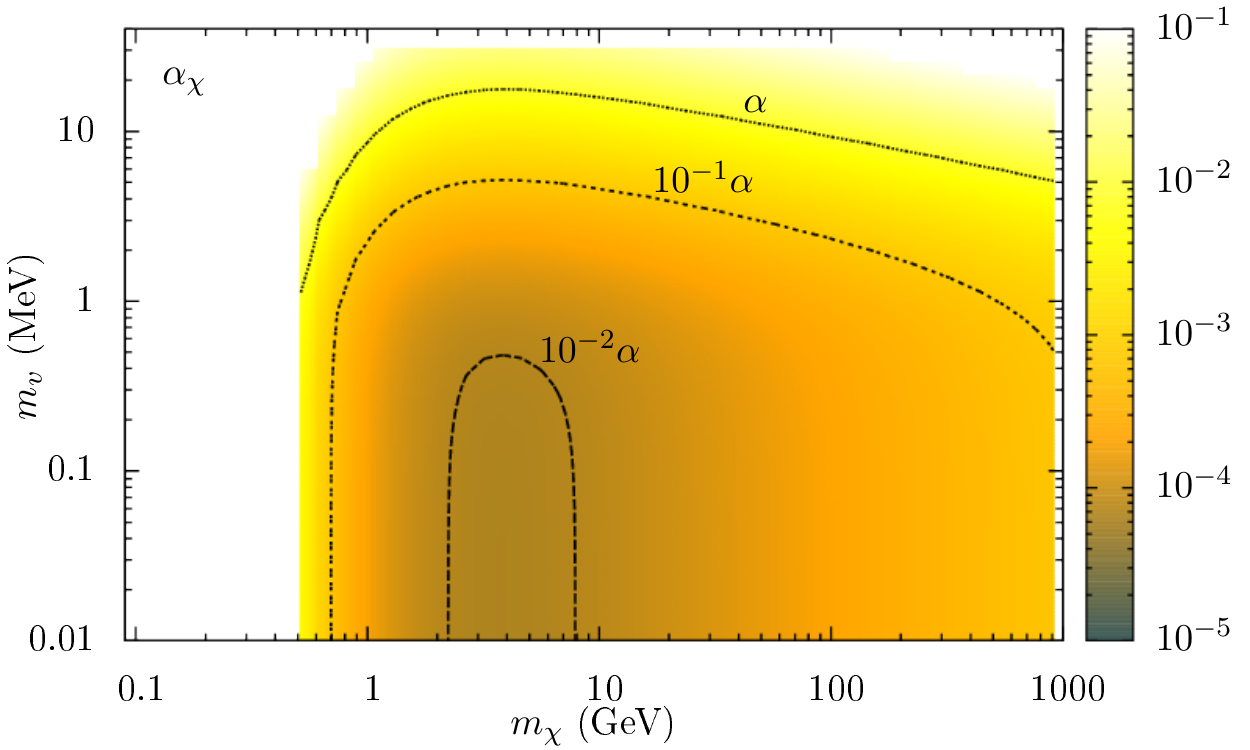}
		\includegraphics[width=0.49\textwidth]{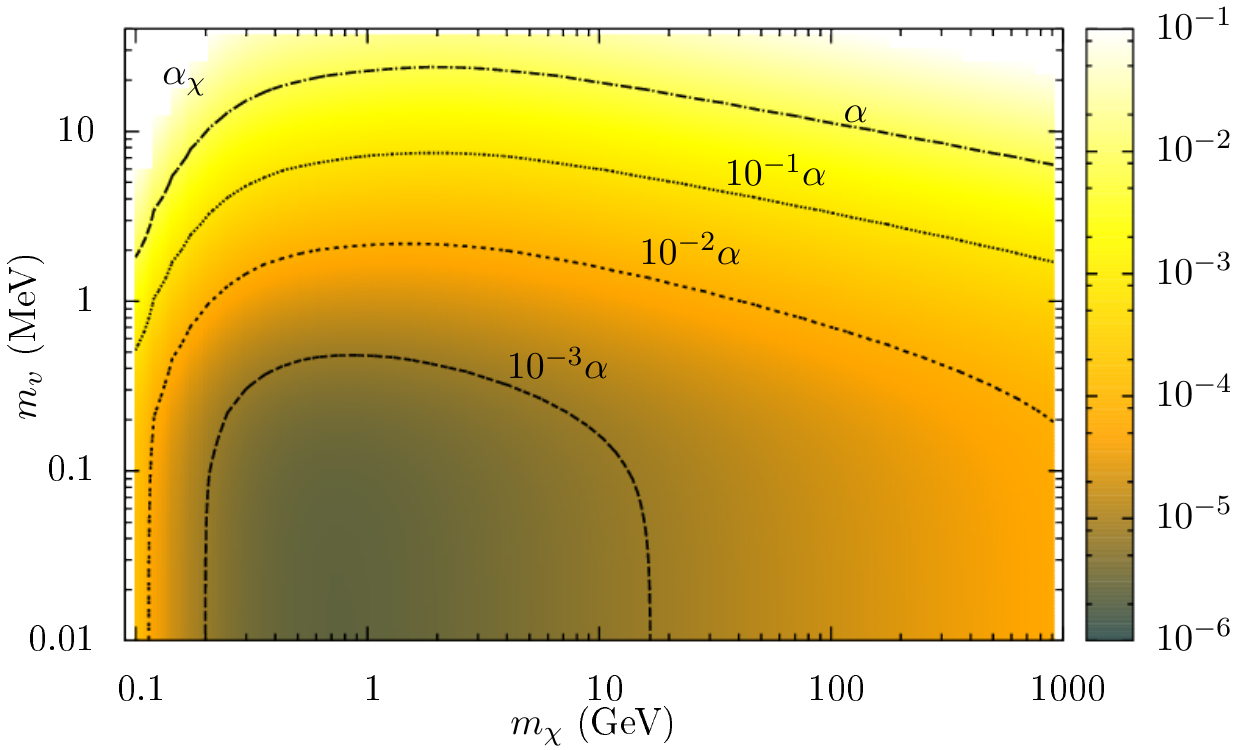}
		\caption{The value that $\alpha_\chi$ must take in order to reproduce the DAMA modulation signal of $0.0112$ cpd/kg/keV in the 2 -- 6 keV interval:
({\sl top}) assuming perfect detector resolution;
({\sl bottom}) including the Gaussian resolution profile (\ref{eq:DAMA-Gaussian}).
}
		\label{fig:axreq}
	\end{center}
\end{figure}

Figure~\ref{fig:sigma-DAMA} shows the dependence of the cross section for the ionization of NaI by DM--electron scattering on the DM particle mass and the mass of the (vector) exchange particle.
The plot is made arbitrarily with $\alpha_\chi=\alpha$; the cross section is linear in $\alpha^2$, so with $\alpha_\chi=10^{-2}\alpha$, for example, the value cross section would be smaller by a factor of $10^{-4}$.
The unmodulated event rate in the energy interval 2--6 keV, relevant to the DAMA experiment, is shown in Fig.~\ref{fig:rates-DAMAR0}.
Shown separately are the event rates calculated assuming a perfect detector resolution, and assuming the Gaussian resolution as in Eq.~(\ref{eq:DAMA-Gaussian}).
Note in particular that the Gaussian profile allows events in this region to be caused by significantly smaller DM masses, and also greatly increases the observed event rate.
This is entirely due to the fact that events originating at smaller energies (which have a much greater amplitude) are allowed to ``leak'' into the detection interval.
As is clear, the dependence on the detector resolution is extreme.
There is a clear favor of low $m_\chi$, and the modulation fraction is large.
The corresponding modulated event rate (including the Gaussian profile) is shown in Fig.~\ref{fig:rates-DAMARm}.

The DAMA collaboration observes a significant modulation in the event rate in this 2--6 keV interval, as described above.
The amplitude of the observed modulation is \cite{Bernabei2013}
\begin{equation}
R_m^{\rm DAMA}=1.12(12)\E{-2}\,{\rm cpd/kg/keV},
\label{eq:DAMA-Rm}
\end{equation}
amongst a background signal of approximately {{1\;cpd/kg/keV}}, which is attributed mostly to noise.
To perform our analysis, we assume this modulation signal can be entirely attributed to ionization of NaI by the scattering of WIMPs on the electrons.
Figure~\ref{fig:axreq} shows the value that the effective DM--electron coupling constant ($\alpha_\chi$) must take in order to give the required modulation amplitude.

In the WIMP--electron scattering scenario, the large modulation fractions (as reported by the DAMA \cite{Bernabei2013}, CoGeNT \cite{CoGeNT2014}, and XENON100 \cite{XENONcollab2015a} collaborations) are reproduced naturally.
The expected modulation fraction $R_m/R_0$ is plotted explicitly for DAMA in Fig.~\ref{fig:oscfrac-DAMA}.
The fraction is very large, over 20\% for large portions of the parameter space, even reaching as high as 50\% for reasonable values.
Note that this is assuming just the standard Maxwellian halo model for the DM velocity distribution (\ref{eq:veldistro}).
The large modulation is due to the fact that the ionization cross section is highly velocity dependent.
This is in contrast to WIMP--nucleon scattering cross section, where exotic DM velocity distributions must be assumed in order to replicate the large modulation fraction (see, e.g., Ref.~\cite{Green2001}).
Our findings in this regard are in agreement with those of Ref.~\cite{Lee2015}.

In order to avoid disproportionally large values of $\alpha_\chi$ the mass $m_v$ of the mediating particle must be light,
as seen from the contour plot in Fig. ~\ref{fig:axreq}.
However, even with sub-MeV masses, the required value of $\alpha_\chi$ may not be small enough for the existing constraints. 
Note that, from constraints on the energy loss in stars, the mass of the mediator cannot be smaller than $\sim200$ keV \cite{An2013}.
Taking $m_v$ close to this boundary, and DM mass close to a GeV, we conclude on the basis of our DAMA signal analysis 
that coupling constant can be as small as $\alpha_\chi \sim 10^{-3}\alpha$. While this is definitely a rather small value, 
it is perhaps not sufficiently small to escape current constraints, as we discuss below.

There are several potential constraints to be considered, some of which are model dependent. 
We write the fine structure constant in terms of its coupling to electrons and DM in the 
following way:
\begin{equation}
\label{alpha_chi}
\alpha_\chi = \alpha \times \left(\frac{g_\chi}{e}\right )\times \left( \frac{g_e}{e} \right),
\end{equation}
where $g_\chi$ and $g_e$ are mediator couplings to DM and electrons. 
There are separate constraints on both $g_e$ and $g_\chi$. From the fact that the visible sector is 
more constrained, one would have to assume a hierarchy $g_e\ll g_\chi$.

From the consistency of the electron $g-2$ with the QED calculations and independent measurement of electromagnetic $\alpha$ one can derive
strong constraints on the value of $g_e$ \cite{Pospelov2009,Davoudiasl2014}. While the constraint would slightly vary depending on whether mediator is a scalar of vector, from the general consistency of electron $g-2$
for $m_v \sim 1$~MeV one expects $|g_e/e| <  10^{-4}$. This is difficult to combine with $\alpha_\chi \sim 10^{-3} \alpha$ 
requirement. Therefore, additional fine-tuning of the $g-2$ may be required by other unspecified new physics. 
Direct constraints on $g_e$ vary depending on how the mediator decay (photons, electron-positrons, or to invisible particles such as 
neutrinos). The range of the mediator masses just below $2 m_e$ may represent a ``blind spot'' for the searches, and couplings 
$|g_e/e| \sim  10^{-3}$ may not be excluded \cite{Izaguirre2015a}. 

The large values of $g_\chi$ are constrained as well, primarily through the 
DM self-interaction, which is known to affect the radial profiles of the 
DM halos. Despite the significant uncertainties involved, it is unlikely that the 
self-scattering cross section per unit mass is allowed to exceed $\sim 10^{-23} {\rm cm}^2/{\rm GeV}$. 
For $m_\chi = 1~{\rm GeV}$, and the mediator mass in the MeV range, this would imply 
$|g_\chi/e| < 0.01$ (see {\em e.g.} Ref.\cite{Tulin2013}), which is also a stringent constraint. To avoid this constraint, one would have to introduce yet additional interaction that is fine tuned to  interfere destructively with  the  WIMP-WIMP scattering amplitude. Thus, we see that the values of $\alpha_\chi$ required to match the level of DAMA modulation signal generally require very light mediators and fine tuning, both in $g_\chi$ and $g_e$.

\begin{figure}
	\begin{center}
		\includegraphics[width=0.49\textwidth]{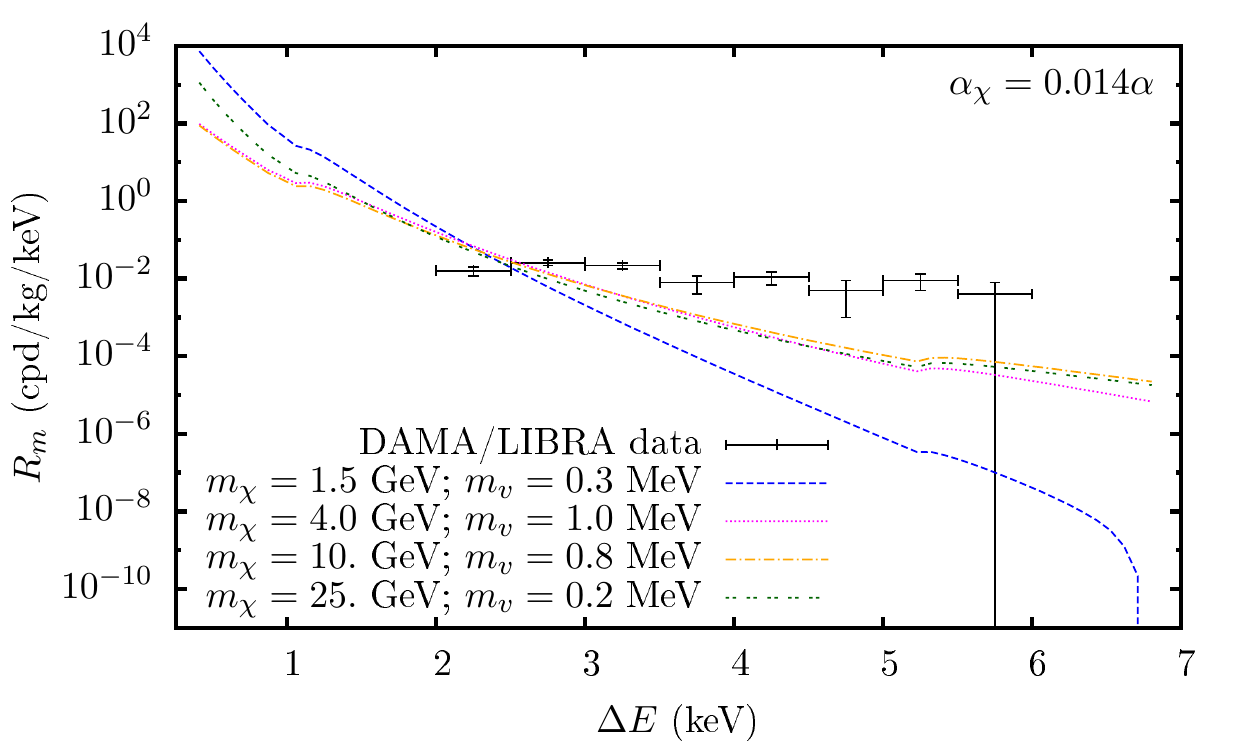}
               \includegraphics[width=0.49\textwidth]{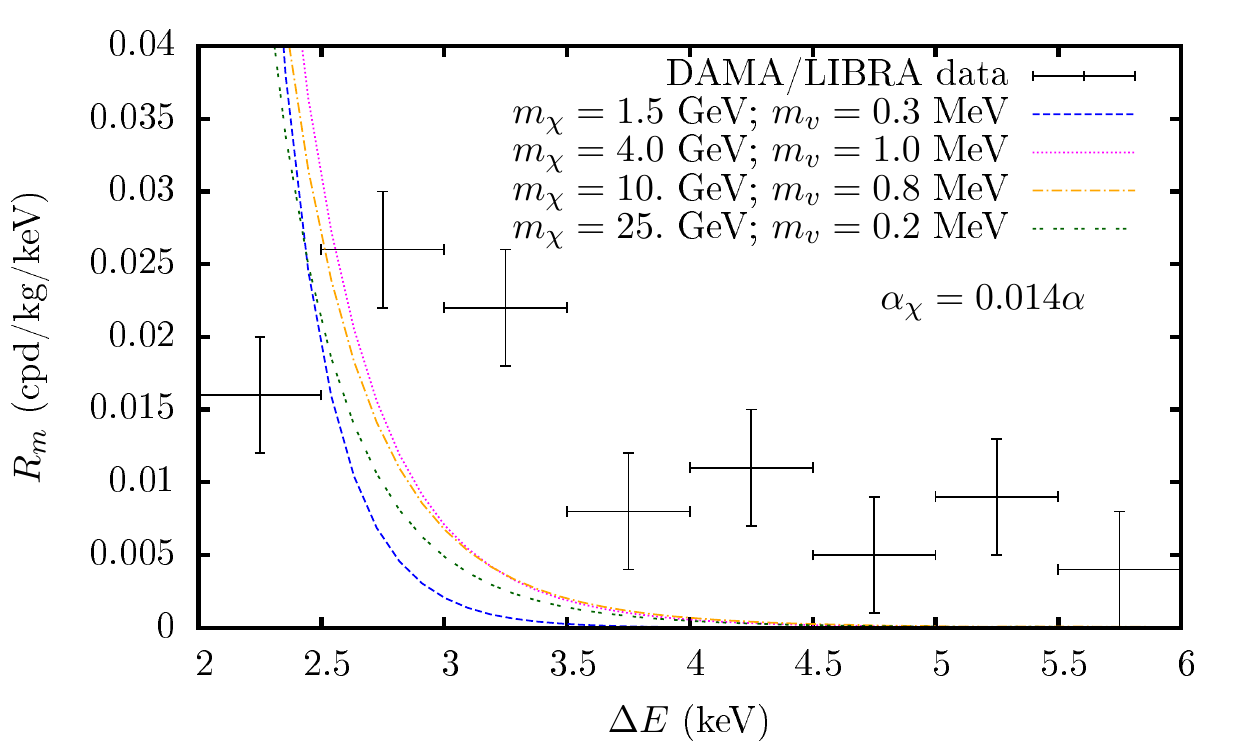}
		\caption{Calculated modulated event rate spectrum $R_m$ for DAMA for a few specific choices of DM parameters which are able to replicate the amplitude of the observed modulation.}
		\label{fig:shapeDAMA}
	\end{center}
\end{figure}

Finally, note that in performing the DAMA signal  analysis, we have paid no attention to the shape of the recoil spectrum, just choosing the parameters to reproduce the total number of counts in the given interval.
This procedure represents the most conservative case; if the detectors were any less efficient, the acceptable values of $\alpha_\chi$ would be forced to be larger.
Taking these factors into account can therefore only strengthen our conclusions.
In Fig.~\ref{fig:shapeDAMA}, the calculated spectrum is compared to the results of the DAMA experiment for a few specific sets of DM parameters that can reproduce the observed modulation amplitude averaged over the 2 -- 6 keV interval.
As to the energy shape of the modulation spectrum, the predictions for the electron recoil are more peaked near the threshold than data would suggest, and have very few events above 3 keV. This is consistent with findings of previous studies \cite{Kopp2009}.

%----------------------------------=============================-----------------------
%----------------------------------=============================-----------------------
\subsection{XENON100 Analysis}\label{sec:results-XE100}

A recent analysis of data from the XENON100 experiment has also investigated WIMP-induced electron-recoil events~\cite{XENONcollab2015,XENONcollab2015a}.
These experiments also observed modest evidence for an annual modulation (at the $2.8\sigma$ level) -- though the phase does not match perfectly with that observed by DAMA \cite{XENONcollab2015a}.
By assuming their result was a positive measurement of an annual modulation, the XENON Collaboration \cite{XENONcollab2015a} (see also Ref.~\cite{XENONcollab2015}) determined the best fit for their data to indicate an unmodulated event rate of
\begin{equation}
R_0^{\rm Xe100}=5.5(6)\E{-3}\,{\rm cpd/kg/keV},
\end{equation}
with a modulation amplitude of
\begin{equation}
R_m^{\rm Xe100}=2.7(8)\E{-3}\,{\rm cpd/kg/keV},
\label{eq:Xe100-Rm}
\end{equation}
with a quoted a background of 5.3\E{-3}\,{\rm cpd/kg/keV} \cite{XENONcollab2015}.
Note that the background (or unmodulated signal) is smaller than the DAMA modulation amplitude by a factor of two.

%-----------------------------------

%====================

The XENON100 Collaboration has performed a detailed analysis of the electron recoil acceptance and efficiency; see, e.g., Refs.~\cite{AprileAxion2014,AprileAP2014,Aprile2012a}, and references therein.
In order to compare the calculated event rate with that observed in XENON100 it first is necessary to convert the calculated event rate as a function of the deposited energy to the rate as a function of the generated photoelectrons (PE), $n$.
The relation between the deposited energy (electron recoil energy), and the produced number of photoelectrons is given in Fig.~2 of Ref.~\cite{AprileAxion2014}.
We model this as a power law:
$N(\Delta E)=\Delta E^x$,
and take $x=1.58$, which gives the best fit at $n=3\;{\rm PE}$ ($\Delta E\simeq2\;{\rm keV}$), noting that the signal is dominated by lower energies.
Then, the generated event rate for $n$ photoelectrons is obtained by applying ``Poisson smearing'' to the calculated differential rate:
\begin{equation}
\mathcal{R}_n = \int_0^\infty \mathcal{R}(\e) P_n(\e)\d\e,
\end{equation}
where the Poisson distribution is
\[
P_n(\e) = \Exp{-N(\e)}\frac{N(\e)^n}{n!},
\]
as in Ref.~\cite{AprileAxion2014}.
Then, to calculate the event rate as a function of the {\sl detected} photoelectrons, $S1$, both the detector resolution and the electron-recoil acceptance must be taken into account.

The electron recoil acceptance, as a function of the observable scintillation photoelectrons $S1$, is given in Fig.~1 (bottom) of Ref.~\cite{AprileAxion2014}.
Roughly, the acceptance rate can be given by the expression
\begin{equation}
A(S1) \approx C_{\rm eff} \left( 1 - \Exp{-S1/3} \right),
\end{equation}
where $C_{\rm eff}$ is an efficiency parameter with a best-fit value around 0.9 \cite{AprileAxion2014}.
To be conservative, we take $C_{\rm eff}=0.85$.
To take the finite resolution of the detectors into account, we convolute the rate with a Gaussian $g_n(S1)$, centred at $S1=n$, and with a standard deviation of $\sqrt{n}\sigma_{\rm PMT}$, where
$\sigma_{\rm PMT}=0.5{\rm PE}$ is the resolution of the XENON100 photomultiplier tube (PMT) detectors \cite{AprileAP2014}.

\begin{figure}
	\begin{center}
		\includegraphics[width=0.49\textwidth]{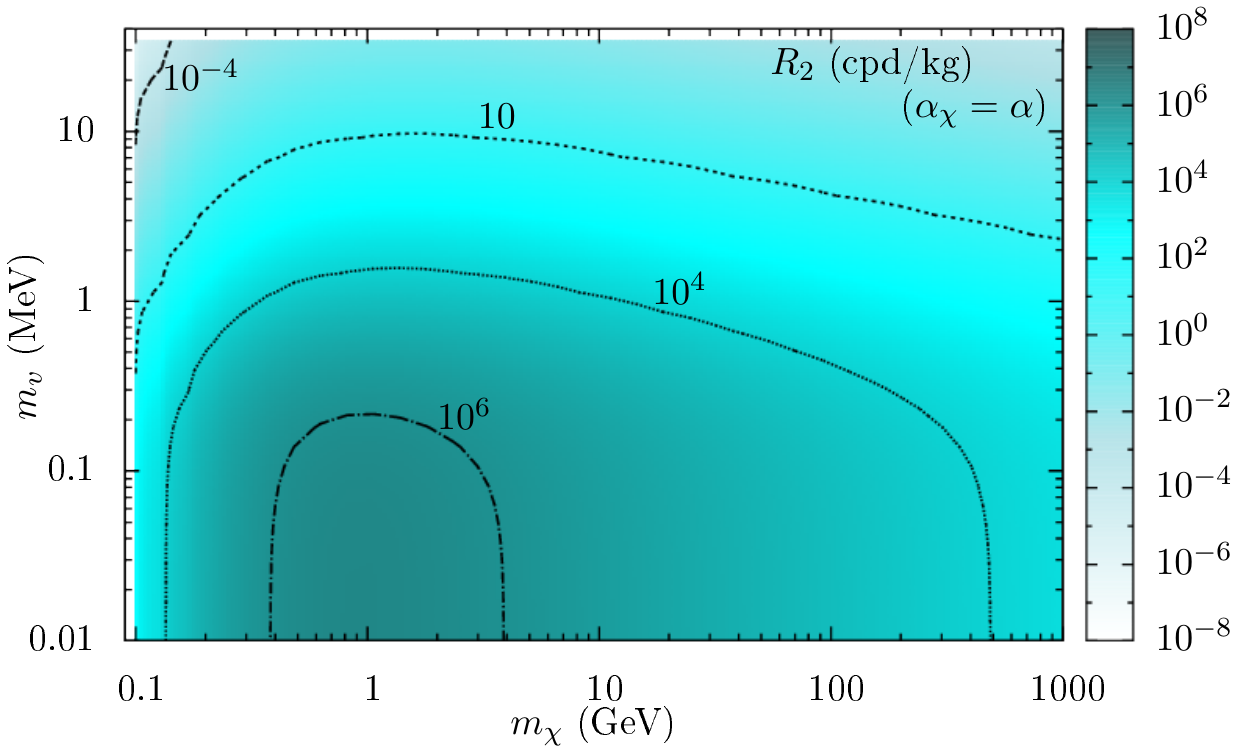}
		\includegraphics[width=0.49\textwidth]{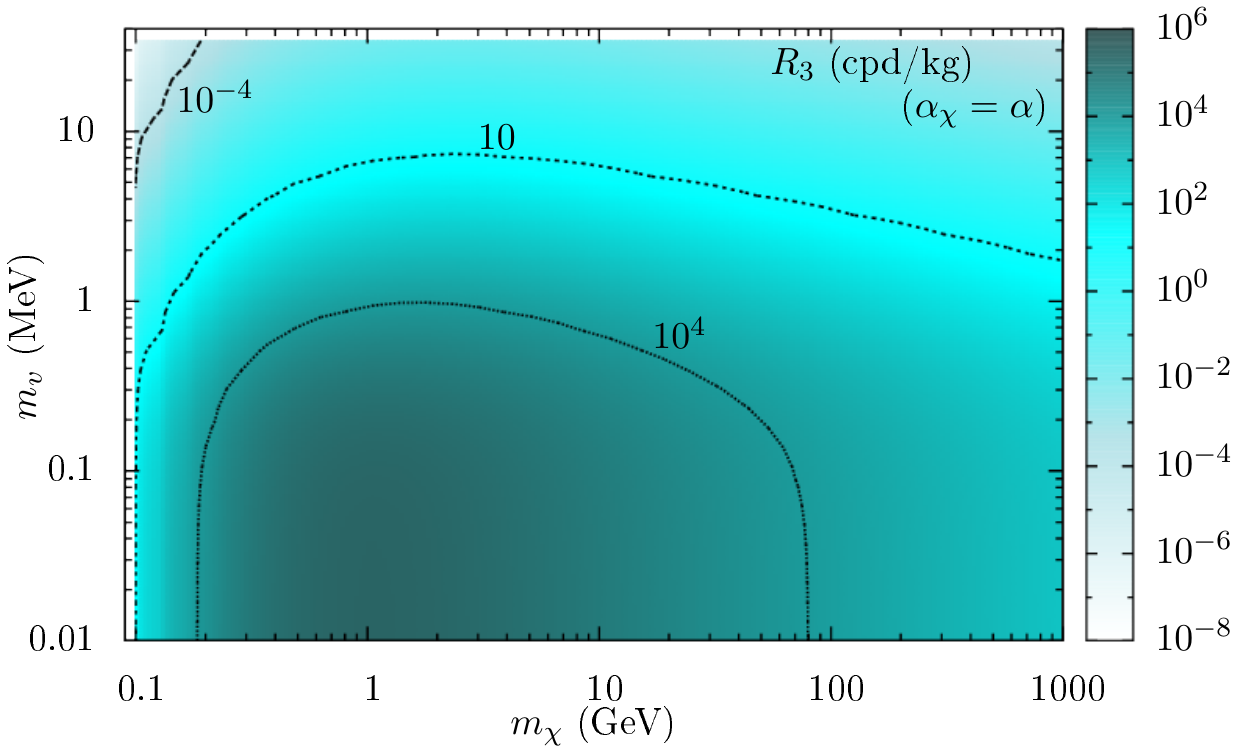}
		\caption{Calculated scintillation event rate for Xe ({\sl top}) for 2 PE, and ({\sl bottom}) for 3 PE.}
		\label{fig:Xe100only-2-3PE}
	\end{center}
\end{figure}

The final detected event rate as a function of observed photoelectrons is thus
\begin{equation}
\widetilde{\mathcal{R}}(S1)=A(S1)\sum_{n=1}^{\infty}g_n(S1)\mathcal{R}_n.
\label{eq:Xe100-R(S1)}
\end{equation}
In order to compare the results with those of the DAMA experiment, we follow Ref.~\cite{XENONcollab2015} and integrate between $S1=3\;{\rm PE}$ and $14\;{\rm PE}$, corresponding roughly to the $2$--$6\;{\rm keV}$ interval. Again, to aid in the comparison with the DAMA results, we divide the result by 4 keV to make the units consistent.
Note that the summation in Eq.~(\ref{eq:Xe100-R(S1)}) converges very quickly, due to the huge enhancement coming from lower energy events, as shown in Fig.~\ref{fig:Xe100only-2-3PE}; we also note that the integration depends strongly on the lower $S1$ bound, but is essentially independent of the upper bound (so long as it's above 5 or 6 PE).

Though the specifics of the way the Gaussian and Poisson ``smearing'' are taken into account for the calculations of the DAMA and XENON100 rates differ, the overall effect is essentially the same. 
The details provided by the XENON100 Collaboration (in, e.g., Refs.~\cite{AprileAxion2014,AprileAP2014,Aprile2012a,XENONcollab2015}) allows us to be rather precise.

%====================

\begin{figure}
	\begin{center}
		\includegraphics[width=0.49\textwidth]{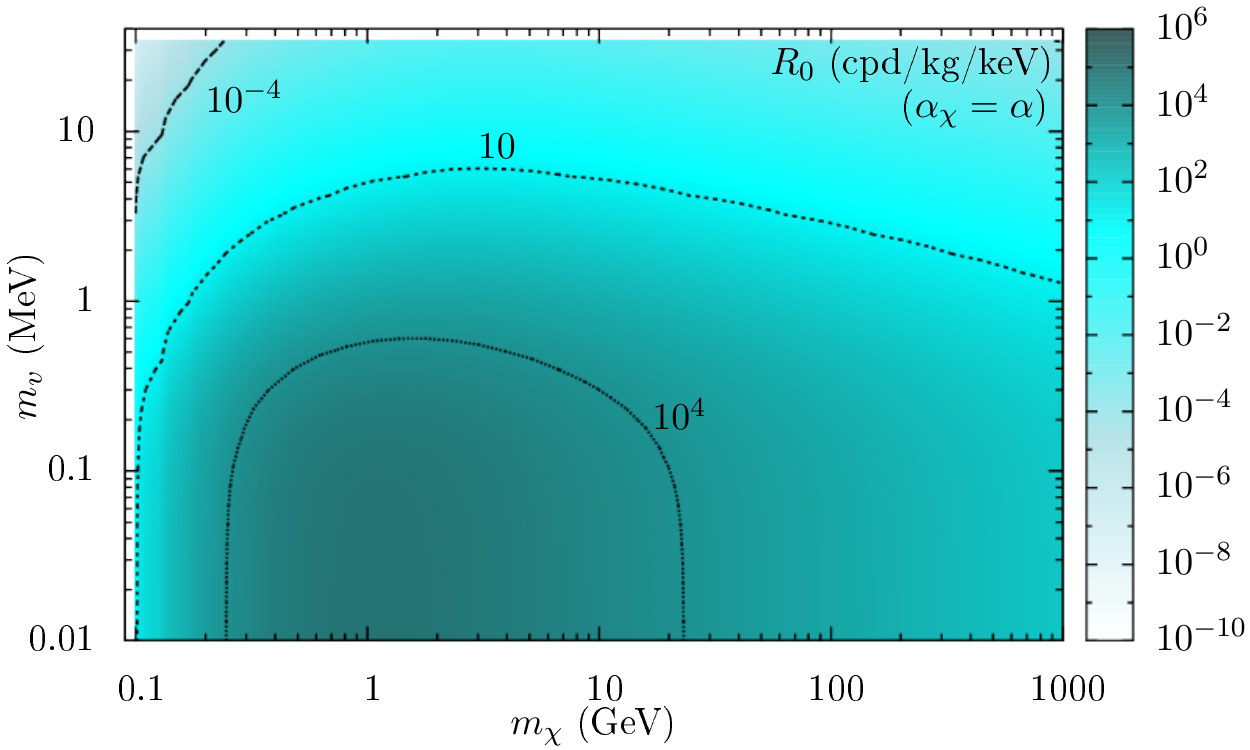}
		\includegraphics[width=0.49\textwidth]{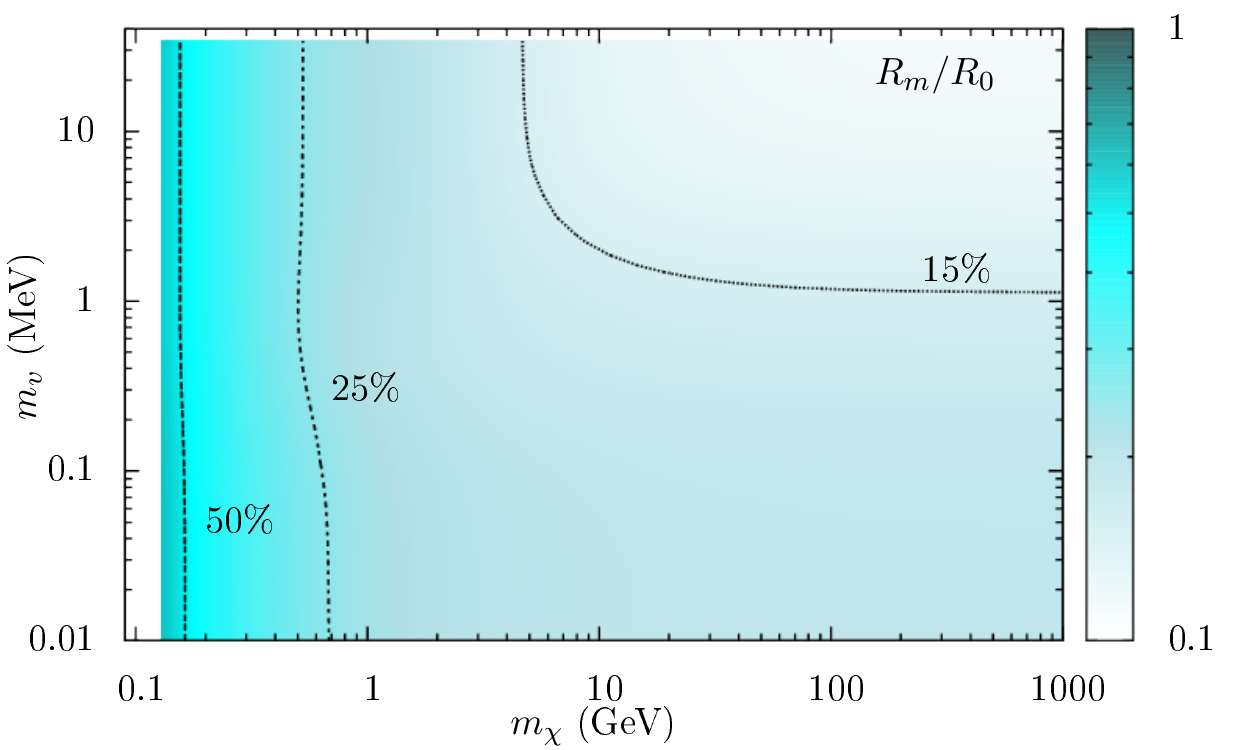}
		\caption{The calculated ({\sl top}) unmodulated event rate (for fixed $\alpha_\chi=\alpha$), and ({\sl bottom}) modulation fraction ($R_m/R_0$), for the scintillation signal in the 3 -- 14 PE interval (corresponding to 2 -- 6 keV) for Xe.}
		\label{fig:Xe100only}
	\end{center}
\end{figure}

%----------

In Fig.~\ref{fig:Xe100only} we present our calculations for the unmodulated event rate $R_0$ (for a fixed coupling $\alpha_\chi=\alpha$) and the modulation fraction $R_m/R_0$ for the XENON100 scintillation experiment, in the 3 -- 14 PE range. 
The modulation fraction observed in the XENON100 experiment (\ref{eq:Xe100-Rm}) is extremely large.
We find, however, that this alone is not enough to discount the WIMP hypothesis as a source for the modulations.
The calculated modulation fraction is very large, easily reaching 50\% for very low values of $m_\chi<1$ GeV.
Note that the oscillation fraction is independent of the coupling constant.

\begin{figure}
	\begin{center}
\includegraphics[width=0.49\textwidth]{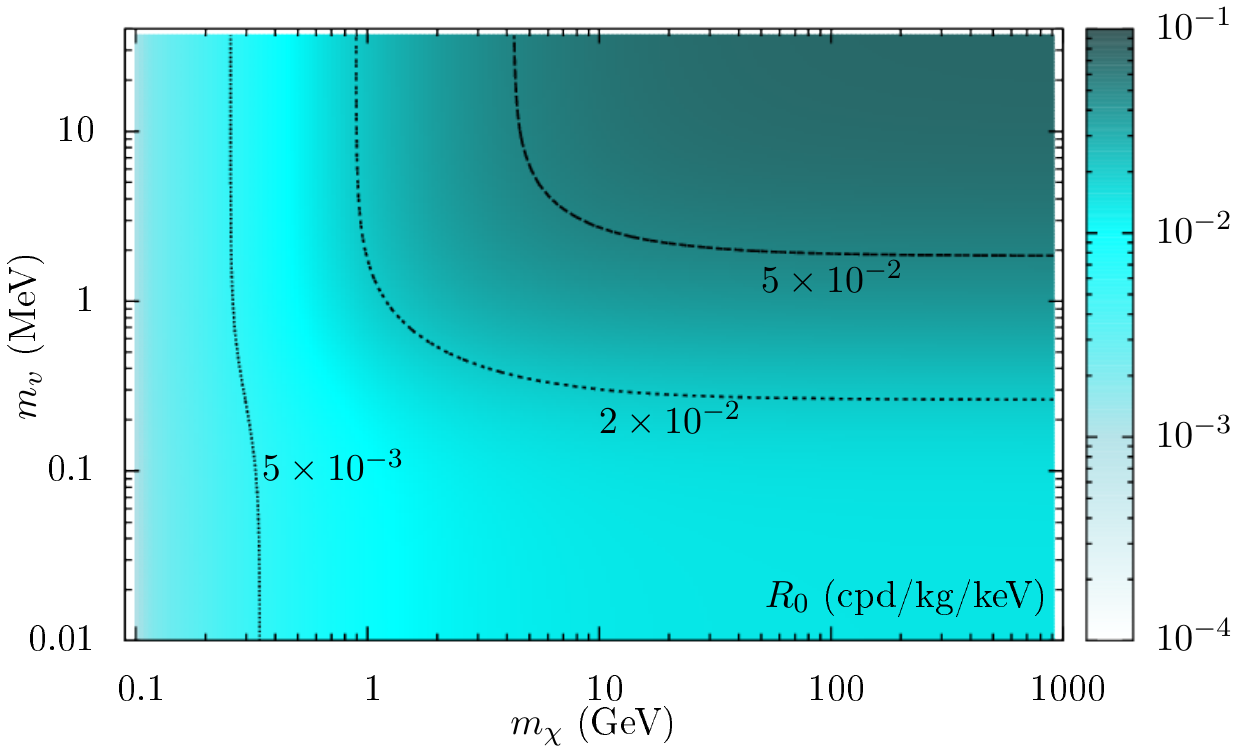}
\includegraphics[width=0.49\textwidth]{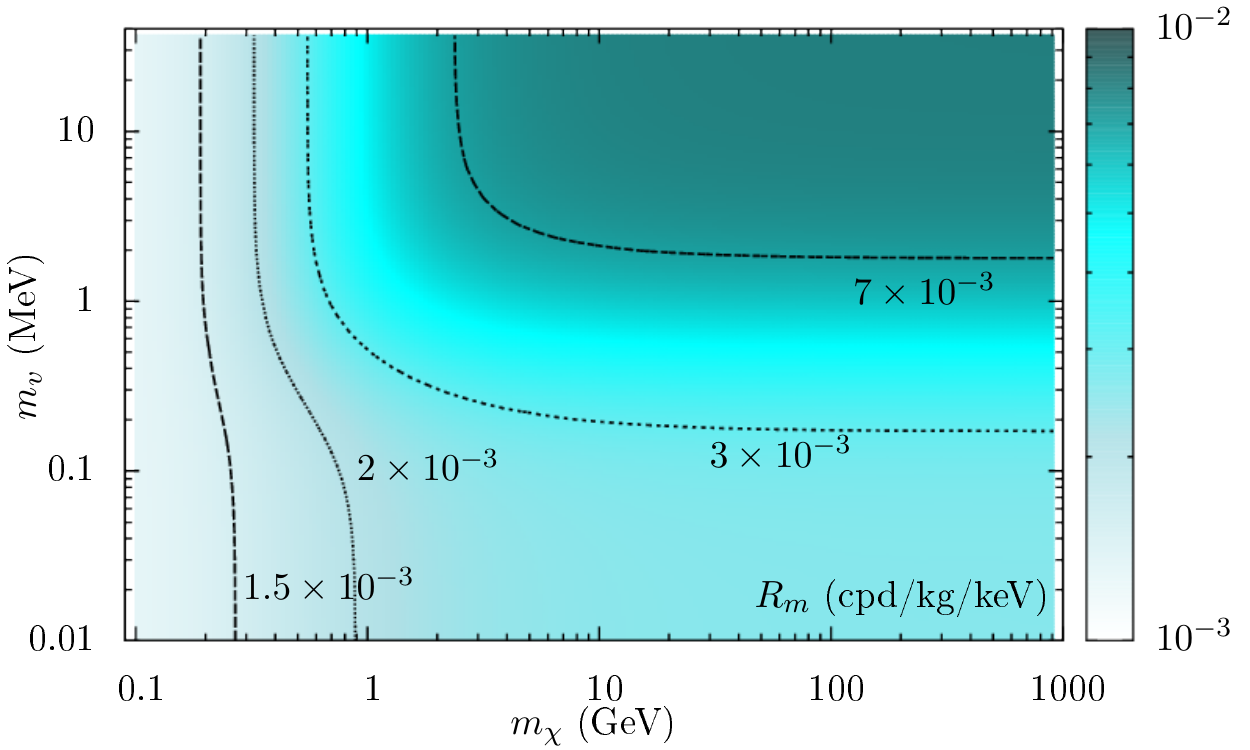}
\caption{({\sl Top}) The umodulated event rate, and ({\sl bottom}) the modulation amplitude, that would be expected in the XENON100 scintillation experiment in the 3 -- 14 PE interval (corresponding to 2 -- 6 keV) assuming the DAMA modulation signal is a positive WIMP detection (the value of $\alpha_\chi$ for each point on the parameter plot is shown in Fig.~\ref{fig:axreq}).
}
		\label{fig:XENON-scint}
	\end{center}
\end{figure}

\begin{figure}
	\begin{center}
		\includegraphics[width=0.49\textwidth]{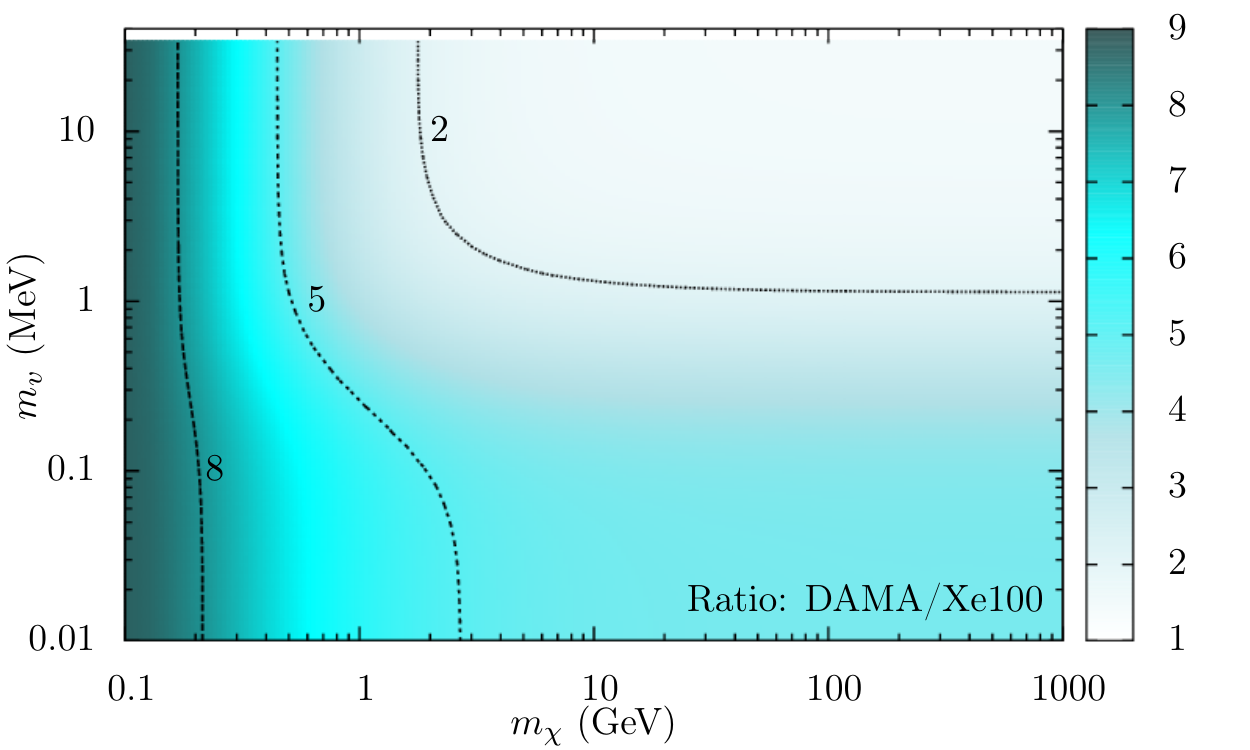}
		\caption{Ratio of the calculated event rate for DAMA (in the 2 -- 6 keV interval) to that expected for XENON100 (in the 3 -- 14 PE interval). Note that the ratio is highly dependent on the detector efficiency and resolution, but is essentially independent of the DM velocity distribution.}
		\label{fig:ratio}
	\end{center}
\end{figure}

By assuming the DAMA result is due to electron-interacting WIMPs, we can calculate the expected scintillation signal in xenon relevant to the XENON100 electron-recoil experiment.
For each set of DM and mediator masses, we calculate the coupling required to reproduce the DAMA modulation signal in the 2 -- 6 keV interval, assuming it is due to WIMP--electron scattering on the NaI crystal.
These couplings, shown in Fig.~\ref{fig:axreq} ({\sl bottom}), are used as inputs into the calculations for xenon.
Figure \ref{fig:XENON-scint} shows the resulting calculated 
 event rates that would be generated in liquid xenon summed between 3 and 14 photoelectrons (PE), as in the XENON100 electron recoil experiment \cite{XENONcollab2015,XENONcollab2015a}.
In Fig.~\ref{fig:ratio}, we also directly plot the ratio of the calculated event rates for DAMA and XENON100 in the relevant energy intervals.

It appears that there is a region below $m_\chi\sim0.5$GeV in which the DAMA result may be compatible with the XENON100 limits.
The unmodulated event rate comfortably sits below the limit of $\sim 5\E{-3}\;{\rm cpd/kg/keV}$, and the modulation fraction is very large, between 25\% and 50\%.
We remind, however, that this is very dependent on the low-energy efficiency and cut-acceptance criteria of the DAMA experiment, which is not detailed in the literature to the same extent as it is for XENON100.
In lieu of a more thorough investigation of the detector efficiency, acceptance, and resolution by the DAMA Collaboration, we employed a simple Gaussian resolution profile (based on resolution measurements of the DAMA Collaboration \cite{Bernabei2008b}). 
This amounts to a very generous assumption for the DAMA modulation, while we take very conservative assumptions for the XENON100 rate.

Nevertheless, tight constraints can be placed upon the considered WIMP models as an explanation for the DAMA modulation based on the XENON100 electron recoil constraints \cite{XENONcollab2015}.
Based on our calculations, for the region above $m_\chi\gtrsim10\un{GeV}$ and $m_v\gtrsim2\un{MeV}$ (corresponding to the $5\times10^{-2}\un{cpd/kg/keV}$ contour of Fig.~\ref{fig:XENON-scint} (top), the exclusion is $7.5\s$, taking into account both the DAMA and XENON100 uncertainties.
For the region above $m_\chi\gtrsim1\un{GeV}$ and $m_v\gtrsim0.3\un{MeV}$ (corresponding to the $2\times10^{-2}\un{cpd/kg/keV}$ contour of Fig.~\ref{fig:XENON-scint} (top), the exclusion is $5.2\s$.
The region below $m_v\lesssim0.2$ MeV is ruled out based on stellar bounds \cite{An2013}, and the region above $m_v\gtrsim2$ MeV is ruled out based on the size of the coupling strength.

\begin{figure}
	\begin{center}
		\includegraphics[width=0.49\textwidth]{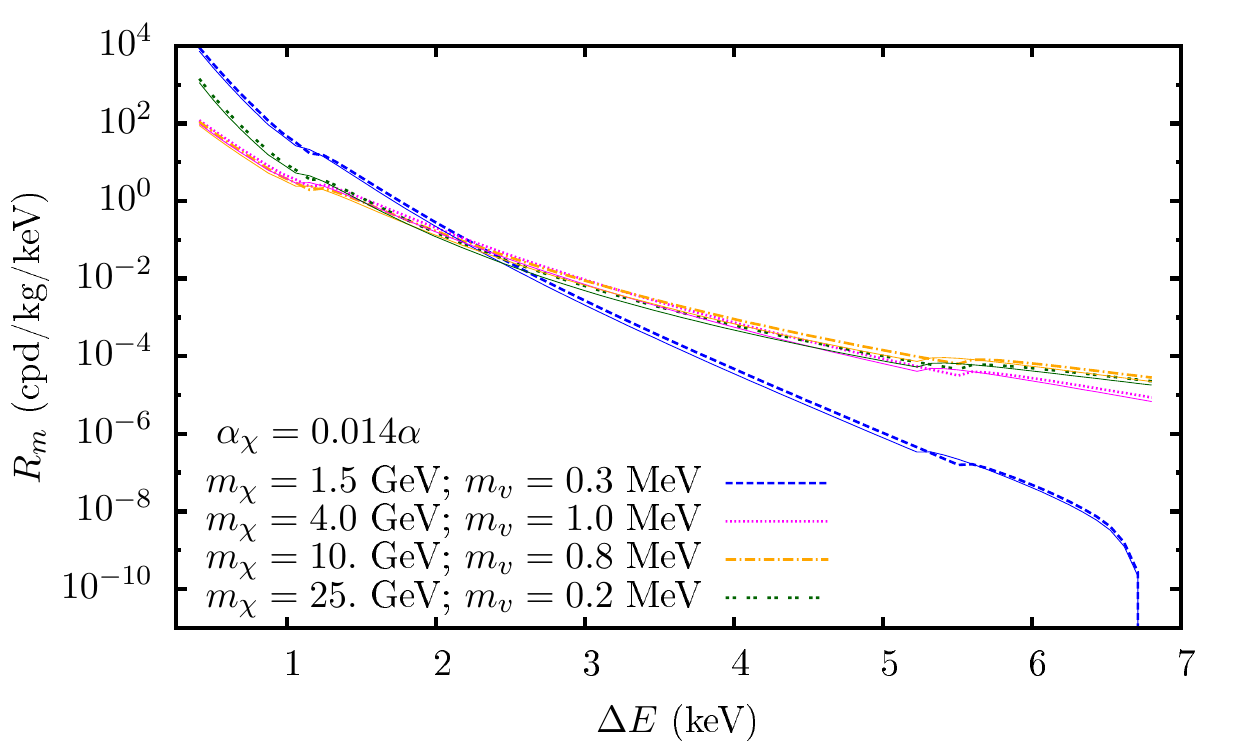}
		\caption{Calculation of the modulated ionisation event rate spectrum $R_m$ for xenon (relevant to XENON100) for a few specific choices of DM parameters which are able to replicate the amplitude of the observed DAMA modulation. The corresponding signals generated in NaI are plotted in thin solid lines (almost indistinguishable from the xenon rates on this scale).}
		\label{fig:shapeXe100}
	\end{center}
\end{figure}

In order to demonstrate the energy-dependence of the event rate, in Fig.~\ref{fig:shapeXe100}, we plot the modulated part of the ionisation event rate for xenon for a few specific choices of DM parameters that are able to reproduce (the amplitude of) the DAMA modulation signal. Note that here we plot the bare event rate as in Eq.~(\ref{eq:diffEventRate}), not taking into account the Poisson smearing or detector resolution.
It is clear that a detailed knowledge of the detector efficiency at very low energies is crucial for interpreting observed scintillation signal in terms of electron interacting DM.
A discussion of the low energy efficiency is presented in Ref.~\cite{AprileAxion2014} (see also Refs.~\cite{Aprile2011,Aprile2012a,Aprile2012,Szydagis2011} and Ref.~\cite{Collar2011}).

%----------------------------------------------------------
\subsection{Massless mediator ($m_v=0$) case}\label{sec:results-mv0}

\begin{figure}
	\begin{center}
		\includegraphics[width=0.49\textwidth]{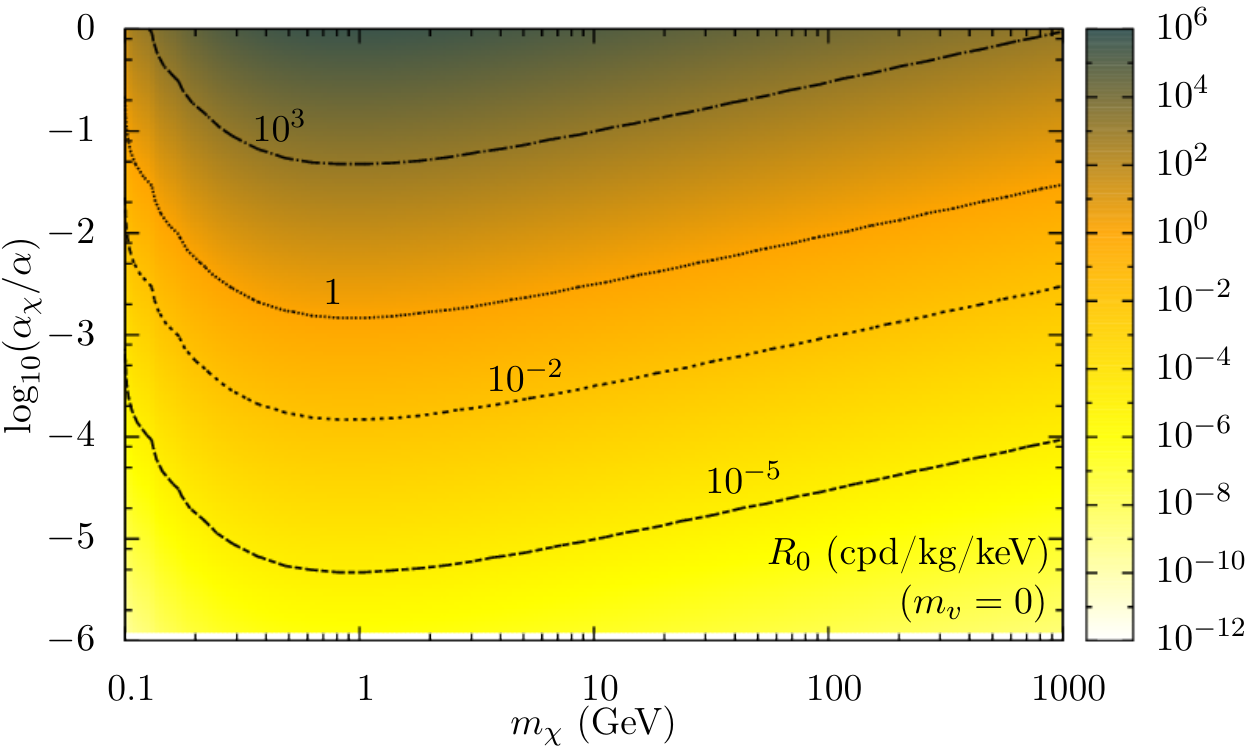}
		\includegraphics[width=0.49\textwidth]{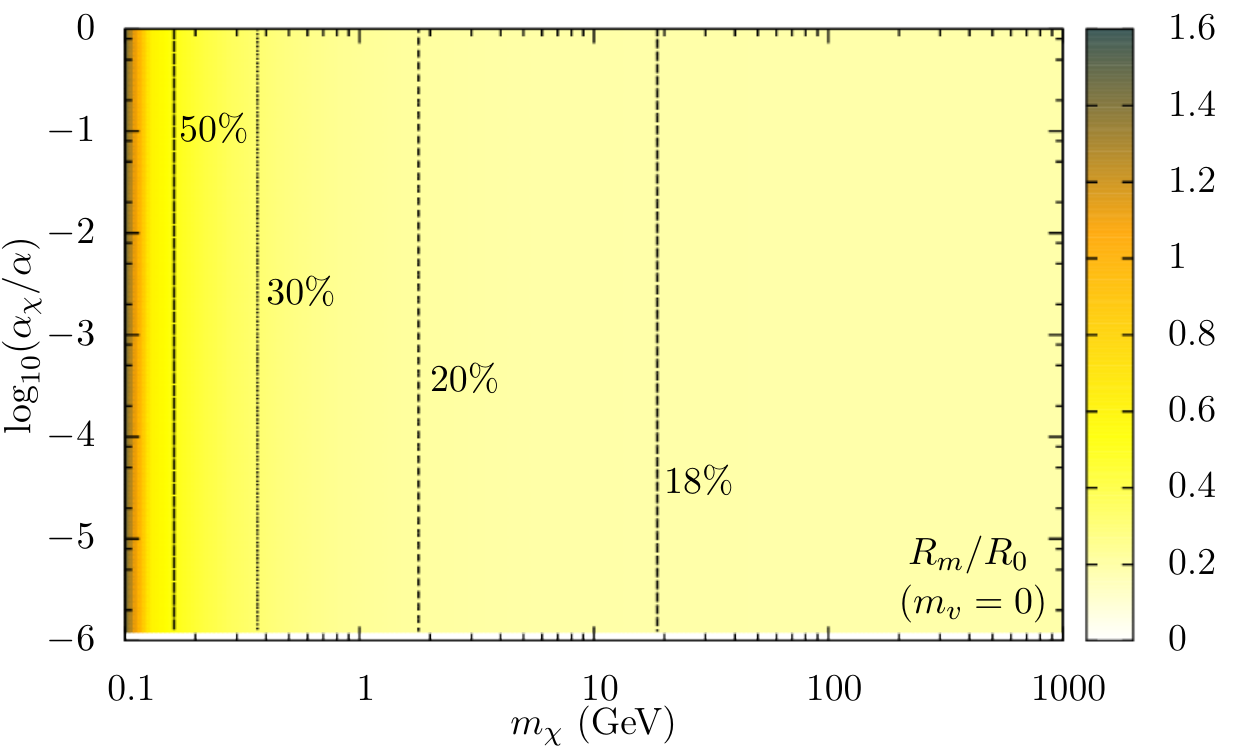}
		\caption{Unmodulated event rate ({\em top}), and modulation fraction ({\em bottom}), for DAMA in the 2--6 keV interval, for the massless mediator case $(m_v=0)$.}
		\label{fig:DAMAmv0}
	\end{center}
\end{figure}

\begin{figure}
	\begin{center}
		\includegraphics[width=0.49\textwidth]{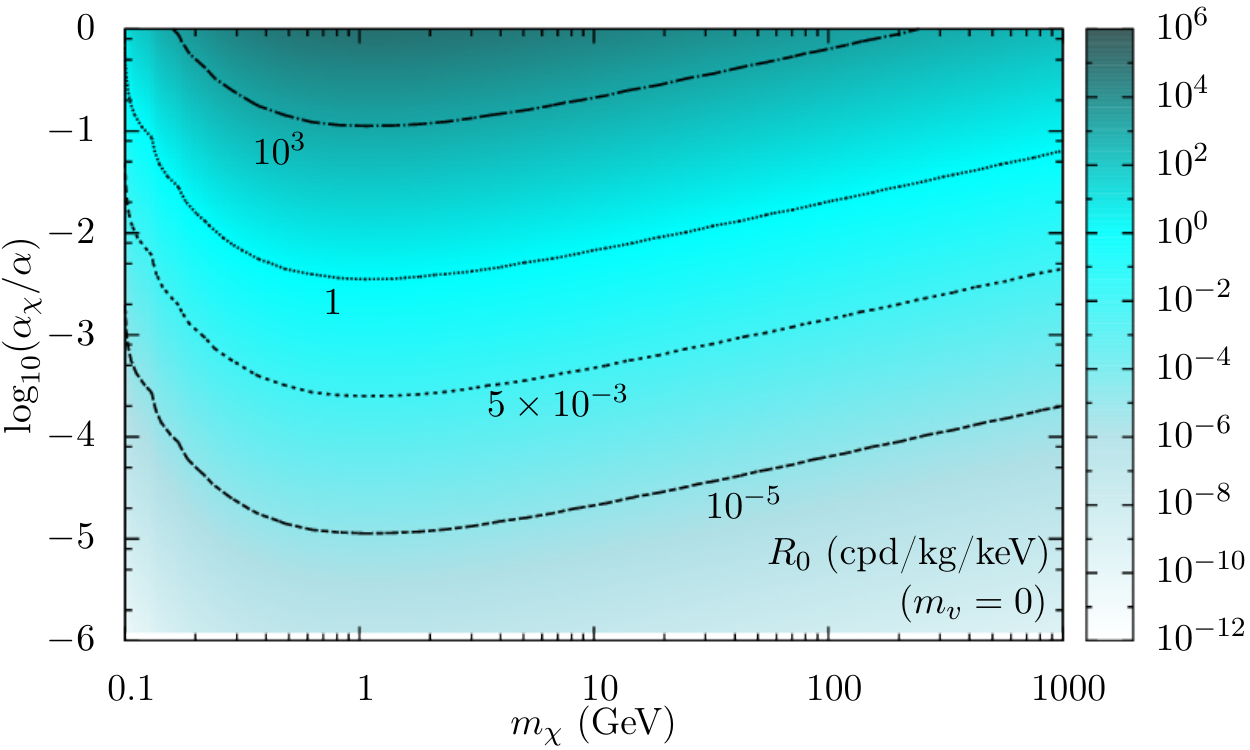}
		\includegraphics[width=0.49\textwidth]{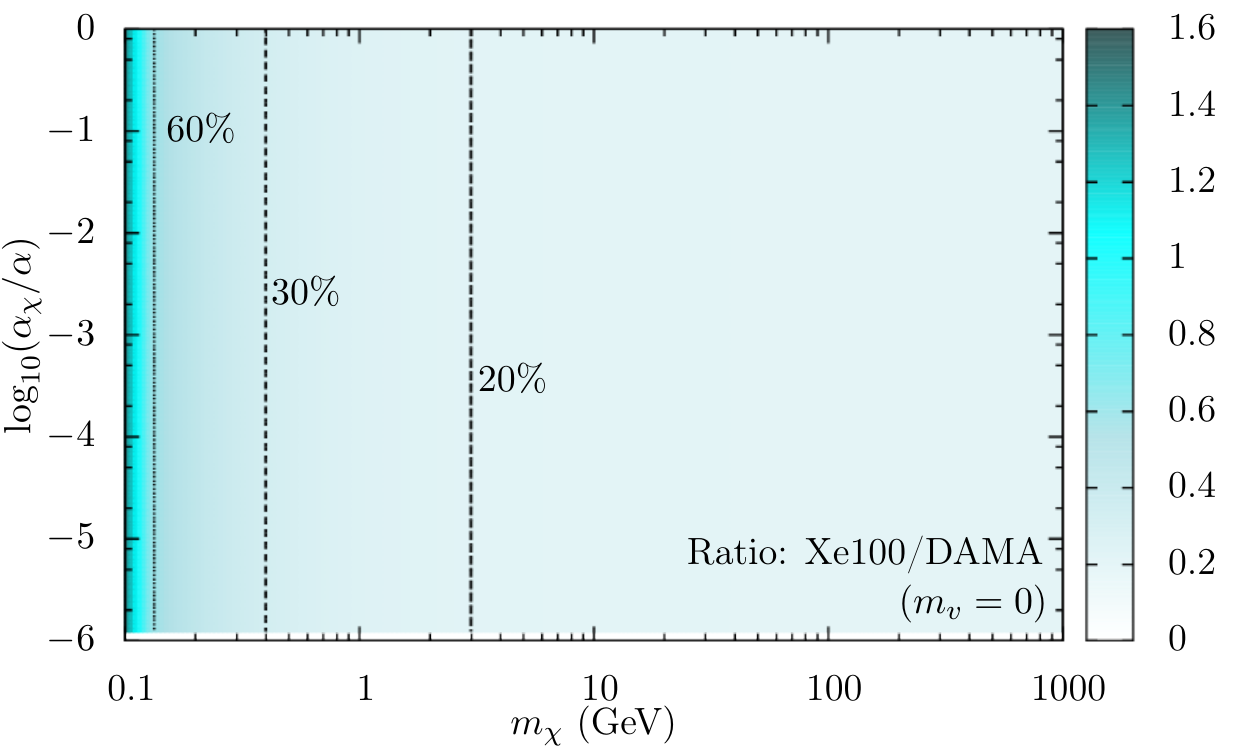}
		\caption{({\em Top}) The unmodulated event rate for XENON100 in the 3 -- 14 PE interval (corresponding to 2 -- 6 keV), for the massless mediator case $(m_v=0)$. ({\em Bottom}) Ratio of the event rate for DAMA to that of Xe100 (in the 2 -- 6  keV/3 -- 14 PE range) for the $m_v=0$ case.}
		\label{fig:Xe100mv0}
	\end{center}
\end{figure}

For the case of the vector mediator, the constraints on its couplings to normal matter 
are significantly weakened as $m_v$ is taken to sub-eV values, as discussed in Ref.~\cite{An2013}.
Therefore it will be useful to discuss the case of a purely massless mediator, $m_v=0$, and  we consider this case separately. 
Figure~\ref{fig:DAMAmv0} shows calculations of the event rate and modulation fraction expected in the relevant energy interval for the DAMA experiment, as a function of the effective coupling constant $\alpha_\chi$, and the WIMP mass, $m_\chi$.
In this case, the large modulation is also present, and the event rates are significantly larger than in the $m_v>0$ case, as expected (see Fig.~\ref{fig:rates-DAMAR0}).
The corresponding calculations relevant to the XENON100 experiment are shown in Fig.~\ref{fig:Xe100mv0}.
Unsurprisingly, the expected event rate is very similar to that for DAMA.

%----------------------------------------------------------
\subsection{XENON10 `ionization only' analysis}\label{sec:results-XE10}

The XENON10 Collaboration \cite{Angle2011} (see also Refs.~\cite{Angle2009,Angle2008,Aprile2011a,Sorensen2009}) has performed an analysis of the ionization-only signal in their liquid xenon detector.
This data has been analysed in terms of low mass electron-interacting WIMPs \cite{Essig2012}, and limits have been set \cite{EssigPRL2012}; see also Refs.~\cite{Armengaud2012,Armengaud2016}.

\begin{figure}
	\begin{center}
\includegraphics[width=0.49\textwidth]{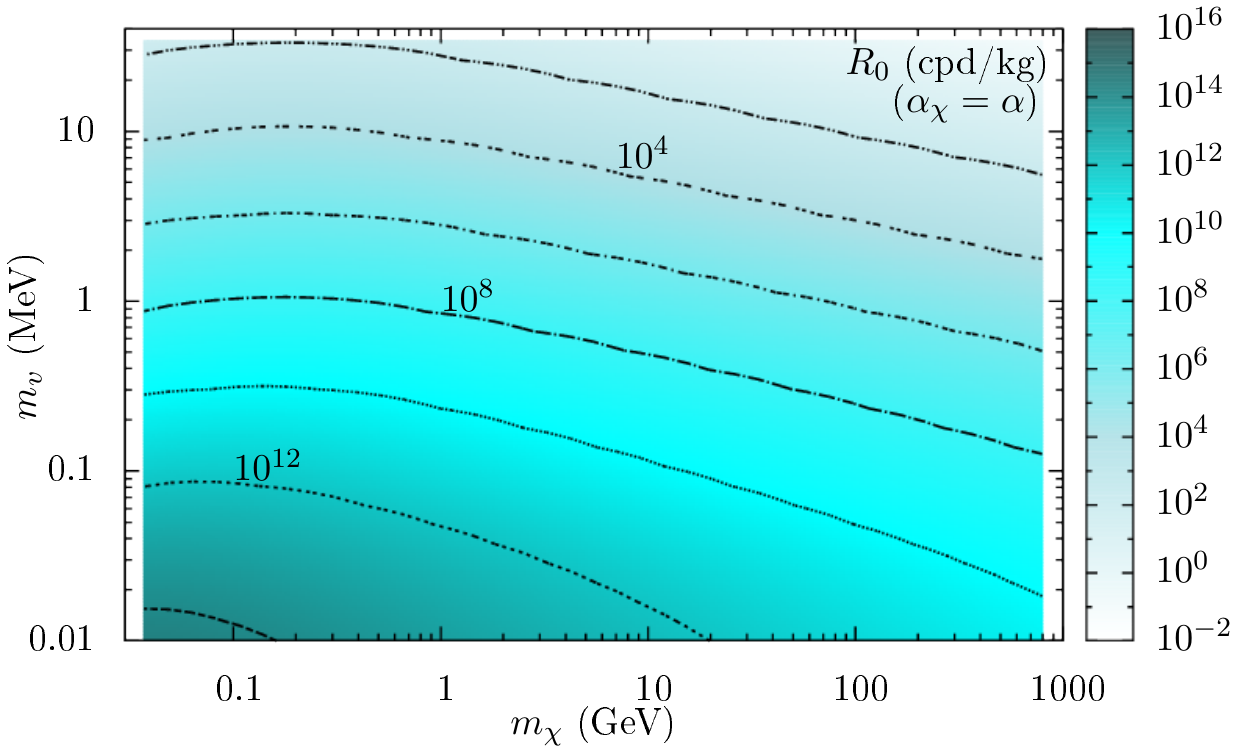}
\includegraphics[width=0.49\textwidth]{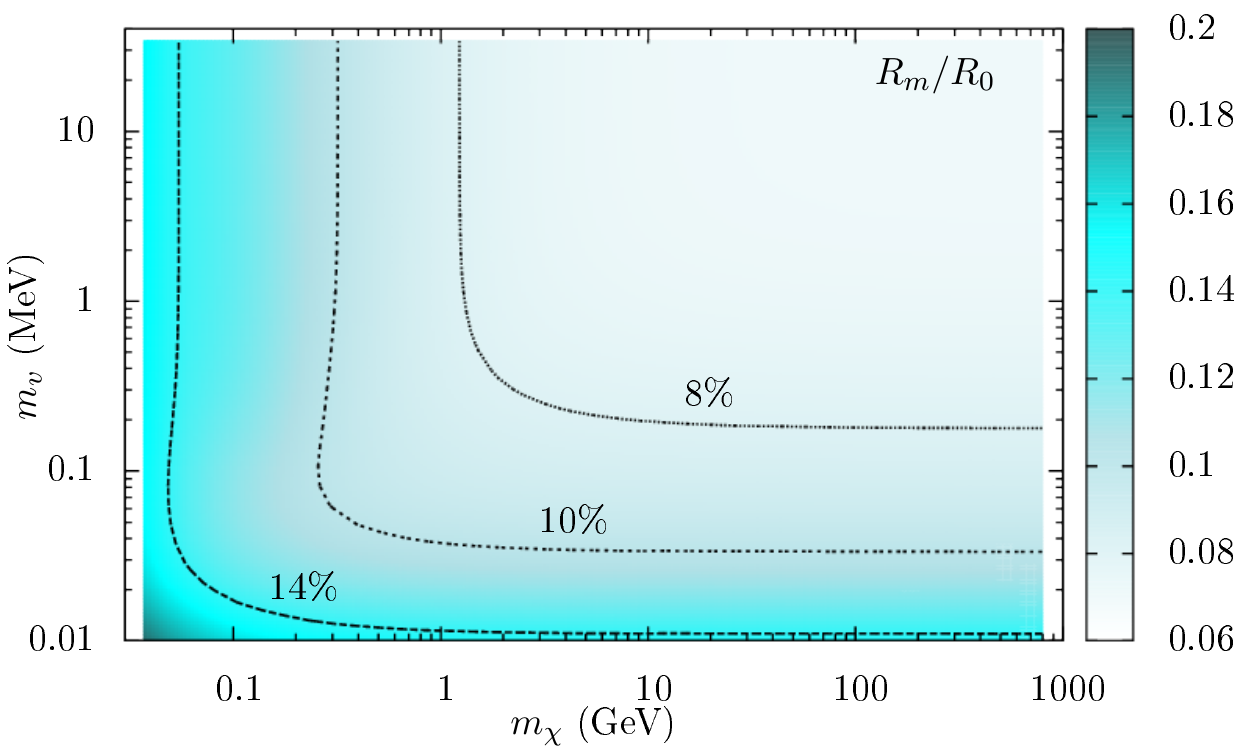}
\caption{Calculation of ({\sl top}) the unmodulated single primary-electron ionization signal in xenon (relevant to the XENON10 experiment \cite{Angle2011}) for a fixed coupling of $\alpha_\chi=\a$, and ({\sl bottom}) the modulation fraction.}
		\label{fig:Xe10}
	\end{center}
\end{figure}

\begin{figure}
	\begin{center}
\includegraphics[width=0.49\textwidth]{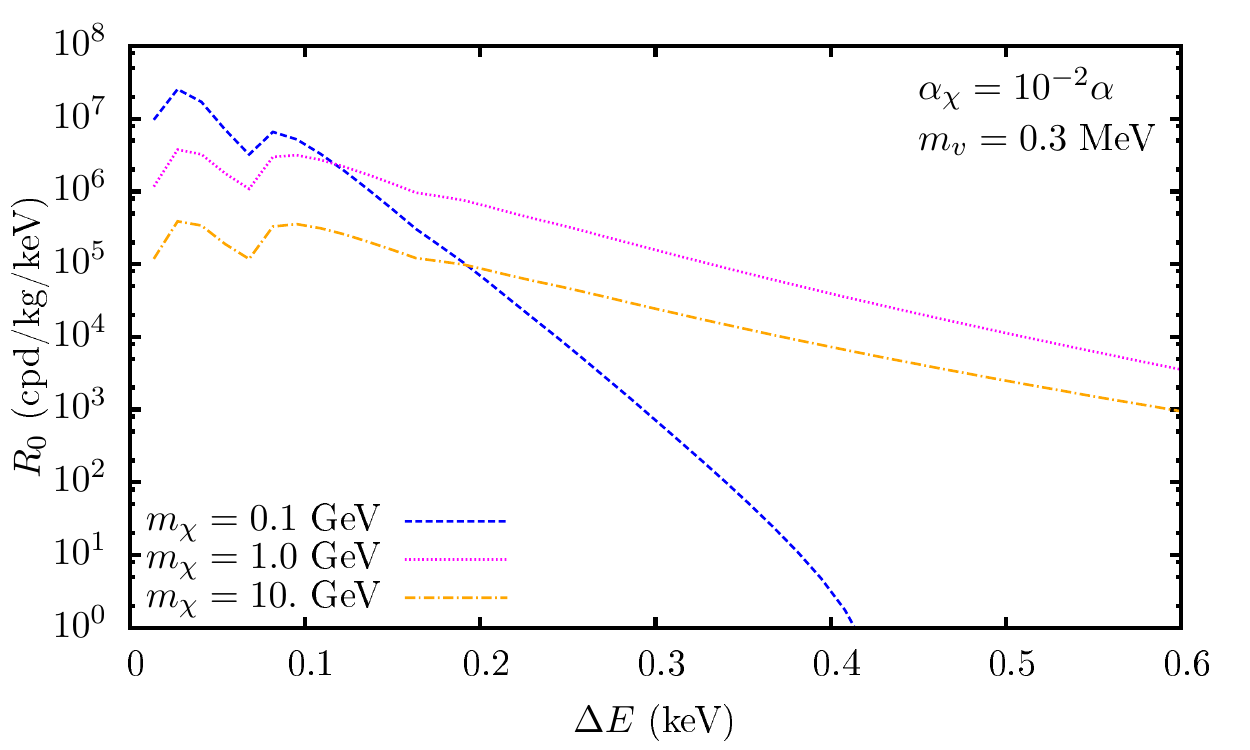}
\caption{Calculation of the spectral shape for the single primary-electron ionization signal (unmodulated) in xenon for a fixed coupling of $\alpha_\chi=10^{-2}\a$ and $m_v=0.3$ MeV, for a few DM masses.}
		\label{fig:shapeXe10}
	\end{center}
\end{figure}

In Fig.~\ref{fig:Xe10} we plot the event rate for the primary ionizations generated in a xenon detector due to the scattering of electron-interacting WIMPs.
Note that this is a lower-bound on the generated events, since the primary ionizations (particularly from lower shells) will also induce secondary ionisations with some probability.
The dominating contribution at low DM masses comes from the upper most shells; this is in agreement with previous calculations~\cite{EssigPRL2012}.
For very large DM masses (and large mediator masses) higher energy ranges play a significant role also.
The modulation fraction for the ionization-only signal is substantially smaller than for the scintillation signal; it is below 10\% for most of the parameter space.
This is because the low-energy cut-off required for the scintillation signal means the observed signal can only originate from the high-energy (and high momentum transfer) tail of the cross section. In this region, the cross section becomes highly velocity dependent, hence the large modulation fraction; see Fig.~\ref{fig:Imx}.
The spectrum of events for a few selected values of the DM mass is given in Fig.~\ref{fig:shapeXe10}.

\begin{figure}
	\begin{center}
\includegraphics[width=0.49\textwidth]{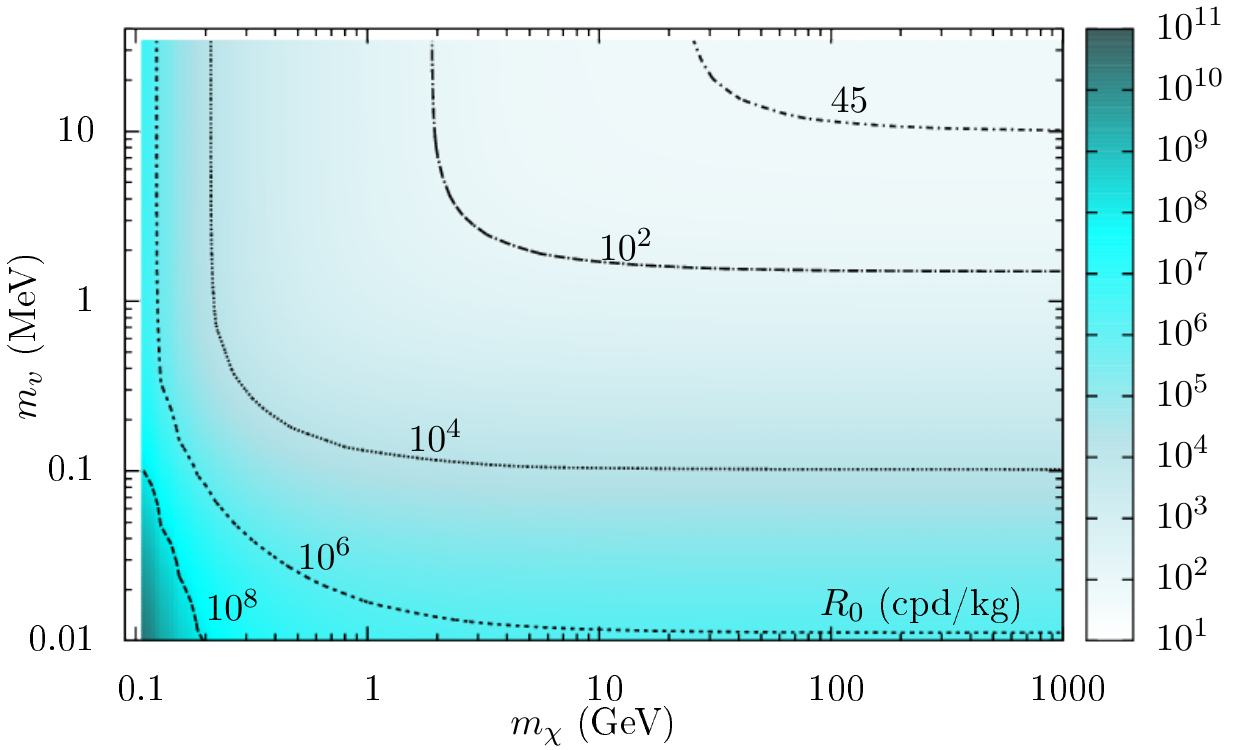}
\caption{Calculation of the expected unmodulated ionization-only signal in xenon (relevant to the XENON10 experiment \cite{Angle2011}) assuming the DAMA modulation signal (\ref{eq:DAMA-Rm}) is a positive WIMP detection.
}
		\label{fig:DaXe10}
	\end{center}
\end{figure}

Figure~\ref{fig:DaXe10} shows calculations of the ionization rate for xenon integrated over all energy depositions (relevant to the ionization-only XENON10 experiment \cite{Angle2011}), assuming the DAMA modulation is due to electron-interacting WIMPs.
Note that presented here is the calculation of primary, or ``first-order'' ionization events only. 
Some fraction of these ionized electrons will recombine emitting photons which may also ionize other atoms. Also, when it is not the outermost electron which is ionized, the decay of the outer electrons to fill the created vacancy will also release photons which will ionize subsequent atoms with some probability.
For a discussion, see Ref.~\cite{EssigPRL2012}.
Therefore, we have actually calculated a lower bound on the expected XENON10 event rate.
Note also that the $m_v=0$ case is already explicitly ruled out here (see Fig.~\ref{fig:DaXe10}), so we do not need to consider it separately.

The XENON10 Collaboration \cite{Angle2011} observes at most 30 cpd/kg; at the 90\% confidence level, the authors of Ref.~\cite{EssigPRL2012} put a bound on the single-electron ionisation rate at 23.4 cpd/kg. The two-electron rate is substantially smaller at $<4.23$ cpd/kg.
Here, it appears as though there may be some part of the parameter space (for very large $m_v$, and $m_\chi\gtrsim5$ GeV) for which the WIMP explanation for the DAMA modulation may be consistent with the XENON10 constraints; note that this is the opposite side (for both $m_v$ and $m_\chi$) that was favored considering the XENON100 comparison.
Still, we are able to place very tight constraints on the DM parameter space.
The region below $m_\chi\lesssim25\un{GeV}$ and $m_v\lesssim10\un{MeV}$ (corresponding to the $45\un{cpd/kg}$ contour of Fig.~\ref{fig:DaXe10}) is excluded at better than the 90\% confidence level.
The region below $m_\chi\lesssim2\un{GeV}$ and $m_v\lesssim1.5\un{MeV}$ (corresponding to the $10^2\un{cpd/kg}$ contour of Fig.~\ref{fig:DaXe10}) can be excluded by many orders of magnitude.

\begin{figure}
	\begin{center}
\includegraphics[width=0.49\textwidth]{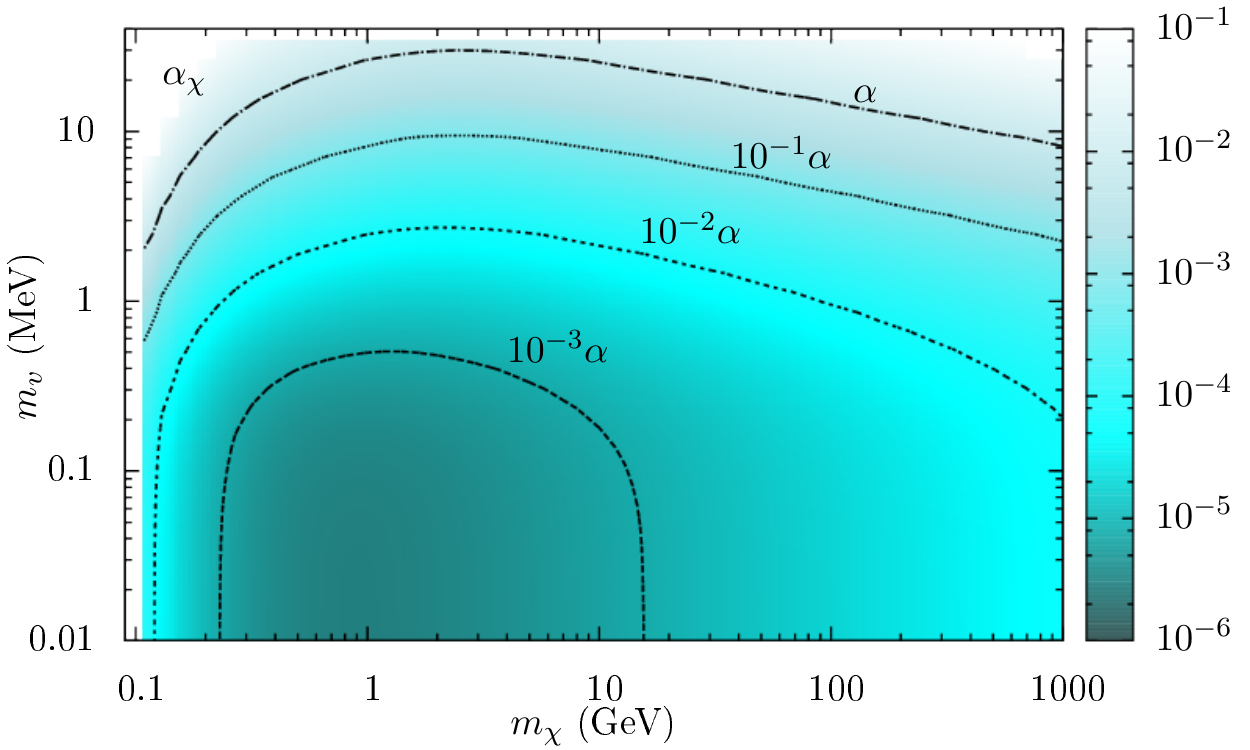}
\includegraphics[width=0.49\textwidth]{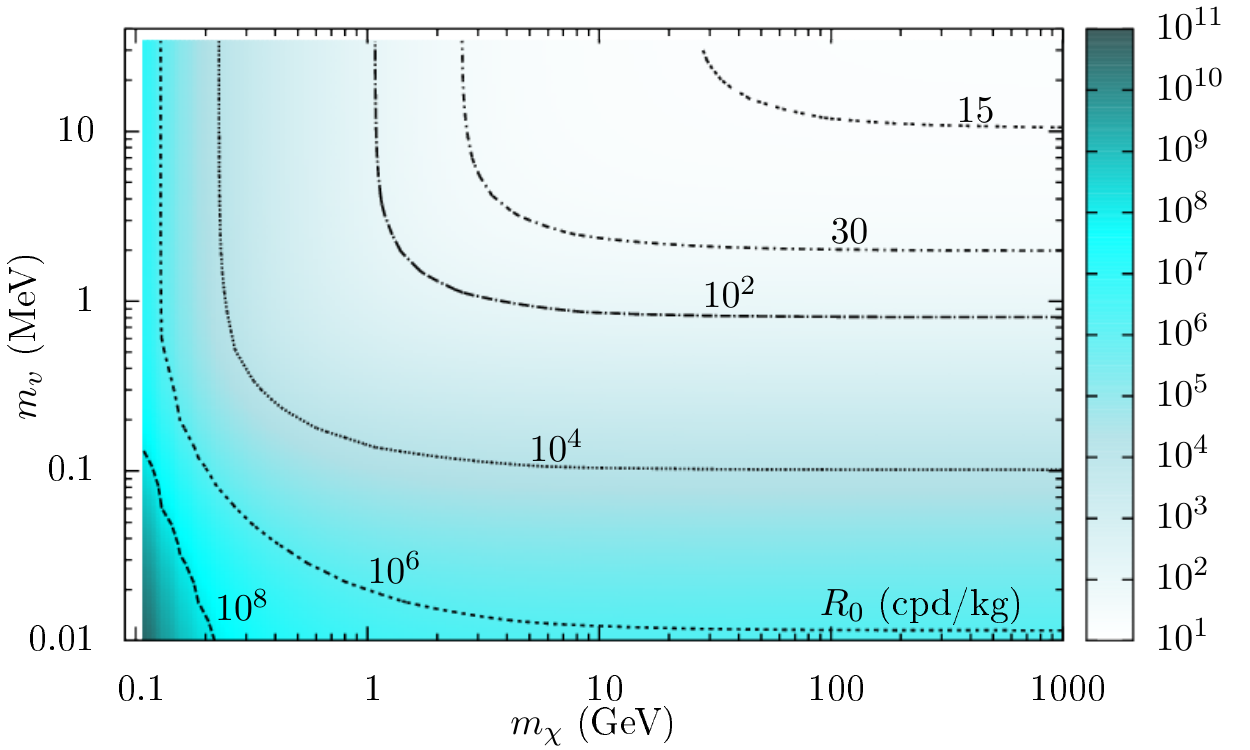}
\caption{Calculations for the expected ionization-only signal in xenon (relevant to the XENON10 experiment \cite{Angle2011}) assuming the modulation signal observed in the XENON100 scintillation experiment \cite{XENONcollab2015a} is due to WIMP--electron scattering. 
({\sl Top}) Value that $\alpha_\chi$ must take to explain the XENON100 modulation (midpoint);
({\sl bottom}) the resulting unmodulated event rate $R_0$ for XENON10 (single-electron primary ionizations only).
}
		\label{fig:XeXe}
	\end{center}
\end{figure}

We can also perform calculations to investigate whether the XENON100 scintillation and XENON10 ionization experiments can be mutually consistent with the electron-interacting WIMP assumption.
Figure~\ref{fig:XeXe} shows calculations of the ``ionization-only'' event rate for xenon (integrated over all energy depositions), assuming the modulation observed in the XENON100 experiment (\ref{eq:Xe100-Rm}) is due to electron-interacting WIMPs.
This shows that for relatively large values of $m_\chi$ and $m_v$ the XENON100 modulation may be compatible with the XENON10 limits (though note that the XENON100 Collaboration does not consider this modulation a positive WIMP detection).

%==============================================
\section{Conclusion}\label{sec:concl}

We have revisited the hypothesis that WIMP-type dark matter scattering on electrons could be an explanation for the anomalous DAMA/NaI and DAMA/LIBRA annual modulation signals.
By performing high-accuracy numerical calculations of atomic ionization, including electron relativistic effects, we have calculated the event rates that would be expected assuming this scenario for several relevant experiments.
Our calculations can be generalized for other existing or planned experiments. 
We have scanned the parameter space consisting of the dark matter particle mass, the dark matter--electron interaction mediator mass, and the effective coupling strength, searching for any region of the parameter space that could potentially explain the DAMA  modulation signal. 
Below, we discuss the main findings and features of our analysis:
\begin{itemize}

\item We find that the modulation fraction of all events with energy deposition above $2$~keV in NaI are quite significant, reaching $\sim 50$\%, 
which could be useful for linking the DAMA modulation signal to electron recoil. This also allows one to tolerate higher 
levels of background in the unmodulated DAMA rate compared to the case of the nuclear recoil. 

\item The shape of the spectrum is necessarily very much enhanced for small values of $\Delta E$, and is a poor fit to the DAMA modulation 
spectrum. However, the overall modulated rate (averaged in the 2--6 keV interval) can be achieved with a very light (e.g., sub-MeV) mediators, and
the WIMP-electron coupling constants as small as $\alpha_\chi \sim (10^{-4}-10^{-3})\times \alpha$.

\item The inferred strength of the coupling $\alpha_\chi$ is in strong tension with known contraints on couplings of light mediators 
both to electrons and to dark matter, and generally requires extra fine tuning in several observables. 

\item Irrespective of this fine tuning, we were able to exclude the DAMA modulation expalanation via the electronic 
recoil using the results of the XENON10 and XENON100 experiments. It is important to note  
that the XENON10 and XENON100 constraints are complementary, in that they each ``favor'' opposite ends of the parameter space (with XENON100 favoring low $m_\chi$ and low $m_v$, and XENON10 favoring large $m_\chi$ and $m_v$).
Therefore, by combining the two sets of constraints, we can exclude the entire parameter space for electron-interacting WIMPs as the source of the DAMA annual modulation; see Fig.~\ref{fig:summary}.

\item We also note that for larger values of $\alpha_\chi$ (that would require even larger tunings of couplings $g_e$ and $g_\chi$) 
the effects of the WIMP slow-down by the earth material (not considered in this paper)
 may reduce fluxes and energies of WIMPs at the location of DAMA experiment, 
further shrinking the parameter space for the explanation of the annual modulation by the DM signal.

\end{itemize}

We consider that these limits are conservative.
For example, we made a number of generous assumptions relevant to the DAMA experiment (e.g., that their detectors were perfectly efficient), while making more conservative assumptions for the XENON100 and XENON10 cases (e.g., we calculated only lower bounds on the expected event rate for the XENON10 experiment).
Taking the DAMA spectrum into account, and including the higher-order processes in the XENON experiments would lead to significantly more stringent limits.
We also note, that our calculations are relatively impervious to systematic uncertainties, since they are based on ratios of calculations performed using the same method and codes (this is particularly true for the XENON100 case, which concerned the same energy range as the DAMA case).
Any DM parameters outside those considered directly in our analysis either cannot account for the DAMA modulation (as demonstrated in Fig.~\ref{fig:axreq}) or have been previously excluded from stellar bounds \cite{An2013}.

We would like to conclude by noting that as the XENON and LUX DM detection programs progress and scale up, one should expect even greater 
sensitivity to the electron recoil. For example, the anticipated background rates in XENON1T \cite{Xe1T2015} are up to two orders of magnitude 
lower than in XENON100, which will provide sensitivity to even smaller scattering cross sections, and eventually probe 
regions of parameter space $\{m_\chi, \alpha_\chi, m_v \}$ unconstrained from other sources.

%=======================================================================
%===============
\acknowledgments
The authors would like to thank 
J.~Berengut,
R.~Budnik,
A.~Derevianko,
G.~Gribakin,
R.~Lang,
M.~Schumann,
and 
I.~Yavin
for helpful discussions.
This work was supported by the Australian Research Council, the Perimeter Institute for Theoretical Physics, and NSF grant PHY-1506424.
BMR, VVF, and MP are grateful to the Mainz Institute for Theoretical Physics (MITP) for its hospitality and support.
BMR is grateful to the Perimeter Institute for Theoretical Physics, where part of this work was completed, for its hospitality and financial support.
MP gratefully acknowledges the support of the Gordon Godfrey fellowship and UNSW Australia. 
Research at the Perimeter Institute is supported by the Government of Canada through Industry Canada and by the Province of Ontario through the Ministry of Economic Development \& Innovation.

%=======================================================================
%=======================================================================
%========================                                ==================================
%=======================================================================
%=======================================================================

%===============
\appendix
%----------------------------------------------------------------------------------------------
\section{Methods for {\em ab initio} Relativistic Atomic Calculations}
\label{sec:methods}

The relativistic Dirac-Coulomb Hamiltonian is given 
\begin{equation}
\hat H=\sum_i\left[c\v{\a}_i\cdot\v{p}_i+m_ec^2(\g^0_i-1)-V^{\rm nuc}_i+\sum_{j<i}\frac{e^2}{r_{ij}}\right],
\label{eq:dirac}
\end{equation}
where, $\v{\a}=\g^0\v{\g}$ and $\g^0$ are Dirac matrices,
\v{p_i} is the relativistic (three-)momentum of the $i$th electron, $e=\abs{e}$ is the elementary charge, $r_{ij}\equiv\abs{\v{r}_i-\v{r}_j}$, and for large distances the nuclear potential is given by $V^{\rm nuc}_i\simeq{Ze^2/r_i}$.
Note that the Eigenvalues of the above Hamiltonian, defined via the equation
$\hat H\ket{n}= E_n\ket{n}$, do not include the electron mass-energy (for ease of comparison with nonrelativistic calculations).
The total relativistic energy is given by
$\widetilde E_n= E_n+m_ec^2$.

In the calculations, we use the Relativistic Hartree-Fock (HF) method, in which Eq.~(\ref{eq:dirac}) is replaced by the single-electron HF Hamiltonian:
\begin{equation}
\hat h^{\rm HF}= c\v{\a}\cdot\v{p}+m_ec^2(\g^0-1)-V^{\rm nuc}+U^{\rm HF}.
\label{eq:Hhf}
\end{equation}
We use a Fermi-type distribution for the nuclear potential,
\begin{equation}
\rho_Z(r)=\frac{Z\rho_0}{1+\Exp{(r-c)/a}},
\label{eq:fermi}
\end{equation}
where $t=a(4\ln3)$ is the skin-thickness and $c$ is the half-density radius, see, e.g., Ref.~\cite{Fricke1995}, and $\rho_0$ is found from the normalization condition $\int \rho(r)\d^3r=1$.
This is important since the effects considered here depend strongly on the form of the wave functions at short distances.
We express the four-component single-electron orbitals (employing the Dirac basis) in the form
\begin{equation}
\psi_{n\k m}(\v{r})=\frac{1}{r}
{\twocomp{f_{n\k}(r)\Omega_{\k m}(\v{n})}{i\a g_{n\k}(r)\Omega_{-\k, m}(\v{n})}},
\label{eq:psi}
\end{equation}
where $f_{n\k}$ and $g_{n\k}$ are the large and small components of the Dirac wave function, respectively, 
$\Omega_{\k m}$
 is a two-component spherical spinor, $\v{n}\equiv\v{r}/r$, and $\a\approx1/137$ is the fine-structure constant.
The continuum-state wave functions, $\psi_{\e \k m}$, take the same form and, for a state with energy $\e$, we denote the large and small Dirac components as $f_{\e \k}$ and $g_{\e \k}$, respectively.
The atomic wave functions are then made of the orbitals $\psi_{n\k m}$, which are found for each of the $N$ states in the core by solving the Dirac equation
\begin{equation}
\hat H^{\rm HF} \psi_{n\k}=\en_{n\k} \psi_{n\k},
\label{eq:hf}
\end{equation}
where $\en_{n\k}$ is the single-electron Hartree-Fock energy corresponding to the orbital $\psi_{n\k}$. 

The Hartree-Fock potential is given by the sum of the local (direct) and nonlocal (exchange) parts of the interaction, {${U^{\rm HF}=U^{\rm dir}+U^{\rm exch}}$}, with
\begin{equation}
\begin{aligned}
U^{\rm dir}\psi_a(\v{r})
&=e^2\sum_{n\neq a}^N\int\frac{\psi_n^\dagger(\v{r}')\psi_n(\v{r}')}
{\abs{\v{r}-\v{r}'}}\d^3r'\psi_a(\v{r})
\\
U^{\rm exch}\psi_a(\v{r})&=
-e^2\sum_{n\neq a}^N\int\frac{\psi_n^\dagger(\v{r}')\psi_a(\v{r}')}
{\abs{\v{r}-\v{r}'}}\d^3r'\psi_n(\v{r}),
\label{eq:Uhf}
\end{aligned}
\end{equation}
where the indices $n$ and $a$ denote core orbitals.
The equations  (\ref{eq:hf}) and (\ref{eq:Uhf}) are solved iteratively until an acceptable level of convergence has been reached.
(To start the iterative procedure, an initial approximation for the potential is required; for this we use a Thomas-Fermi potential or a simple parametric potential.)
Then, the HF potential is kept constant and the wave functions for the continuum states are calculated for a specified energy in this ``frozen core'' potential.

To calculate the matrix elements for the atomic kernel, defined in Eq.~(\ref{eq:AK}), we expand the exponential operator in terms of spherical harmonics and spherical Bessel functions, and use the Wigner-Eckart theorem and orthogonality conditions to perform the angular integrations and the summations over the magnetic quantum numbers analytically.
The full formulas for the atomic kernel are given in Appendix~\ref{sec:appendix}.

The atomic kernel, dominated by $s$ states for both the continuum and bound electrons, is proportional to the radial integral
\[R=\int \psi_{ns}(r)\psi_{\e s}(r)\frac{\sin(qr)}{qr}r^2\d r.\]
Note that, in general, $qr$ is not small, and the integrand oscillates rapidly (typical values of $q$ are on the order of $10^{3}$--$10^{4}$ au, with $r\sim1$ au).
Therefore, in doing numerical calculations on a grid, where the above integral becomes,
\[R\to \sum_i \psi_{ns}(r[i])\psi_{\e s}(r[i])\frac{\sin(qr[i])}{qr[i]}r[i]^2\delta r,\]
care is needed.
At high $q$, where the atomic ionization can occur, the integral is dominated by low $r$ contributions.
We must ensure that the separations in the grid spacings, $\delta r$, are significantly smaller than the width of the oscillations for all relevant values of $q$: $\delta r \ll {\pi}/{q_{\rm max}}$.
In other words, we can safely integrate over $q$ up to a value of $q^{\rm max}\simeq \pi/\delta r$.
We use a non-uniform grid, which has exponentially more points close to the nucleus than far away, to ensure sufficient numerical accuracy for the important low-$r$ part of the wave functions.
The non-uniform grid $r[i]$ (for $i=1,2,...N$, where $N$ is the number of grid points), is written as a function of a uniformly spaced grid, $s[i]$, with separations $s[i+1]-s[i]=h$,
then $\delta r = \frac{\d r}{\d s} h$. There are various ways to do this; we chose a simple parametrization in which
$\frac{\d r}{\d s}={r[i]}/{(b+r[i])},$ and take $b=4$, which means the grid is roughly exponential for $r\lesssim4$ {au}, and linear when $r\gtrsim4$ {au}.
We have checked that in all cases of interest, the $q$ integral converges well within the region of stability.
There is also an integral over $q$ (and $\Delta E$) in determining the cross-section; these integrals are of relatively smooth functions, and are much simpler. 
Convergence and stability are easily checked by varying the grid density and cut-offs, and we have checked that they are attained in all cases.

We note that the methods we use are accurate for deep atomic shells, but not necessarily for the valence electrons. 
This is because we are performing calculations for atoms, whereas in the detectors these atoms form molecules or crystals, which will affect the outer electron wave functions.
{\sl Ab initio} relativistic solid state and molecular calculations can also be done, but this is outside the scope of the current work (though we note that for the lower energy depositions involved in the ionization of the outer shells, the relativistic effects are not so important).
The calculations for xenon, a noble gas, are accurate for all shells.
Note that we only consider low-energy ionization signals (where the valence electrons are important) for xenon, therefore we do not consider any case for which our calculations are not accurate.

%=========================
\section{Angular Decomposition and Evaluation of the Atomic Kernel}
\label{sec:appendix}

To evaluate the sum of matrix elements in Eq.~(\ref{eq:AK}), we first write $\Exp{i\v{q}\cdot\v{r}}=\sum_{L=0}^\infty\sum_{M=-L}^L T_{LM}$, where 
\begin{equation}
T_{LM}=4\pi(i)^L j_L(qr)Y_{LM}(\theta_r,\phi_r)Y^*_{LM}(\theta_q,\phi_q)
\end{equation}
is an irreducible (spherical) tensor operator, with $Y_{LM}$ the spherical harmonics, and $j_L$ the spherical Bessel functions.
Then, using the standard angular momentum summation rules (see, e.g.,~\cite{Varshalovich1988}), we express Eq.~(\ref{eq:AK}) as
\begin{align}
K_{n\k}(\Delta E,q)
&=\sum_{\k'}\sum_{m,m'}\sum_{L,M}
\abs{\bra{\e\k'm'}T_{LM}\ket{n\k m}}^2
\notag\\
&=\sum_{\k'}\sum_L\abs{\bra{\e\k'}|{T_{L}}|\ket{n\k}}^2x(n,j),
\label{eq:AKred}
\end{align}
where $x(n,j)$ is the fractional occupation number for a given shell (for the shells of interest here, $x=1$, however, $x<1$ for open shells).
The factor $\bra{p\k'}|{T_{L}}|\ket{n\k}$ is known as the reduced matrix element, and is defined via the Wigner-Eckart theorem:
\begin{align}
\bra{\e\k'm'}&T_{LM}\ket{n\k m}
=\notag\\&
(-1)^{j'-m'}
\threej{j'}{L}{j}{-m'}{M}{m}
\bra{\e\k'}|{T_{L}}|\ket{n\k},
\label{eq:WE}
\end{align}
where $\threej{j'}{L}{j}{-m'}{M}{m}$ is a $3j$~symbol.
Importantly, the reduced matrix elements are independent of the quantum numbers $m$ and $m'$, as well as the index $M$.

Therefore, the atomic kernel is reduced to a summation over reduced matrix elements, which are found from Eq.~(\ref{eq:WE}) with, e.g., $M=0$ and $m=m'=1/2$:
\begin{align}
\abs{\bra{\e\k'}|{T_{L}}|\ket{n\k}}^2
&=\threej{j'}{L}{j}{-\frac{1}{2}}{0}{\frac{1}{2}}^{-2}\abs{\bra{\e\k'\tfrac{1}{2}}T_{L0}\ket{n\k \tfrac{1}{2}}}^2
\notag\\&
={C_{\k\k'}^L}\left(R_f^2+2\a^2R_fR_g+\a^4R_g^2\right),
\end{align} 
where $R_f$ and $R_g$ are the radial integrals,
\begin{align}
R_f&=\int f_{\e\k'}f_{n\k}j_L(qr)\d r\\
R_g&=\int g_{\e\k'}g_{n\k}j_L(qr)\d r,
\end{align}
and the angular coefficient is
\begin{widetext}
\begin{align}
{C_{\k\k'}^L} = 
\frac{1}{4}(-1)^{j+j'-l-l'}&(2L+1)\threej{l'}{l}{L}{0}{0}{0}^2
\threej{j'}{L}{j}{-\frac{1}{2}}{0}{\frac{1}{2}}^{-2}
\Bigg[
(-1)^{j+j'-l-l'}(2j+1)(2j'+1)\threej{l'}{l}{L}{0}{0}{0}^2
\notag\\&
+8\sqrt{l'(l'+1)l(l+1)}\threej{l'}{l}{L}{0}{0}{0}\threej{l'}{l}{L}{-1}{1}{0}
-4(\k'+1)(\k+1)\threej{l'}{l}{L}{-1}{1}{0}^2
\Bigg].
\end{align}
\end{widetext}
For $s_{1/2}$ and $p_{1/2}$ states, this reduces simply to
${C} = 2$ (with $L=0$ for $\k=\k'=\pm1$, and $L=1$ for $\k=-\k'=\pm1$). 
Note that, since the reduced matrix elements do not depend on $M$, $m$, or $m'$, we can choose any values for these indices that leave the $3j$~symbol in (\ref{eq:WE}) nonzero, however the minimal values are typically the simplest to compute.

Similarly, if instead we consider a scalar, pseudoscalar, or (spin-independent) pseudovector electron coupling, the relevant electron operator is replaced with $T_{LM}\g^0$, $T_{LM}\g^0\g_5$, or $T_{LM}\g_5$, respectively.
Then the atomic structure factors can be expressed as
\begin{align}
\abs{\bra{\e\k'}|{T_{L}\g^0}|\ket{n\k}}^2
& ={C_{\k\k'}^L}\left(R_f^2-2\a^2R_fR_g+\a^4R_g^2\right),
\\
\label{eq:rint-PS}
\abs{\bra{\e\k'}|{T_{L}\g^0\g_5}|\ket{n\k}}^2
& ={D_{\k\k'}^L}\a^2\left(R_{fg}^2+2R_{fg}R_{gf}+R_{gf}^2\right),
\\
\label{eq:rint-PV}
\abs{\bra{\e\k'}|{T_{L}\g_5}|\ket{n\k}}^2
& ={D_{\k\k'}^L}\a^2\left(R_{fg}^2-2R_{fg}R_{gf}+R_{gf}^2\right),
\end{align} 
where the radial integrals are
\begin{align}
R_{fg}&=\int f_{\e\k'}g_{n\k}j_L(qr)\d r\\
R_{gf}&=\int g_{\e\k'}f_{n\k}j_L(qr)\d r.
\end{align}
The angular coefficient $D$ is related to $C$ via the transformation $\k\to\tilde\k=-\k$ and $l\to\tilde l=\abs{\tilde\k+1/2}-1/2$ ($\k'$ and $l'$ remain unchanged). For $s_{1/2}$ and $p_{1/2}$ states we also have $D=2$.
The calculations for the pseudovector case should be approached with particular care due to the possibility of large cancellations in the radial integrals, see Eq.~(\ref{eq:rint-PV}).
Shown here is the temporal (spin independent, zero component) contribution to the pseudovector coupling case only.
To lowest-order, the spin-dependent components for the pseudovector case behave like the scalar case or the temporal component of the vector case.

\begin{figure*}
	\begin{center}
		\includegraphics[width=0.49\textwidth]{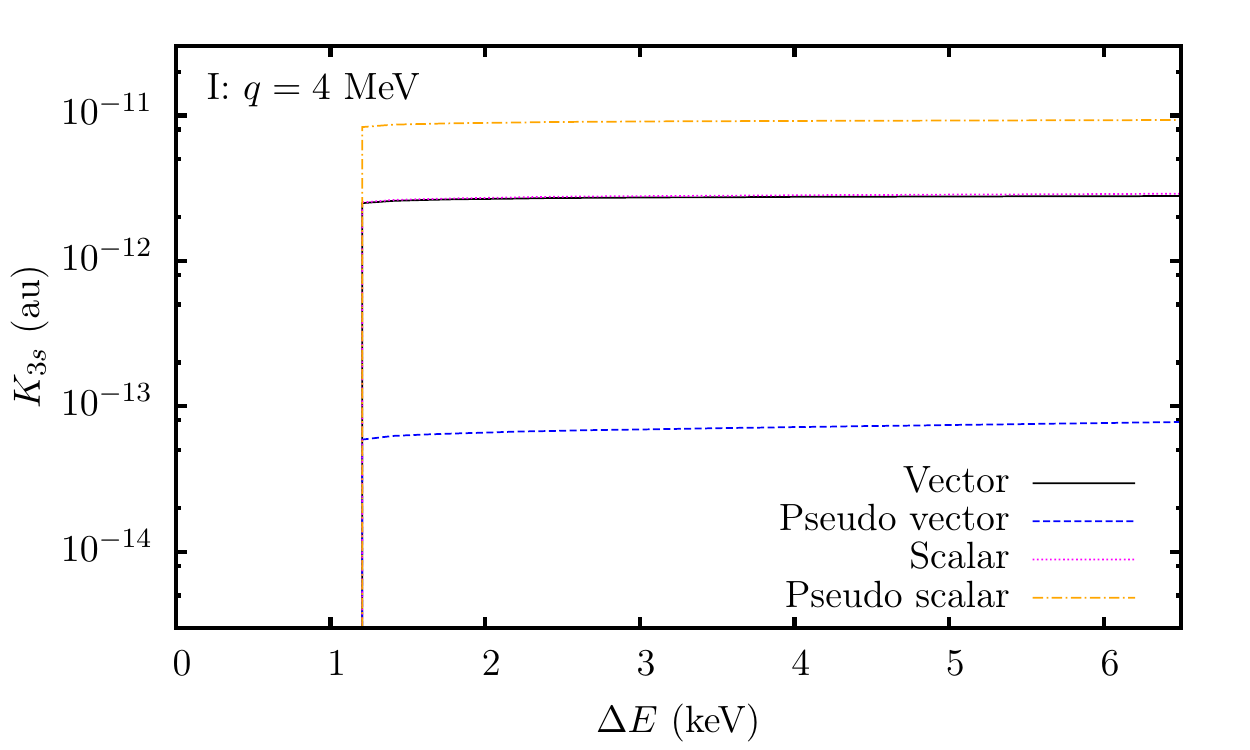}
		\includegraphics[width=0.49\textwidth]{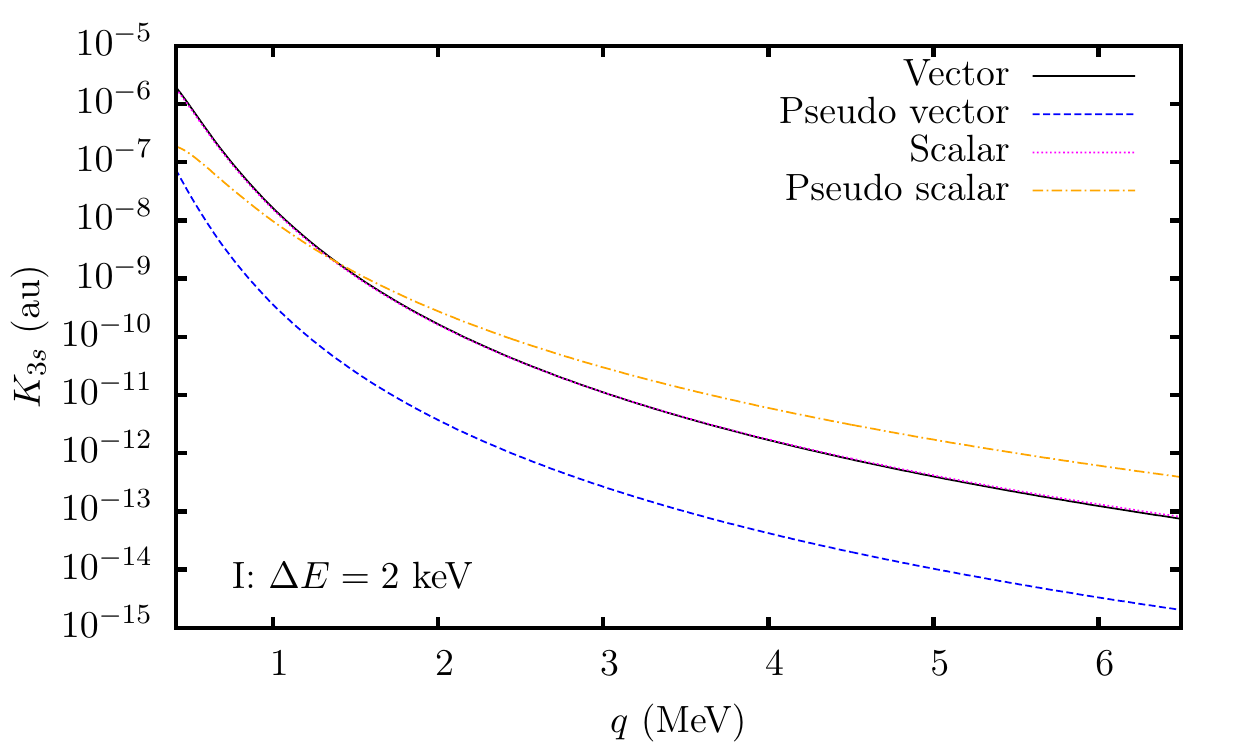}
		\caption{Comparison of different Lorentz structures for the $3s$ core-state contribution to the atomic kernel for I as a function of the energy deposition, $\Delta E$ (with $q\simeq4$ MeV), and of the momentum transfer, $q$ (with $\Delta E\simeq2$~keV).
		The pseudoscalar case gives a large effect (at higher $q$), since in this case the  radial integrals include a contribution from initial and final $s$-states with $L=1$ ($L=0$ for the $s$-$s$ contribution to the vector and scalar cases); the pseudovector case (temporal contribution) gives by far the smallest contribution due to very large cancellations in the radial integrals, see Appendix~\ref{sec:appendix}.
The pseudovector case here includes only the temporal part of the interaction; to lowest-order, the spatial components for the pseudovector case behave like the vector/scalar case.
}
		\label{fig:I-Lorentz}
	\end{center}
\end{figure*}

In Fig.~\ref{fig:I-Lorentz}, we present calculations of the atomic structure factors for the vector, scalar, pseudovector, and pseudoscalar electron interactions. 
It is evident here that the electron pseudoscalar interaction gives the largest result (for very high momentum transfer), while the temporal part of the pseudovector case gives by far the smallest.
The largeness of the pseudoscalar case can be understood as follows.
The Factor $(Z\a)^2$ in the numerator of Eq.~(\ref{eq:rint-rel2}) comes from the expansion of the gamma function in the denominator of Eq.~(\ref{eq:rint-rel1}), which approaches infinity as $\g$ approaches unity for $L=0$.
For the case where $L=1$, however, this denominator is nonzero even in the $Z\a\to0$ limit. 
Considering an initial (bound) $s$-state, there appears a contribution for the pseudoscalar and pseudovector cases that comes from the final $s_{1/2}$ continuum state with $L=1$. 
In this case, the $(Z\a)^2$ suppression from Eq.~(\ref{eq:rint-rel2}) is removed, instead it is replaced by just a $\sim Z\a$ suppression which comes from the small Dirac component that appears in the radial integral for the pseudoscalar case (\ref{eq:rint-PS}). 
There is another enhancement by a factor of $\sim4$ due to the few roughly equal terms in Eq.~(\ref{eq:rint-PS}).
In the pseudovector case, on the other hand, this situation does not lead to an enhancement. Instead there is huge suppression, which comes from the very large cancellation of terms in Eq.~(\ref{eq:rint-PV}).
This means that calculations of the electron structure for the pseudovector case are very susceptible to numerical instabilities and must be treated with great care (if high accuracy is to be achieved).

~\\
%==========================================
\section{Scaling of the analytic results}
\label{sec:scaling}

\begin{figure*}
	\begin{center}
		\includegraphics[width=0.49\textwidth]{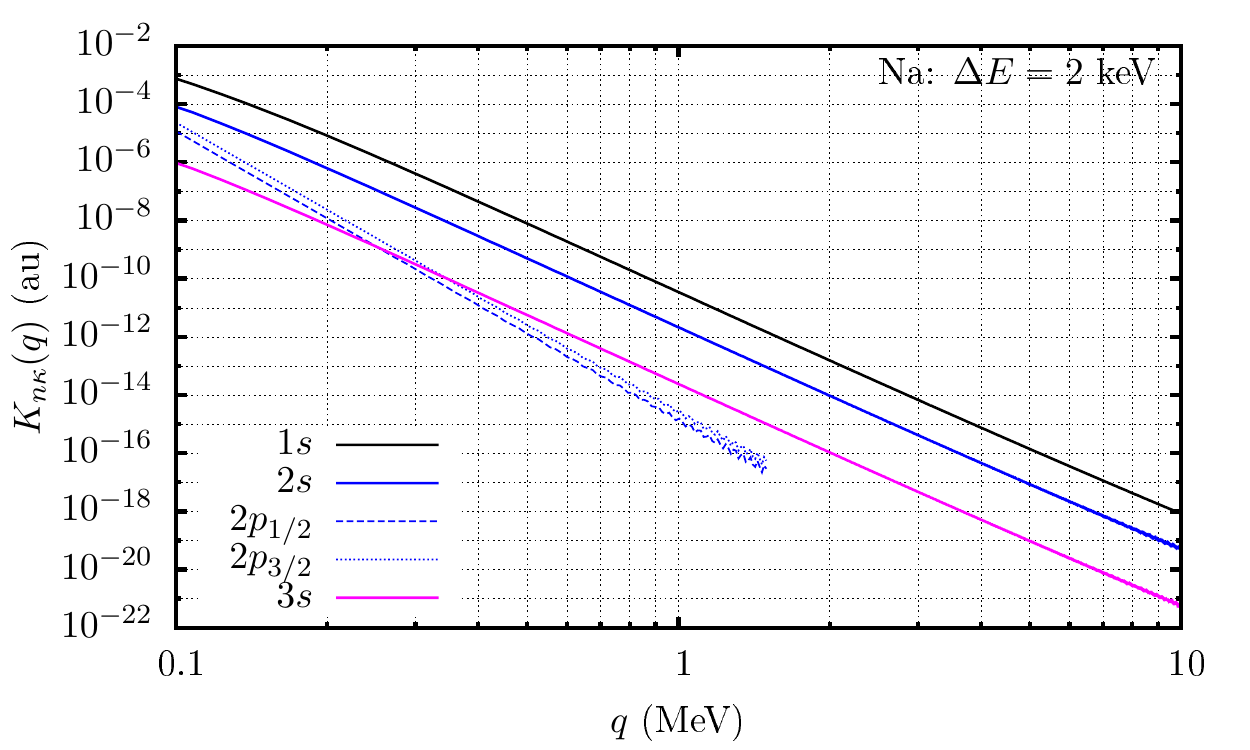}
		\includegraphics[width=0.49\textwidth]{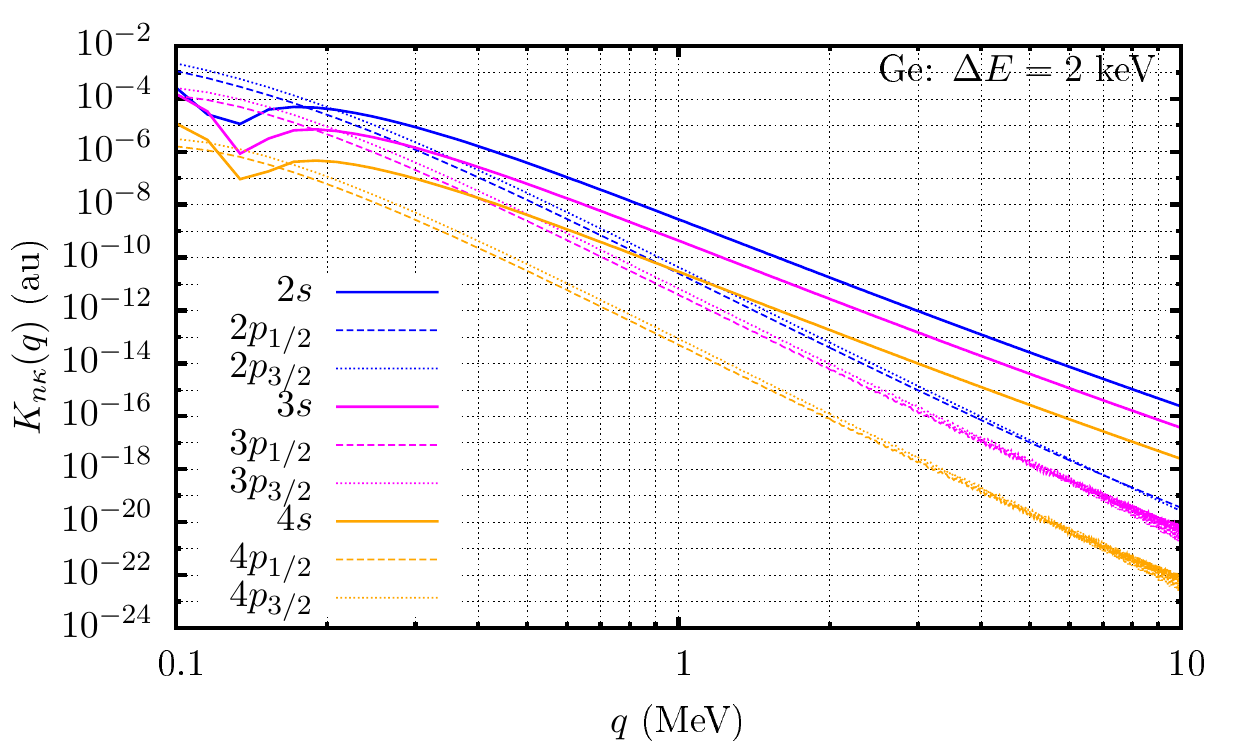}
		\includegraphics[width=0.49\textwidth]{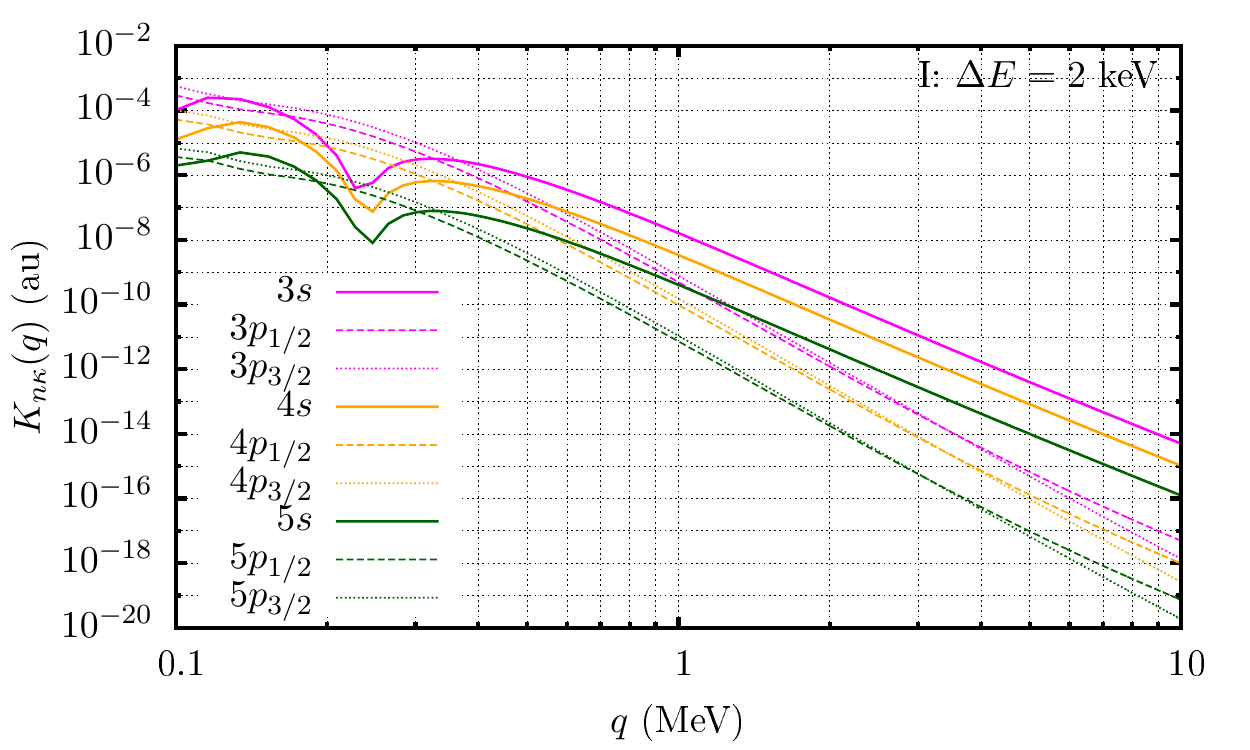}
		\includegraphics[width=0.49\textwidth]{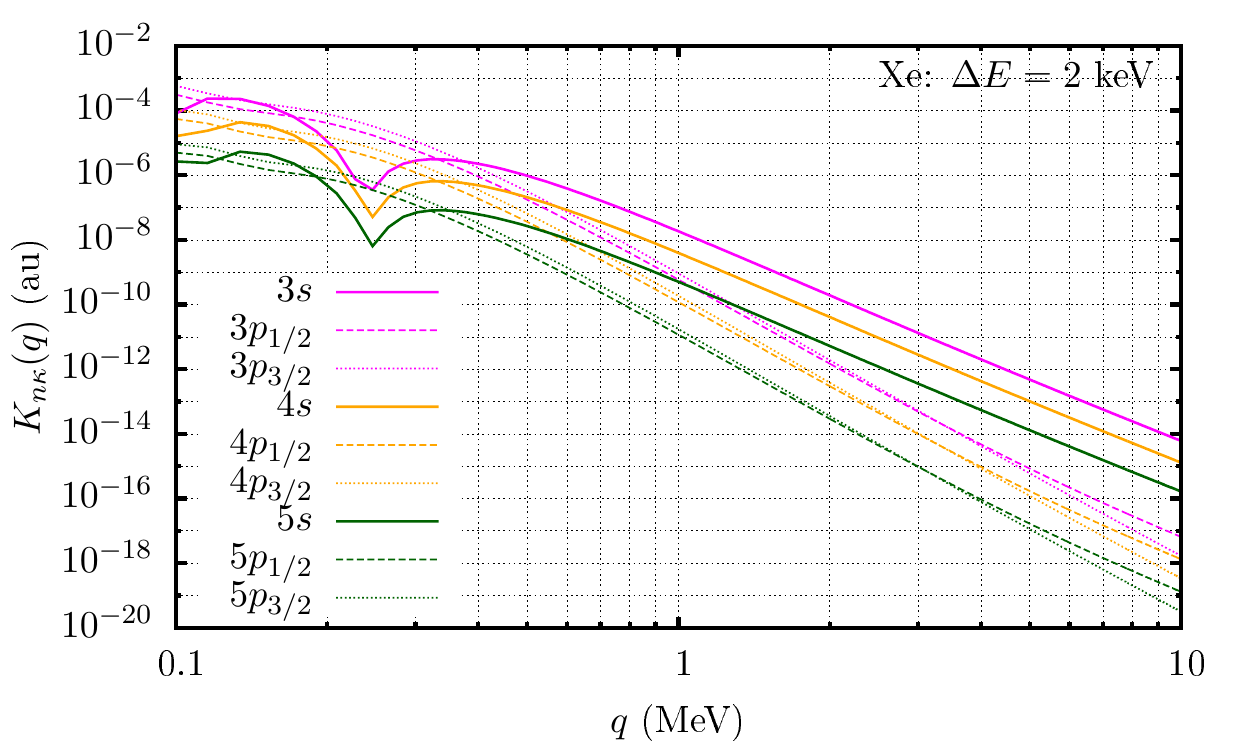}
		\caption{Plots of the atomic kernel (\ref{eq:AK}) for several dominating core states of Na, Ge, I, and Xe, as a function of momentum transfer $q$, for a fixed energy deposition $\Delta E=2.0$ keV.}
		\label{fig:AKq-NGIX}
	\end{center}
\end{figure*}

\begin{figure}
	\begin{center}
		\includegraphics[width=0.49\textwidth]{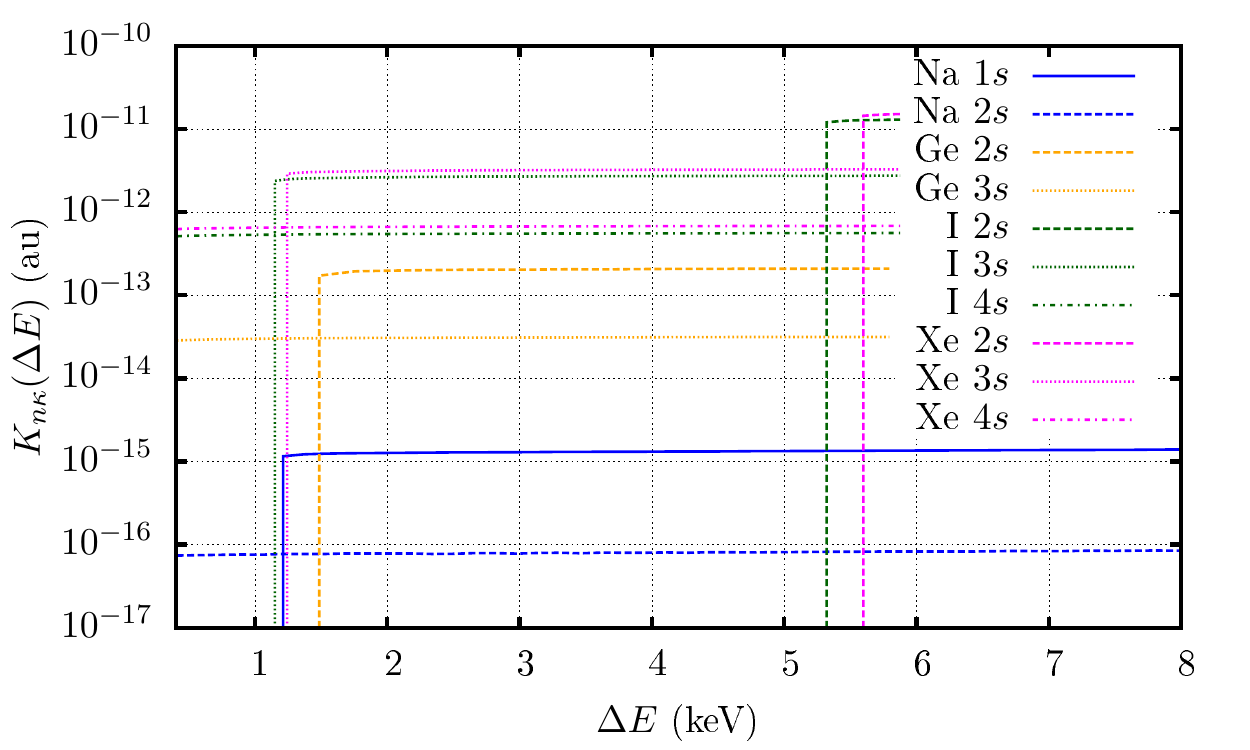}
		\caption{Plots of the atomic kernel (\ref{eq:AK}) for a few dominating core states of Na, Ge, I, and Xe, as a function of energy deposition $\Delta E$, for a fixed momentum transfer $q=3.73$ MeV.}
		\label{fig:AKdE}
	\end{center}
\end{figure}

In Figs.~\ref{fig:AKq-NGIX} and \ref{fig:AKdE} we plot the contribution of several dominating core states to the atomic kernels (\ref{eq:AK}) for Na, Ge, I, and Xe.
Several orders-of-magnitude enhancement of the Xe/I atomic kernel compared to that of Na is observed, which is expected from the high-power of the $Z$-scaling of the electron matrix element, and the larger relativistic factor.
Using the simple expression given in Eq.~(\ref{eq:rint-rel2}) to formulate the momentum transfer dependence of the atomic kernel for high values of $q$, one may use simple $Z$-dependent scaling factors to reproduce our full-scale calculations.
For values below $q=1$ MeV, the non-relativistic calculations using screened hydrogen-like wave functions are sufficient, though it should be noted that the usual notion of the effective nuclear charge
$\widetilde Z_{nl}=n\sqrt{2I_{nl}}$ is not valid. This value is chosen to reproduce the correct energies, and gives a reasonable approximation of the wave functions at medium distances.
The ionization cross section, however, is dominated by the wave function at very small distances.
Instead, the correct value for $\widetilde Z_{nl}$ should be chosen to reproduce the curves in Fig.~\ref{fig:AKq-NGIX}, and will be fairly close to the true $Z$.

Note that the cross section contains energy dependent terms, meaning the atomic kernel cannot be summed (over the bound atomic states) independently; see Eq.~(\ref{eq:dsv-de}). Each partial contribution must be calculated individually, and then summed over (though, there is typically a single dominating contribution).

%=============================================
\bibliography{dama}

\end{document}